\documentclass[11pt,final,peerreview]{IEEEtran}
\usepackage [cmex10]{amsmath}
\usepackage{epsfig,psfig}
\usepackage{graphicx}
\usepackage{cite}
\usepackage{amssymb}
\usepackage{setspace,ulem}
\usepackage{graphics}
\usepackage{algorithm}
\usepackage{algpseudocode}

\usepackage{color}   
\usepackage{graphicx}
\usepackage[english]{babel}

\newcommand{\cjbb}[1]{#1}
\newcommand{\Sphi}{{\bf S_{\Phi}}}
\newcommand{\Xphi}{{\bf X_{\Phi}}}
\newcommand{\Wphi}{{\bf W_{\Phi}}}

\begin{document}

\title{Sparsity and adaptivity for the blind separation of partially correlated sources}

\author{J. Bobin, J. Rapin, A. Larue and J-L Starck \thanks{J. Bobin and J.-L. Starck are with CEA, IRFU, Service d'Astrophysique, 91191 Gif-sur-Yvette Cedex, France.}% <-this % stops a space
\thanks{A. Larue and J. Rapin are with CEA, LIST, 91191 Gif-sur-Yvette Cedex, France.}}

%\address[1]{DSM /IRFU/SEDI-Sap, CEA/Saclay, F-91191 Gif-sur-Yvette, France}

\maketitle

\begin{abstract}
Blind source separation (BSS) is a very popular technique to analyze multichannel data. In this context, the data are modeled as the linear combination of sources to be retrieved. For that purpose, standard BSS methods all rely on some discrimination principle, whether it is statistical independence or morphological diversity, to distinguish between the sources. However, dealing with real-world data reveals that such assumptions are rarely valid in practice: the signals of interest are more likely partially correlated, which generally hampers the performances of standard BSS methods. In this article, we introduce a novel sparsity-enforcing BSS method coined Adaptive Morphological Component Analysis (AMCA), which is designed to retrieve sparse and partially correlated sources. More precisely, it makes profit of an adaptive re-weighting scheme to favor/penalize samples based on their level of correlation. Extensive numerical experiments have been carried out which show that the proposed method is robust to the partial correlation of sources while standard BSS techniques fail. The AMCA algorithm is evaluated in the field of astrophysics for the separation of physical components from microwave data.\\
EDICS: MLR-SSEP
\end{abstract}

%========================================================
% Keywords


%========================================================
% Keywords

%------------------------------------------------------------------------------------
\section{Introduction}\label{sec:intro}
%------------------------------------------------------------------------------------

With the rapid development of multi-wavelength or multi-sensor instruments, it is crucial to design analysis tools to efficiently extract meaningful information. This is especially the case in the field of astrophysics where the development of dedicated component separation methods has been recently of prime importance for the processing of microwave surveys such as NASA-WMAP or ESA-Planck  \cite{PR1_compsep,WMAP9_LGMCA,Bobin_13a}. For that purpose, Blind source separation (BSS) is a particularly well-suited framework to analyze multichannel data. In this context and following the so-called linear mixture model, each observation or channel $\{x_i\}_{i=1,\cdots,m}$ is modeled as the linear combination of $n \leq m$ sources $\{s_j\}_{j=1,\cdots,n}$:
$$
\forall i=1,\cdots,m; \quad x_i = \sum_{j=1}^n a_{ij} s_j + z_i,
$$
where $a_{ij}$ quantifies the contribution of the $j$-th source in the observation $x_i$. Each datum $\{x_i\}_{i=1,\cdots,m}$ and source $\{s_j\}_{j=1,\cdots,n}$ are assumed to have $T$ entries. The term $z_i$ stands for additive noise or model imperfections. The well-known linear mixture model is more conveniently recast in the following matrix form:
\begin{equation}
\label{eq:mmodel}
{\bf X} = {\bf AS} + {\bf Z},
\end{equation}
where $\bf X$ is a $m \times T$ matrix the rows of which are composed by the observations, $\bf A$ is the so-called mixing matrix and $\bf S$ is the $n \times T$ matrix which contains the sources. The goal of blind source separation is to blindly estimate both the mixing matrix $\bf A$ and the sources $\bf S$ from the knowledge of $\bf X$ only. If one neglects the noise $\bf Z$, there exists an infinite number of couples $({\bf A},{\bf S})$ which satisfy ${\bf X} = {\bf AS}$. This is an ill-posed problem which requires using prior information about either the mixing matrix or the sources to be recovered. For that purpose, classical approaches rely on information which promote some discriminant information or diversity between the sources. Since the beginning of the 90s, Independent Component Analysis or ICA (see and \cite{JutBook,hyvarinen2001ica,Choi05} references therein) has taken the lion share. In the framework of ICA, the sources are assumed to be statistically independent. Most methods proposed so far in this context generally differ in the way they promote and measure independence. In the remaining of this paper, we will primarily focus on the separation of sparse sources.

\subsection{Sparse blind sources separation}
Sparkled by advances in harmonic analysis and applied mathematics, the sparse modeling of signals \cite{EladBook} has attracted a lot of interest and has proved to be effective at solving a very large range of inverse problems: denoising, deconvolution \cite{StarckBook10}, compressive sensing reconstruction \cite{CandesReview2}, inpainting \cite{starck:jalal06} to only name a few. In this context, it is assumed that the each source $\{s_j\}_{j=1,\cdots,n}$ can be sparsely decomposed in some basis, waveform dictionary or signal representation $\bf \Phi$:
\begin{equation}
s_j = \alpha_j {\bf \Phi}.
\end{equation}
The sparsity of $s_j$ in the signal representation $\bf \Phi$ is customarily described with two different models:
\begin{itemize}
\item{\it Exact sparsity model:} each source $\{s_j\}_{j=1,\cdots,n}$ is assumed to have a small number of nonzero coefficients in $\bf \Phi$. The set of nonzero coefficients, or support of the source $j$, is denoted by $\Omega_j$.\\
\item{\it Approximate sparsity model:} natural signals do not generally have an exactly sparse distribution in $\bf \Phi$. They rather exhibit an approximately sparse distribution: most of the entries of the expansion coefficient vectors $\{\alpha_j\}_{j=1,\cdots,n}$ are zero or with negligible amplitudes and only a few take significant amplitudes.\\
\end{itemize}
Most natural signals exhibit such sparsity property in adequately chosen signal representations $\bf \Phi$. Examples of standard signal representations include the wavelets \cite{ima:mallat98,SFM:unde}, the curvelets \cite{starck:fadili08} or even adaptively learned signal representations \cite{ksvd:elad}. The sparsity property is particularly appealing since few significant expansion coefficients of $s$ in $\bf \Phi$ efficiently or compressively encode its information content. Sparsity has been exploited in recent sparse blind source separation methods (see \cite{ica:zibu_relnewton,ica:zibu_bronst,ica:cicho06_2,starck:bobin07}).\\

In the framework of ICA, sparse and independent sources have been studied in \cite{ica:zibu_relnewton,koldo06}. The RNA method \cite{ica:zibu_relnewton} looks for sources by enforcing their sparsity in an orthonormal basis. The Efficient FastICA (EFICA) algorithm described in \cite{koldo06} is an improvement of the acclaimed FastICA algorithm \cite{Hyvarinen:1999it} which is specifically adapted to retrieve sources with generalized Gaussian distribution; this includes sparse sources.\\
The compressibility of the signal allowed by sparse representations has been exploited in the Sparse Component Analysis (SCA) algorithm introduced in \cite{GribonSCA06,Lesage06a,ica:cicho06_2}. In the framework of SCA, the sources are assumed to have nearly or exactly disjoint supports $\Omega_j$: for each source, a few entries are nonzero and each active entry is active in only a single source. Relaxing this assumption has been proposed assuming that there exist entries where only one source is active. Assuming that the sparsity patterns of the sources are fully or partially disjoint has inspired several sparse BSS techniques for the processing of time series and audio signals \cite{DUET00,Yilmaz04,Saab05,Abrard05}.\\
However these assumptions are seldom verified with more complex data, especially in imaging : i) signals or images are generally not exactly but rather approximately sparse in $\bf \Phi$, thus entailing that there is no sample where the sources vanish, ii) there is generally no sample where each source is active alone. These assumptions have been further relaxed to approximately sparse sources in \cite{starck:bobin07} with the introduction of morphological diversity (MD). This concept rather states that sources with different morphologies are unlikely to have similar large amplitude coefficients in $\bf \Phi$. For instance, sparse and independently distributed sources are very likely to verify the MD assumption. Since high-amplitude entries of the sources in $\bf \Phi$ encode the most salient morphological features of the sources, sources which do not share similar morphologies will not therefore activate similar expansion coefficients in $\bf \Phi$. Building on the concept of morphological diversity, Generalized Morphological Component Analysis (GMCA) has been showed to be an effective BSS method \cite{starck:bobin07,bobin08_aiep}.\\
\cjbb{Recent developments in sparse BSS have mainly emphasize on solving non-negative matrix factorization problems with sparsity constraints on the sources \cite{Kim_12_GroupSparsityin,Gillis_12_SparseandUnique,Gillis_SemiNMF14,Rapin_13_SparseandNon,Rapin_14_AnalysisandSynthesis}. In the present paper, we focus on the general setting where the mixing matrix and the sources are not necessarily non-negative.}\\

\subsection{Real-world data are often partially correlated}

\begin{figure}[htb]
\centerline{\includegraphics [scale=0.25]{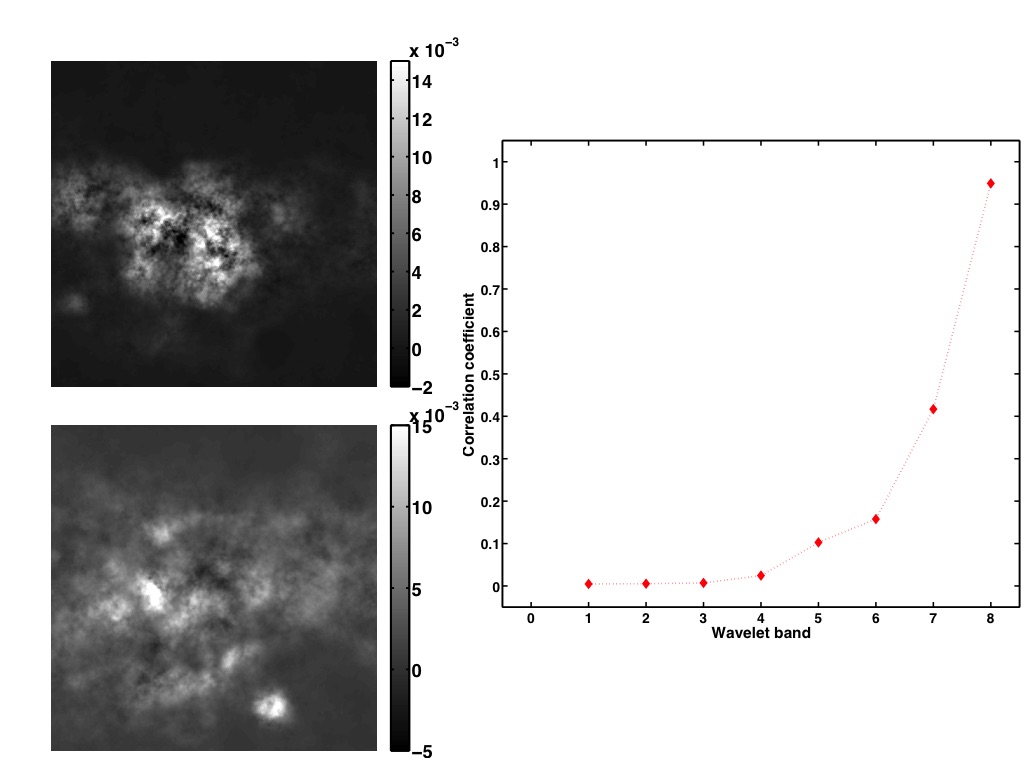}}
\caption{Left : samples of partially correlated sources in astronomical imaging: simulations free-free and spinning dust emissions as observed by the WMAP probe. Right : correlation coefficient of astrophysical components per wavelet scale.}
\label{fig:astro_corr}
\end{figure}

Dealing with real-life signals reveals that the actual sources are seldom perfectly uncorrelated: neither statistical independence nor morphological diversity are valid assumptions. Most classical source separation methods may not work properly for dependent sources and would likely yield erroneous results.\\
In the field of spectroscopic data processing, the sources of interest are often partially correlated. In this particular field of research, the development of dedicated source separation techniques has already been considered assuming strong assumptions about the sources to be retrieved \cite{Naanaa05,Sun11}. In \cite{Naanaa05}, the separation of partially correlated signals is made possible by assuming that for each source there exists at least one non-zero entry that vanishes for the other sources; this is obviously reminiscent of the disjoint source support introduced in the framework of SCA. \cjbb{Together with the non-negativity of the sources and the mixing matrix, the fact that the sources are assumed to have partially disjoint supports dramatically simplifies the separation task. Indeed, in such a setting, each column of the matrix are colinear with some of the columns of the data $\bf X$. Thanks to the non-negativity assumption, the data are positive linear combinations of positive sources. From a geometrical point of view, the columns of the mixing matrix $\bf A$ are the vertices of the conical hull generated by the columns of $\bf X$ (see \cite{Donoho_03_Whendoesnon,Gillis_WhyHow14} for more precisions). This geometrical property has been exploited to design algorithms that seek the simplex with minimal volume that encloses the data samples \cite{MVES}. This framework has been showed to be well adapted to the blind separation of very sparse and non-negative sources. This assumption has been somehow relaxed to model sources that are composed of sparse spikes and a non-negative baseline in \cite{Sun11}. In this case, the sources are modeled as a combination of two components: i) a sparse spikes train and ii) a smooth continuum, where the sparse spike component verify the source support assumption made in \cite{Naanaa05}.}\\
\cjbb{Partially correlated sources are often found in other seemingly different scientific domains such as astrophysics. As an illustration, the left panels of Figure \ref{fig:astro_corr} show simulations of two components which originate from distinct astrophysical phenomena. From a mathematical perspective, these two components are characterized by rather different columns of the mixing matrix (in physics, this translates into different electromagnetic spectra), and by somehow different spatial distributions. As a consequence, they can be physically and mathematically considered as sources. These components are generated by different physical mechanisms ({\it e.g.} synchrotron emission, Bremsstrahlung, thermal emission, etc.) and are therefore considered as distinct components or sources by astrophysicists. However these signals depend on common physical parameters ({\it e.g.} gas density, temperature, etc.) which may make them partially correlated. We refer the interested reader to \cite{Bouchet99,PSM12} for more details about the astrophysics in the microwave wavelengths.}\\
The right panel of Figure \ref{fig:astro_corr} reports the values taken by the correlation coefficients between these two astrophysical components in the wavelet domain. This particular example also highlights a second origin for signal partial correlation: chance-correlations. Indeed, in the finite sample case, single realizations of theoretically independent or uncorrelated sources are likely to have non-zero correlation. In Figure \ref{fig:astro_corr}, the number of available samples decreases at low-frequency, which plays a non-negligible role in the increase of the correlation coefficients at large-scale.\\
\cjbb{In the general setting, such as in imaging science, the limitations of standard methods are the following:
	\begin{itemize}
		\item{Sparsity:} standard methods assume that the sources are sparse and non-negative in the sample domain. However, it is more generally appropriate to model the sparsity of the sources in some transformed domain $\bf \Phi$, where non-negativity does not hold.\\
		\item{Non-negativity:} the non-negativity of both the sources and the mixing matrix, which is a valid assumption in a wide range of applications, does not hold in some applications (for instance in astrophysics, see \cite{Bobin_07_SZandCMB}).\\
		\item{Sources with partially disjoint supports:} assuming that, for each source, there exist at least one data sample where a single source is active is a strong assumption in general. In the general setting, the sources are approximately sparse in a transformed domain. As a consequence, all the sources have non-vanishing entries.
	\end{itemize}
	The goal of the present paper is to introduce a novel blind source separation algorithm that relaxes these assumptions in the case of sparse and partially correlated sources.}

\subsection*{Contribution}
In this article, building upon sparsity and the concept of morphological diversity, we introduce a novel blind source separation technique for the separation of sparse and partially correlated sources from noisy measurements. This new technique coined Adaptive Morphological Component Analysis (AMCA) makes no assumption about the non-negativity of the sources or their mixtures. This makes this method suitable for a very wide range of separation problems especially in imaging.\\
The paper is organized as follows: Section \ref{sec:pcor} reviews the basics of the GMCA algorithm, its limitations in the case of partially correlated sources, and how they can be alleviated. Section \ref{sec:amca} introduces the AMCA algorithm. Section \ref{sec:numerics} reports results of extensive evaluations of performances the AMCA algorithm with respect to standard sparse BSS algorithms. Finally, Section~\ref{sec:appli_astro} illustrates the performances of AMCA in the context of astrophysical source separation.

%------------------------------------------------------------------------------------
\section{Adaptive sparsity and separation of partially correlated sources}
%------------------------------------------------------------------------------------
\label{sec:pcor}

\subsection{Morphological diversity and the separation of partially correlated sources}
\label{sec:pc_sep}
Building on the concept of morphological diversity, the GMCA algorithm (Generalized Morphological Component Analysis - see \cite{starck:bobin07}) aims at retrieving the sparse sources and the mixing matrix by solving the following optimization problem~:
\begin{equation}
\label{eq:min_gmca}
\min_{{\bf A},{\bf S}} \quad \sum_{j = 1}^n \lambda_j \|s_j {\bf \Phi}^T\|_p + \frac{1}{2}\|{\bf X} - {\bf AS}\|_F^2,
\end{equation}
where the first term penalizes non-sparse solutions; the quadratic part is a classical data-fidelity term. The Frobenius norm is defined as follows: $\| {\bf Y}\|_F^2 = \mbox{Trace}\left( {\bf Y}{\bf Y}^T\right)$. The sparsity level is measured by the $\ell_p$ norm of the sources with $p \leq 1$. We generally choose either $p=1$ or $p=0$; we refer the interested reader to \cite{starck:bobin07,bobin08_aiep} for a detailed discussion of the choice a particular $\ell_p$ norm for the sparsity penalty. The $\ell_1$ norm is particularly appealing since it makes the estimation of $\bf S$ a convex optimization problem when $\bf A$ is fixed. However, as discussed in \cite{starck:bobin07}, it tends to produce biased estimates of the sources. In the remaining of this paper, we will therefore choose $p=0$ which has been showed to provide the best separation results in \cite{starck:bobin07}.\\
The problem in Equation \eqref{eq:min_gmca} is classically tackled by using an alternate projected least-square algorithm. The GMCA algorithm processes by iteratively alternating the following steps~:
\begin{itemize}
\item{\it Step 1 :  Estimation of $\bf S$ for fixed $\bf A$:} 
$$
\min_{{\bf S}} \quad \sum_{j = 1}^n \lambda_j \|s_j {\bf \Phi}^T\|_{\ell_0} + \frac{1}{2}\|{\bf X} - {\bf AS}\|_F^2.
$$
\item{\it Step 2 :  Estimation of $\bf A$ for fixed $\bf S$:}
$$
\min_{{\bf A}} \|{\bf X} - {\bf AS}\|_F^2.
$$
\end{itemize}
Unless $\bf A$ and $\bf \Phi$ are orthogonal matrices, the solution to the problem in Step 1 does not admit a closed-form expression. In the GMCA algorithm, this step is rather approximated with a projected least-square estimator ~:
$$
\forall j=1,\cdots,n; \quad s_j = \mathcal{H}_{\sqrt{\lambda_j}} \left( [{\bf A}^+{\bf X}]_j {\bf \Phi}^T \right) {\bf \Phi},
$$
where ${\bf A}^+ = ({\bf A}^T{\bf A})^{-1}{\bf A}^T$ is the Moore pseudo-inverse of the matrix $\bf A$ and $[{\bf Y}]_j$ stands for the $j$-th row of the matrix $\bf Y$. The operator $\mathcal{H}_{\sqrt{\lambda_j}}$ is the well-known hard-thresholding operator with threshold $\sqrt{\lambda_j}$. For any vector $y$, this operator performs on each entry $y[t]$ as follows:
\begin{equation*}
\mathcal{H}_{\mu}(y[t]) = \left \{ 
\begin{array}{cc}
y[t] & \mbox{ if }  |y[t]| > \mu  \\
0 & \mbox{ otherwise }
\end{array}
\right.
\end{equation*}
The second step of the GMCA algorithm is a more classical least-square problem which admits a closed-form solution~: ${\bf A} = {\bf X}{\bf S}^T({\bf S}{\bf S}^T)^{-1}$.\\

\paragraph{Sparse BSS and partially correlated sources}
\label{sec:model}
In this section, we discuss how sparse and partially correlated (s.p.c.) sources can be estimated using sparse blind source separation. For the sake of clarity, the transform $\bf \Phi$ is fixed to the identity matrix. In essence, it has been emphasized in \cite{starck:bobin07,bobin08_aiep} that, in the low-noise limit, solving the problem in Equation~\ref{eq:min_gmca} with GMCA is somehow similar to finding the sources which are jointly the sparsest possible. More precisely, assuming that the $\ell_1$ norm is used to measure the sparsity level of the sources, sparse BSS tends to estimate the sources and the mixing matrix so that the sources are enclosed in the $\ell_1$ ball with the smallest radius. It is important to notice that, for sparse  sources with independently distributed entries, any non-trivial mixture of the sources has, with high probability, a higher $\ell_1$ norm (see \cite{bobin08_aiep}); it will therefore be enclosed in a $\ell_1$ ball with larger radius. This observation motivates the design of a separation procedure that looks for the sparsest sources as performed by the GMCA algorithm.\\

\begin{figure}[tb]
\centering \includegraphics [scale=0.18]{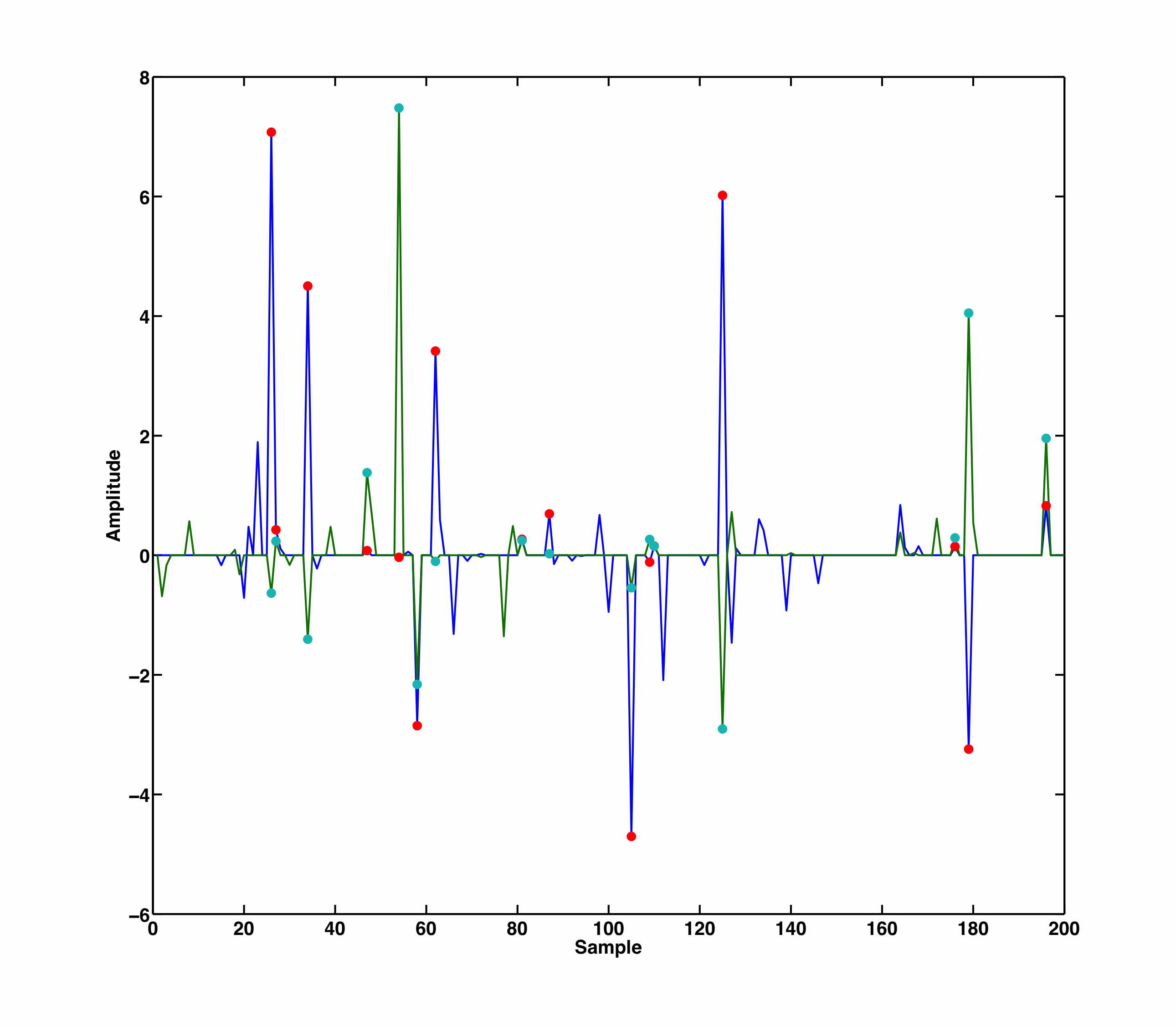}
\caption{A realization of $2$ partially correlated sources displayed in blue and green. Dots indicate the active entries of the sources which are common to the sources.}
\label{fig:example}
\end{figure}

\begin{figure}[tb]
\centering \includegraphics [scale=0.18]{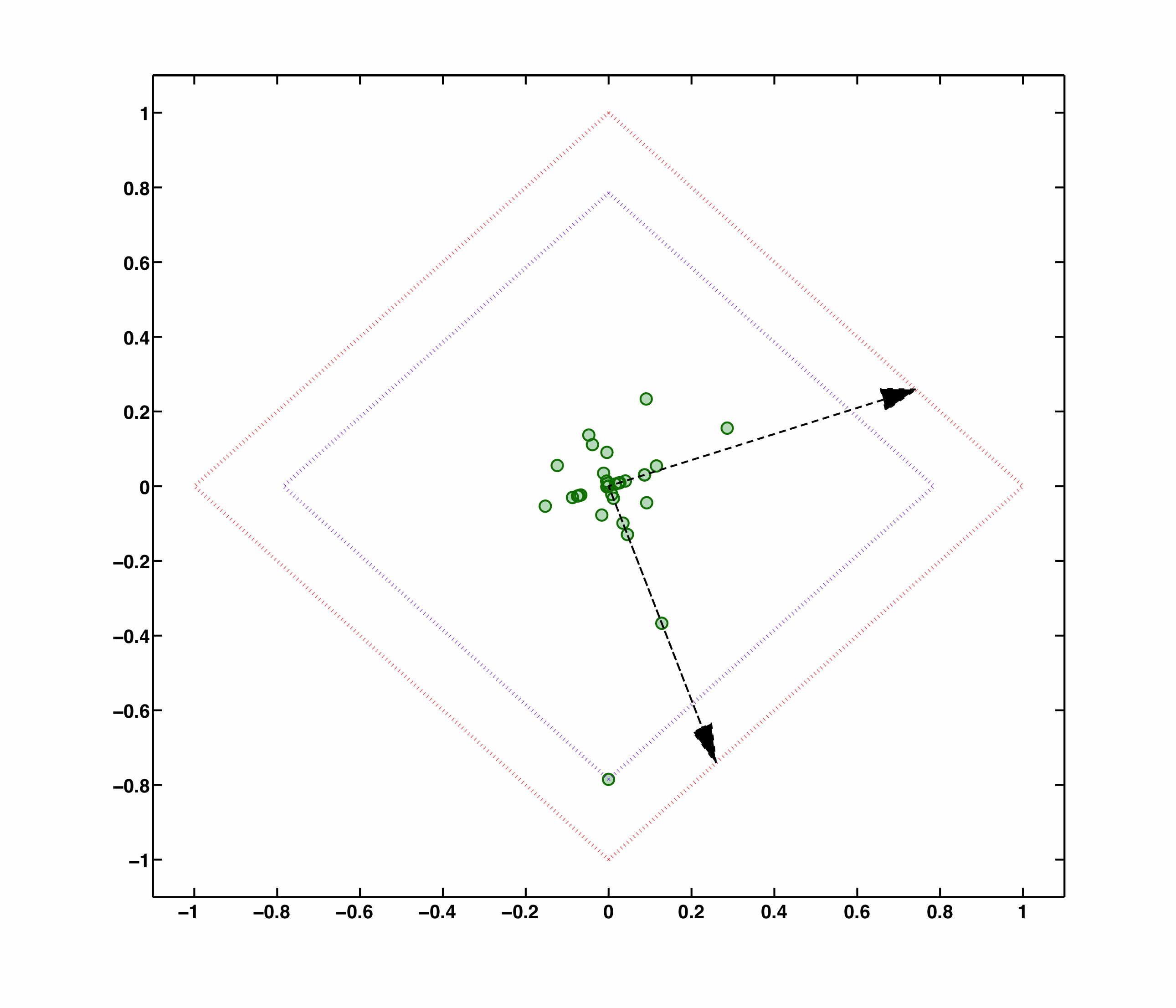}
\caption{Scatter plot of the sources with the sparsest distribution. The boundaries of the $\ell_1$ ball of the true sources (\textit{resp.} with the lowest $\ell_1$ norm) is displayed in purple (\textit{resp.} in ref). The black arrows show the actual directions along which the scatter plot of the true sources should be aligned.}
\label{fig:scatter_bestl1}
\end{figure}

In the case of s.p.c. sources, this separation procedure may no longer be viable. As an illustration, Figure~\ref{fig:example} shows an example of two s.p.c. sources. In brief, the nonzero entries of the sources are independently distributed with the exception of a few samples -- labeled with red and blue dots -- which are shared by the two sources.  Figure~\ref{fig:scatter_bestl1} displays the scatter plot ({\it i.e.} this plane features the samples of $s_1$ in abscissa and $s_2$ in ordinates) of the two sources estimated by looking for the sparsest sources from two mixtures as is done in the GMCA algorithm. The estimated sources are enclosed in the $\ell_1$ ball with the smallest radius (dotted purple $\ell_1$ ball). However they do not match the sought after sources; these would actually be enclosed in the dotted red $\ell_1$ ball. The black arrow features the axis along which the abscissa and ordinates should be aligned if the right sources were correctly estimated. This simple toy-example highlights the limits of any standard sparse BSS method: the presence of significantly active samples in more than one source entails that the sparsest solution might not be the correct solution.\\
  
These empirical observations raise the following crucial question:  {\it which are the sources' samples that are the most relevant or informative for separation ?} Sparse BSS methods will generally be handicapped by the presence of samples which simultaneously have a significant amplitude in more than one source. For instance, in Figure~\ref{fig:scatter_bestl1}, the sample located at the position $(-0.8,0)$  takes a large amplitude in both sources. These samples are deemed non-discriminant and therefore of poor relevance for the estimation of the sources and the mixing matrix.\\
In the rest of the paper, we will assume that the samples of the sources can be decomposed into two distinct subsets: $\Omega^c$ will define the set of the sources' samples where morphological diversity does not hold: these samples are likely to take significant amplitudes in more than one source. The set $\Omega^\star$ will denote the complement of $\Omega^c$; morphological diversity is assumed to hold true in this set. $\Omega^\star$ will therefore contain the samples that are the most discriminant and, subsequently, the most relevant for the separation.\\
If the subset $\Omegaˆ\star$ were known, the sources could be estimated in a straightforward way by restricting the application of any sparse BSS method to $\Omega^\star$. In the framework of GMCA, one would trivially substitute step 2 with the following~:
\begin{equation}
\min_{{\bf A}} \|[{\bf X} - {\bf AS}]^{\Omega^\star}\|_F^2,
\label{eq:OmegaS}
\end{equation}
where the operator $[{\bf Y}]^{\Omega^\star}$ denotes the submatrix made of the column of ${\bf Y}$ indexed by $\Omega^\star$.\\
This assumption is unfortunately unrealistic in practice. The set $\Omega^\star$ has therefore to be learned along with the sources and the mixing matrix. It is important to note that attempting to estimate $\Omega^\star$ as an extra unknown in Equation~\eqref{eq:min_gmca} would lead to a cumbersome combinatorial problem. Since it cannot be reasonably estimated directly as an extra parameter, we rather advocate the use of a more flexible way to privilege discriminant samples, which are likely to belong to $\Omegaˆ\star$.\\ 
The problem in Equation~\ref{eq:OmegaS} is equivalent to a weighted least-square problem of the form:
\begin{equation}
\min_{{\bf A}} \, \, \mbox{Trace} \left\{({\bf X} - {\bf AS}) {\bf W} ({\bf X} - {\bf AS})^T \right\},
\label{eq:weightingA1}
\end{equation}
where $\bf W$ is a diagonal matrix with ones for samples in $\Omega^\star$ and zeros otherwise. Since $\Omega^\star$ is unknown, we propose relaxing this strategy with an adaptive re-weighting procedure. For that purpose, the mixing matrix will be estimated in a similar manner as in Equation \eqref{eq:weightingA1}. Each sample is given a weight which measures their discriminant level. These weights are encoded in the diagonal matrix $\bf W$. We highlighted previously that standard BSS methods are hindered by samples with large amplitudes in more than one source. Conversely, the most discriminant samples will be the ones that mainly take a large amplitude in a single source. These properties suggest that discriminant samples will be traced by the sparsity level of the columns of the matrix $\bf S$. Consequently, we define the weight vector $w_q \in \mathbb{R}^T$ as follows:
\begin{equation}
\label{eq:wqdef}
\forall t=1,\cdots,T; \quad w_q[t] = \frac{1}{\|[{\bf S}]^t\|_{\ell_q} + \epsilon},
\end{equation}
where $[{\bf S}]^t$ is the $t$-th column of $\bf S$. The scalar $\epsilon$ may be required when there exist columns of $\bf S$ with vanishing $\ell_q$ norm; its value is generally small, typically $\epsilon = 10^{-6}$. 
%Since it is set to a small value, $\epsilon$ has a minor impact on the estimation of the mixing matrix since it will mainly act on columns with small or vanishing $\ell_q$ norm which, by the way, have a very slight  contribution in the weighted least-square problem of Equation \eqref{eq:weightingA1}. 
The parameter $q$ will be chosen in the range $[0,1]$ so as to measure the sparsity level of the samples across the sources. Other choices for the samples' weights can be considered. However, this choice is particularly interesting as~: i) it accounts for the natural sparsity of the sources, ii) the samples in $\Omega^\star$ are indeed sparser across the sources than samples which are common to all sources and iii) this weighting strategy is inversely proportional to the amplitude of the columns of $\bf S$. The former entails that it penalizes more samples of the sources which are common to several sources and have a high amplitude in the meantime. This type of samples are generally the most detrimental to sparse BSS methods when the sources are partially correlated.\\

\subsection{AMCA: Adaptive Morphological Component Analysis}
\label{sec:amca}
\cjbb{In this section, we introduce a novel blind source separation algorithm, which we called AMCA (Adaptive Morphological Component Analysis), to specifically deal with s.p.c. sources, that builds upon the GMCA algorithm.\\
In the previous section, we advocated the use of a weighting strategy to privilege/penalize data samples based on their ability to discriminate between s.p.c. sources. This formally amounts to estimating the mixing matrix $\bf A$ and the sources $\bf S$ by solving the following minimization problem:
\begin{equation}
\label{eq:min_amca}
\min_{{\bf A},{\bf S}} \quad \sum_{j = 1}^n \lambda_j \|s_j {\bf \Phi}^T\|_{\ell_p} + \frac{1}{2}\mbox{Trace} \left\{({\bf X} - {\bf AS}) {\bf W} ({\bf X} - {\bf AS})^T \right\},
\end{equation}
where the matrix $\bf W$ stands for the weighting matrix. This problem takes a simpler formulation in the transformed domain. For the sake of simplicity, we will first consider that $\bf \Phi$ is an orthogonal matrix. We will further define ${\bf X}_{\bf \Phi} = {\bf X}{\bf \Phi^T}$ and ${\bf S}_{\bf \Phi} = {\bf S}{\bf \Phi^T}$, which respectively stand for expansion coefficients of the data and the sources in the transformed domain. Thanks to this re-parameterization, and assuming that $p=1$, the problem in Equation~\eqref{eq:min_amca} can be recast as follows:
\begin{equation}
\label{eq:min_amca2}
\min_{{\bf A},{\bf S_{\Phi}}} \quad \|{\bf \Lambda} \Sphi \|_{\ell_1} + \frac{1}{2}\mbox{Trace} \left\{(\Xphi - {\bf A}\Sphi) \Wphi (\Xphi - {\bf A}\Sphi)^T \right\},
\end{equation}
where $\Wphi = {\bf \Phi W }{\bf \Phi}^T$, and ${\bf \Lambda} = \mbox{ diag}(\lambda_1,\cdots,\lambda_j,\cdots,\lambda_T)$ is a diagonal matrix that stores the regularization parameters. According to the sparse modeling developed in the previous section, the partial correlation of the sources is modeled as a correlation of certain entries of the sources in the transformed domain. Subsequently, the matrix $\Wphi$ is diagonal and it can be defined as $\Wphi = \mbox{diag } (w_q \odot w_q)$, where the operator $\odot$ is the Hadamard product. From Equation~\eqref{eq:wqdef}, the weight vector is defined as:
\begin{equation}
\label{eq:wqdef2}
\forall t=1,\cdots,T; \quad w_q[t] = \frac{1}{\|[\Sphi]^t\|_{\ell_q} + \epsilon}.
\end{equation}
In the next, we will write denote the weighting matrix as $\Wphi^{(q)}$ to highlight its dependency on the parameter $q$. \\
The {\it optimal} weight vector $w_q$ clearly depends on the unknown true sources. In practice, one has generally only access to an imperfect estimate $\hat{\bf S}_{\bf\Phi}$; the weight vector $w_q$ and the weight matrix $\Wphi^{(q)}$ will be computed from some current estimate $\hat{\bf S}_{\bf\Phi}$ in the proposed algorithm.\\
For a fixed weight matrix $\Wphi$, minimizing the problem in Equation~\ref{eq:min_amca2} can be tackled by alternately updating $\bf A$ and $\Sphi$ in the spirit of Block Coordinate Relaxation(BCR - see \cite{text:Bruce98,Tseng_01_ConvergenceBlockCoordinate}). This procedure then turns to alternately solving sequences of convex minimization problems in place of a globally non-convex problem.}

\cjbb{\paragraph*{Updating $\Sphi$ for a fixed mixing matrix $\bf A$}
Assuming that the mixing matrix $\bf A$ is fixed, estimating the source matrix $\Sphi$ is performed by looking for the unique minimizer of the following convex problem:
\begin{equation}
\label{eq:min_amca2_S}
\min_{{\bf S_{\Phi}}} \quad \|{\bf \Lambda} \Sphi \|_{\ell_1} + \frac{1}{2}\mbox{Trace} \left\{(\Xphi - {\bf A}\Sphi) \Wphi (\Xphi - {\bf A}\Sphi)^T \right\},
\end{equation}
The equation to be minimized can be decomposed into two terms: i) a quadratic and differentiable data fidelity term, and ii) a convex but non-differentiable $\ell_1$ penalty. Fortunately, the $\ell_1$ is a simple and proper convex function that admits a proximal operator (we refer the interested reader to \cite{CombettesWajs05,Boyd_Proximal14} for more details about proximal calculus). Furthermore, the quadratic term is differentiable and its gradient is $L$-Lipschitz with constant $L = \|{\bf A}\Wphi {\bf A}^T\|_2$. This entails that the problem in Equation~\eqref{eq:min_amca2_S} can be solved exactly using techniques such as the Forward-Backward splitting algorithm \cite{CombettesWajs05}. This optimization strategy has been used in \cite{Rapin_13_SparseandNon} for solving sparse non-negative matrix factorization problems. However, it has the strong disadvantage of dramatically increasing the computational cost of the algorithm: update $\Sphi$ would require resorting to efficient but costly iterative algorithms. Furthermore, each time the source matrix $\Sphi$ is updated in the algorithm AMCA algorithm, it is fully re-estimated; therefore, it may not be necessary to update with high precision $\Sphi$ at each step of AMCA. Subsequently, we rather resort to a simpler but approximate minimization scheme in the spirit of the GMCA algorithm.\\
More precisely, finding the critical point Equation~\eqref{eq:min_amca2_S} can be done by deriving its subdifferential with respect to $\Sphi$ \cite{Rockafellar:1970px} and finding its roots:
\begin{equation}
\label{eq:subdiff}
{\bf 0} \in {\bf \Lambda} \: \partial \| {\bf \Lambda} {\Sphi}  \|_{\ell_1} - {\bf A}^T \Xphi\Wphi +{\bf A}^T {\bf A} \Sphi \Wphi,
\end{equation}
where we used the fact that $\Wphi$ is diagonal. The term $\partial \| {\bf \Lambda} {\Sphi} \|_{\ell_1}$ stands for the subdifferential of the matrix $\ell_1$ norm taken at ${\bf \Lambda} \Sphi$; it is well known that the subdifferential of the matrix $\ell_1$ norm is defined as follows:
\[
\partial \|{\boldsymbol Y}\|_{\ell_1} = \left\{{\bf Z} \in \mathbb{R}^{n \times T} \Bigg| 
\begin{array}{cccc}
{\bf Z}{j,k} & = &\mbox{ sign}({\bf Y}_{j,k}),&  ~ {\bf Y}_{j,k} \neq 0 \\
{\bf Z}{j,k} & \in & [-1,1], & ~ {\bf Y}_{j,k} = 0
\end{array} \right\}.
\]
The root of the subdifferential in Equation~\eqref{eq:subdiff} does not admit an explicit formulation unless ${\bf A}^T{\bf A}$ is diagonal, which would make Equation~\eqref{eq:subdiff} separable with respect to each individual source. In such a case, each source $\Sphi_j$ can be estimated independently and takes the following closed-form expression:
\begin{equation}
\forall j=1,\cdots,n; \; \Sphi_j = \mathcal{S}_{\mu_j} \left( \frac{1}{\|a^j\|_{\ell_2}^2} {a^j}^T\Xphi \right),
\end{equation}
where $\mathcal{S}_{\mu_j}$ stands for the soft-thresholding operator ({\it i.e.} the so-called proximal operator of the $\ell_1$ norm) and $\mu_j = \frac{\lambda_j \|a^j\|_{\ell_2}^2}{\Wphi_{j,j}}$. It is important to note that, in this setting, this is equivalent to: 
\begin{equation}
\label{eq:UpdateS_ML1}
\Sphi = \mathcal{S}_{\boldsymbol \mu} \left( ({\bf A}^T{\bf A})^{-1} {\bf A}^T\Xphi \right),
\end{equation}
where ${\boldsymbol \mu} = [\mu_1,\cdots,\mu_n]$ and $({\bf A}^T{\bf A})^{-1} {\bf A}^T\Xphi$ is precisely the least-squares estimate of the sources. In other, this procedure amounts to thresholding the least-square estimate ({\it i.e.} the minimum of the quadratic data fidelity term), which thus dramatically reduces the computational burden of this step in the AMCA algorithm. However, this equivalence holds whenever ${\bf A}$ is orthogonal ({\it i.e.} ${\bf A}^T{\bf A}$ is diagonal), which is not always a valid assumption. As a consequence, updating $\Sphi$ according to Equation~\eqref{eq:UpdateS_ML} provides a rough proxy in the general case which might prevent the algorithm from convergence. This will be discussed later on in this section.
}
\cjbb{\paragraph*{Updating $\bf A$ for a fixed source matrix $\Sphi$}
The mixing matrix $\bf A$ is updated assuming the source matrix $\Sphi$ is fixed, which is performed by tackling the following convex problem:
\begin{equation}
\label{eq:min_amca2_A}
\min_{{\bf A}} \quad \mbox{Trace} \left\{(\Xphi - {\bf A}\Sphi) \Wphi (\Xphi - {\bf A}\Sphi)^T \right\}.
\end{equation}
This is a standard reweighted quadratic problem, which admits the following unique minimizer:
\begin{equation}
\label{eq:UpdateA_ML}
{\bf A} = \Xphi \Wphi \Sphi^T \left( \Sphi \Wphi \Sphi^T  \right)^{-1}
\end{equation}}

\cjbb{\paragraph*{The AMCA algorithm}
In the spirit of BCR, the AMCA algorithm mainly alternates by updating $\Sphi$ and $\bf A$ according to Equations~\eqref{eq:UpdateS_ML} and ~\eqref{eq:UpdateA_ML}. However, the overall estimation problem of Equation~\eqref{eq:min_amca} is nonconvex; this means that this estimation procedure is likely to be prone to trapping in local minima. This will highly depend on the initialization of the AMCA algorithm. In \cite{bobin08_aiep}, we showed that thanks to the sparsity of the sources and more specifically to the morphological diversity assumption, starting from large regularization parameters $\{\mu_j \}_{j=1,\cdots,n}$ (or equivalently large thresholds) and decreasing their values towards a final noise-dependent threshold greatly helped improving the robustness of the algorithm GMCA with respect to local minima. The same procedure is implemented in the AMCA algorithm: the $\{\mu_j\}_{j=1,\cdots,n}$ are first set up to some large initial values $\{\mu_j^{(0)}\}_{j=1,\cdots,n}$ and then decreased towards final values $\{\mu_j^{(f)}\}_{j=1,\cdots,n}$.\\
As pointed out earlier in this Section, the weight matrix $\Wphi$ is computed from some guess for the true sources $\Sphi$. Since one has only access to estimated sources, and these estimates are expected to improve at each step of the algorithm, the weight matrix is updated at each iteration $k$ according to the current estimate of the sources $\Sphi^{(k)}$. The choice of the parameters are discussed in Section~\ref{sec:parameters}.\\
In \cite{bobin08_aiep}, different thresholding strategies have been considered, namely soft and hard-thresholding. In contrast to hard-thresholding, the main drawback of the soft-thresholding is that it is prone to biasing the estimated sources. In practice, substituting the soft-thresholding with a hard-thresholding in the GMCA algorithm provided a substantial improvement of the separation performances. As a consequence, the update step described in Equation~\eqref{eq:UpdateS_ML} is replaced by : 
\begin{equation}
\label{eq:UpdateS_ML}
\Sphi = \mathcal{H}_{\boldsymbol \mu} \left( ({\bf A}^T{\bf A})^{-1} {\bf A}^T\Xphi \right),
\end{equation}
where $\mathcal{H}_{\boldsymbol \mu}$ stands for the hard-thresholding operator with thresholds ${\boldsymbol \mu} = [\mu_1,\cdots,\mu_n]$.\\
The resulting algorithm has been named Adaptive Morphological Component Analysis (AMCA). The adaptivity of AMCA stems from its attempt to estimate a weight matrix $\Wphi$ that better favors discriminant entries. In other words, this can be interpreted as adaptively learning which samples are more likely to verify the morphological diversity.
}
The AMCA algorithm is detailed below:

\begin{center}
\vspace{0.3in}
\centering
\begin{tabular}{|c|} \hline
\begin{minipage}[hbt]{0.95\linewidth}
\vspace{0.025in}
\footnotesize{

\textsf{I. Transform the data ${\bf X}_{\bf \Phi} = {\bf X}{\bf \Phi}^T$}\\

\textsf{II. Initialize ${\bf A}^{(0)},{\bf S}^{(0)}_{\bf \Phi}$, $\{\mu_j^{(0)}\}_{j=1,\cdots,n}$ and $q^{(0)} = 1$}\\

\textsf{III. While $k < P_{\max}$, } \\

\hspace{0.1in} \textsf{1 - Update $\bf S$ assuming $\bf A$ is fixed :} \\

\hspace{0.3in}$ {\Sphi^\prime} = \left({{\bf A}^{(k)}}^T{{\bf A}^{(k)}}\right)^{-1}{{\bf A}^{(k)}}^T{\Xphi}$\\

\hspace{0.3in}$ \forall j=1,\cdots,n;\quad [{\Sphi}^{(k+1)}]_j = \mathcal{H}_{\mu_j^{(k)}}\left({\Sphi^\prime}\right)$, \\

\hspace{0.5in} where $[\Sphi^{(k+1)}]_j$ is the $t$-th column of $\Sphi^{(k+1)}$ \\

\hspace{0.1in} \textsf{2 - Update the weight matrix :} \\

\hspace{0.3in}$ \Wphi_q^{(k)} = \mbox{diag} \left( \frac{1}{\left\|\left[{{\Sphi}^{(k+1)}}\right]^t \right\|_{\ell_{q^{(k)}}} + \epsilon}  \right)$, \\

\hspace{0.5in} where $\left[{\Sphi}^{(k+1)}\right]^t$ is the $t$-th column of $\Sphi^{(k+1)}$ \\

\hspace{0.1in} \textsf{3 - Update $\bf A$ assuming $\bf S$ is fixed :} \\

\hspace{0.3in} ${\bf A}^{(k+1)} = \Xphi {\Wphi}^{(k)}_q{\Sphi^{(k+1)}}^T\left({\Sphi^{(k+1)}}\Wphi^{(k)}_q{\Sphi^{(k+1)}}^T\right)^{-1}$ \\

\hspace{0.1in} \textsf{4 - Decrease each $\{\mu_j^{(k)}\}_{j=1,\cdots,n}$ and the parameter $q$} \\

\textsf{IV. Reconstruct the sources ${\bf S} = \Sphi {\bf \Phi}$.}}\\

\vspace{0.05in}
\end{minipage}
\\\hline
\end{tabular}
\vspace{0.3in}
\end{center}

\cjbb{\paragraph*{On the convergence of the AMCA algorithm}
The problem in Equation~\eqref{eq:min_amca2} is not convex; therefore, only convergence to a critical point can be expected. The design of the AMCA algorithm has been inspired by Block Coordinate Relaxation \cite{text:Bruce98}. In \cite{Tseng_01_ConvergenceBlockCoordinate}, Tseng precisely proved the convergence of BCR for the minimization of nondifferentiable and nonconvex cost functions, with a special application to sparsity-constrained blind source separation problem. According to \cite{Tseng_01_ConvergenceBlockCoordinate}, alternating the minimization steps in Equations~\eqref{eq:min_amca2_S} and \eqref{eq:min_amca2_A}, for fixed parameters $\{\mu_j \}_{j=1,\cdots,n}$ and $\Wphi$, converges to a critical point of the problem in Equation~\eqref{eq:min_amca2}.\\
Firstly, decreasing the thresholds $\{\mu_j \}_{j=1,\cdots,n}$ is a strategy that has been proposed in \cite{starck:bobin07} to improve the robustness of the GMCA algorithm to spurious local minima. In the field of optimization, this procedure is reminiscent of the fixed point continuation technique \cite{Hale_07_FixedPointContinuation}, which has been proposed to speed up the minimization of $\ell_1$-penalized least-squares problems. The convergence of the AMCA algorithm would be guaranteed as long as steps \eqref{eq:min_amca2_S} and \eqref{eq:min_amca2_A} are alternated until convergence for each value of $\{\mu_j \}_{j=1,\cdots,n}$. The thresholds are however updated at each iteration of the AMCA algorithm, which helps speeding up the algorithm but might prevent it from convergence.\\
Secondly, weight matrix $\Wphi$ is updated at each iteration, which also might prevent the AMCA algorithm from convergence. Lastly, in the spirit of reweighted $\ell_1$ techniques \cite{Candes_07_EnhancingSparsityby}, the weights $\Wphi$ are updated based on the current estimate of the sources $\Sphi$. If this strategy is a well motivated heuristic, the convergence of the AMCA algorithm is not theoretically grounded.\\
In numerical experiments, the AMCA tends to stabilize after a given number of iteration. As an illustration, Figure~\ref{fig:amca_convergence} displays the evolution of the signal-to-distortion ratio (SDE - see Section~\ref{sec:numerics} for more details) during a run of the AMCA algorithm on simulated sources, which are described in Section~\ref{sec:numerics}. Our intuition is that, after a given number of iterations, the parameters $\{\mu_j \}_{j=1,\cdots,n}$ and $\Wphi$ are likely to evolve slowly enough to make the AMCA stabilize. In practice, the AMCA algorithm converges in a few hundreds of iterations. Subsequently, choosing the number of iterations $P_{\max} = 500$ provided good results.\\
\begin{figure}[tb]
	\centerline{\includegraphics [scale=0.35]{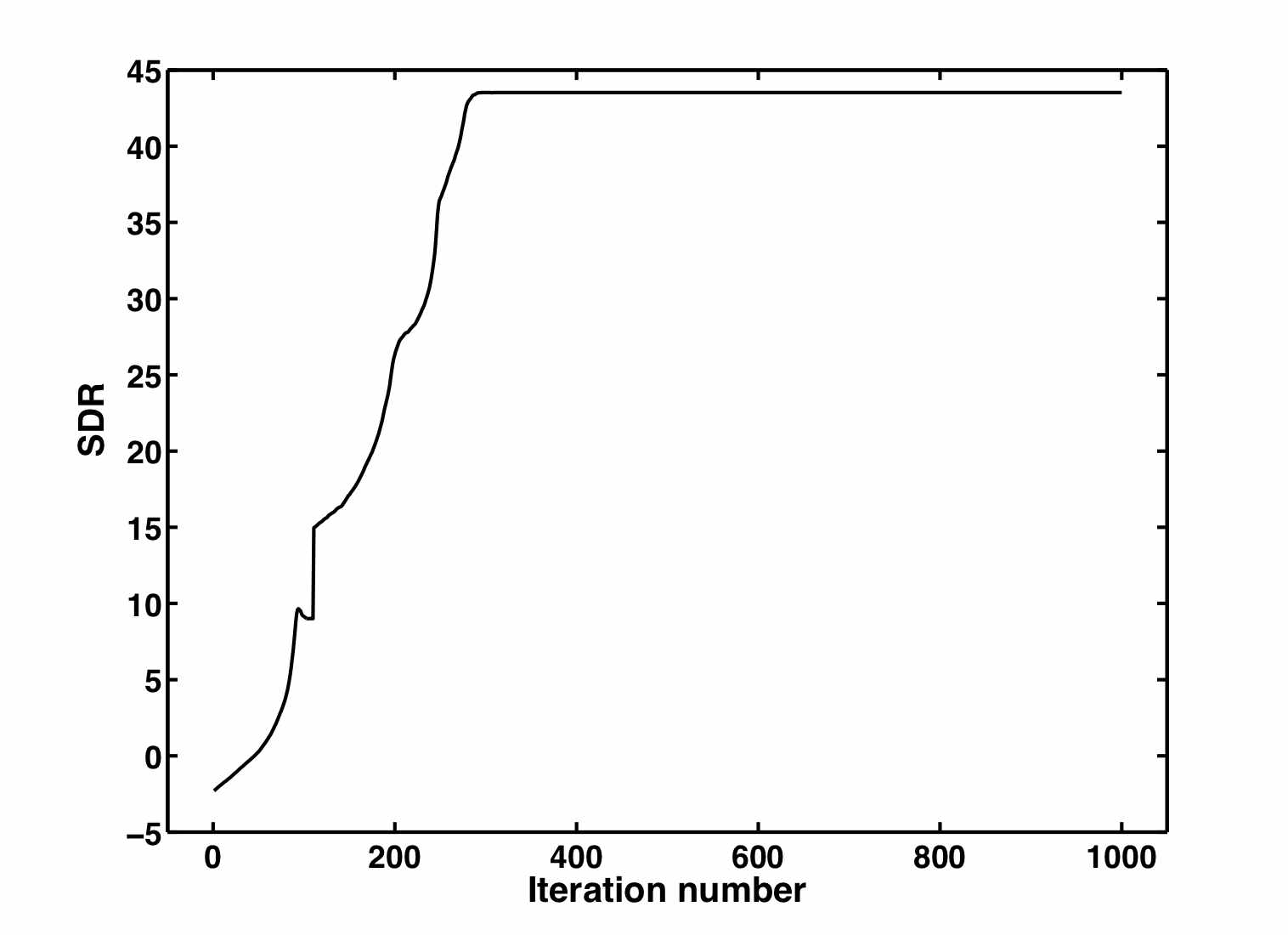}}
	\caption{Evolution of the SDR of simulated sources during a run of the AMCA algorithm. In abscissa: iteration number. In ordinate: valued of the SDR.}
	\label{fig:amca_convergence}
\end{figure}
}

\paragraph*{From orthogonal transforms to tight frames}
Previously, the derivation of the AMCA algorithm has been carried out assuming that the signal representation $\bf \Phi$ is an orthogonal transform; \textit{i.e.} it verifies: ${\bf \Phi}^T {\bf \Phi} = {\bf \Phi}{\bf \Phi}^T = {\bf I}$. This property allows to perform AMCA in the transformed domain. However, it has long been emphasized that appropriately chosen redundant transforms generally allow for extra degrees of freedom which help producing sparser representations. In the context of sparse BSS, the use of redundant transforms has been shown to yield substantial separation improvements (see \cite{bobin08_aiep,starck:bobin07}). In the numerical experiments of Section~\ref{sec:numerics}, we make use of redundant wavelets to sparsely represent the sources. Such type of wavelets have the property of translation-invariance \cite{Coifman_95_Translationinvariantde,starck:sta05_2}. These transforms are non-orthogonal; they are tight frames which verify the exact reconstruction property but for which ${\bf \Phi} {\bf \Phi}^T \neq {\bf I}$. In practice, this means that ${\bf S}_{\bf \Phi} = {\bf S}{\bf \Phi}^T$ is a sparse representation of the signal $\bf S$ but not generally the sparsest, if one exists (see \cite{EladBook}). Fortunately, since the Gram matrix ${\bf \Phi} {\bf \Phi}^T$ is diagonally dominant for most tight frames of interest in imaging, a first order approximation amounts to resorting to the same changes of variables ${\bf S}_{\bf \Phi} = {\bf S}{\bf \Phi}^T$ and ${\bf X}_{\bf \Phi} = {\bf X}{\bf \Phi}^T$. This allows to apply the AMCA algorithm in the transform domain in the tight frame case which greatly lower the computational cost of the algorithm. 

\subsection{Choosing the parameters in AMCA}
\label{sec:parameters}

\paragraph*{Choice of the thresholds $\{\mu_j\}_{j=1,\cdots,n}$} It has been emphasized in \cite{bobin08_aiep} that the choice of the thresholds $\{\mu_j\}_{j=1,\cdots,n}$ plays a crucial role in the separation process. From the concept of morphological diversity, the high-amplitudes entries of the sources are likely to belong to a single source. In the case of statistically independent sources, this assumption holds true, with high probability, for high enough amplitudes. This makes the high-amplitude entries the most discriminant samples of the sources.\\
The hard-thresholding operator $\mathcal{H}_{\mu_j}$ selects entries with amplitudes higher than $\sqrt{\mu_j}$ and rejects the others. In the AMCA algorithm, the thresholds $\{\mu_j\}_{j=1,\cdots,n}$ are first initialized to high values; generally each parameter $\mu_j$ is set to the maximum amplitude of the initial guess of the $j$-th source $s_j^{(0)}$. Their values then decrease towards a final threshold which depends on the noise level. In practice, the final thresholds are fixed to $3 \, \sigma_j$ where $\sigma_j$ stands for the standard deviation of the noise which contaminates the $j$-th sources. Whenever the level of the noise is unknown, it is estimated empirically using the Median Absolute Deviation estimator (MAD). Assuming that the noise is additive white Gaussian, this procedure guarantees, with high probability, that no noise sample is selected.\\
Such kind of {\it coarse-to-fine} procedure has some interesting properties : i) the sources are first estimated from the most discriminant entries ({\it i.e.} the high-amplitude entries of the sources), ii) in the spirit of simulated annealing, the variation of the thresholds makes AMCA less prone to trapping in local minima and iii) the thresholding helps preventing AMCA from incorporating noise entries which makes it more robust to noise.\\

\paragraph*{Choice of the weight parameter $q$} The AMCA algorithm rely on a re-weighting procedure that penalizes/favors certain entries of the estimated sources. The weights are function of the $\ell_q$ norm of the columns of $\bf S$. They somehow measure the activity of each sample across the sources. Intuitively, choosing a low value for $q$ seems quite natural since it yields more contrast between sparse and non-sparse columns of $\bf S$. This argument would make perfect sense if the true sources were known. A trade-off has to be made between the two following options : i) large values for $q$ might lead to an under-penalization of less discriminant entries and ii) small values for $q$ provide a larger penalization of non-sparse entries of $\bf S$, which is desirable to efficiently separate s.p.c. sources. However, at the beginning of the AMCA algorithm, one has only access to imperfect, if not erroneous, estimates of $\bf S$. In this case, small values of $q$ might mis-penalize/mis-favor entries of $\bf S$ which can eventually hamper the separation process. Alleviating this dilemma is made by starting with a high value for $q$ -- typically $1$ -- and then decreasing it, at each iteration, towards some final value $q_f$. Several values for $q_f$ have been tested; it turns out that choosing $q_f = 0.01$ leads to a good trade-off for all the experiments we carried out. Smaller values for $q_f$ did not bring any noticeable improvement.

%------------------------------------------------------------------------------------
\section{Numerical experiments}\label{sec:numerics}
%------------------------------------------------------------------------------------

\subsection{Experimental protocol}
In this section, we evaluate the performances of the AMCA algorithm with respect to standard sparse BSS methods~: i) the GMCA algorithm \cite{starck:bobin07}, ii) RNA (Relative Newton Algorithm) \cite{ica:zibu_relnewton} and iii) EFICA (Efficient FastICA) \cite{koldo06}. It is important to recall that, to our knowledge, the blind separation of s.p.c. sources has never been dealt with in the general case where: i) neither the mixing matrix nor the sources are assumed to be non-negative, ii) the sources do not have (quasi)-disjoint supports. Relaxing these assumptions is interesting so as to tackle the separation of complex data which are approximately sparse in general signal representations $\bf \Phi$ ({\it e.g.} Fourier, wavelets, curvelets, etc.) but where non-negativity is certainly not a valid assumption.\\ \\

\begin{figure}[tb]
\centerline{\includegraphics [scale=0.27]{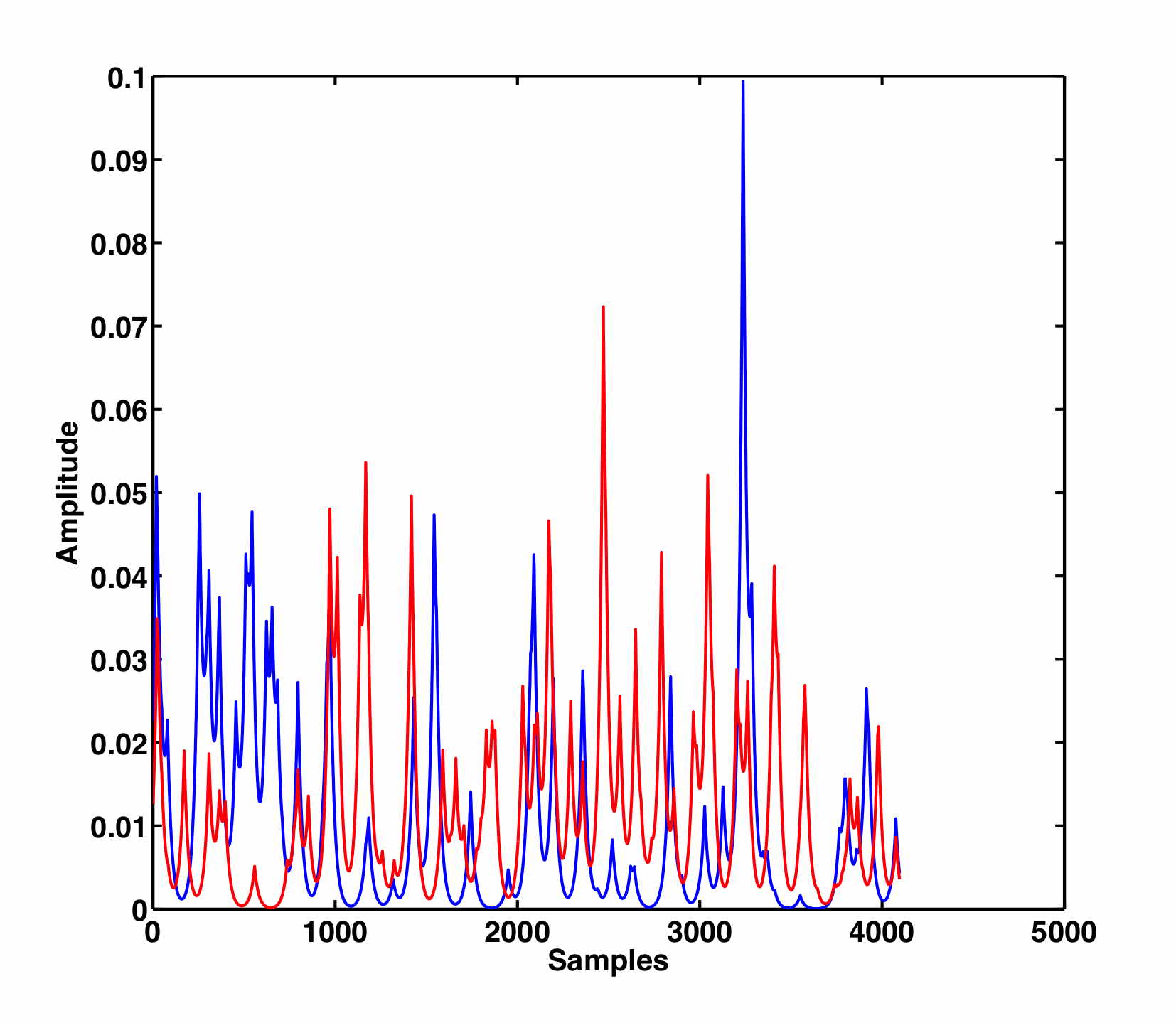}
\hfill
\includegraphics [scale=0.27]{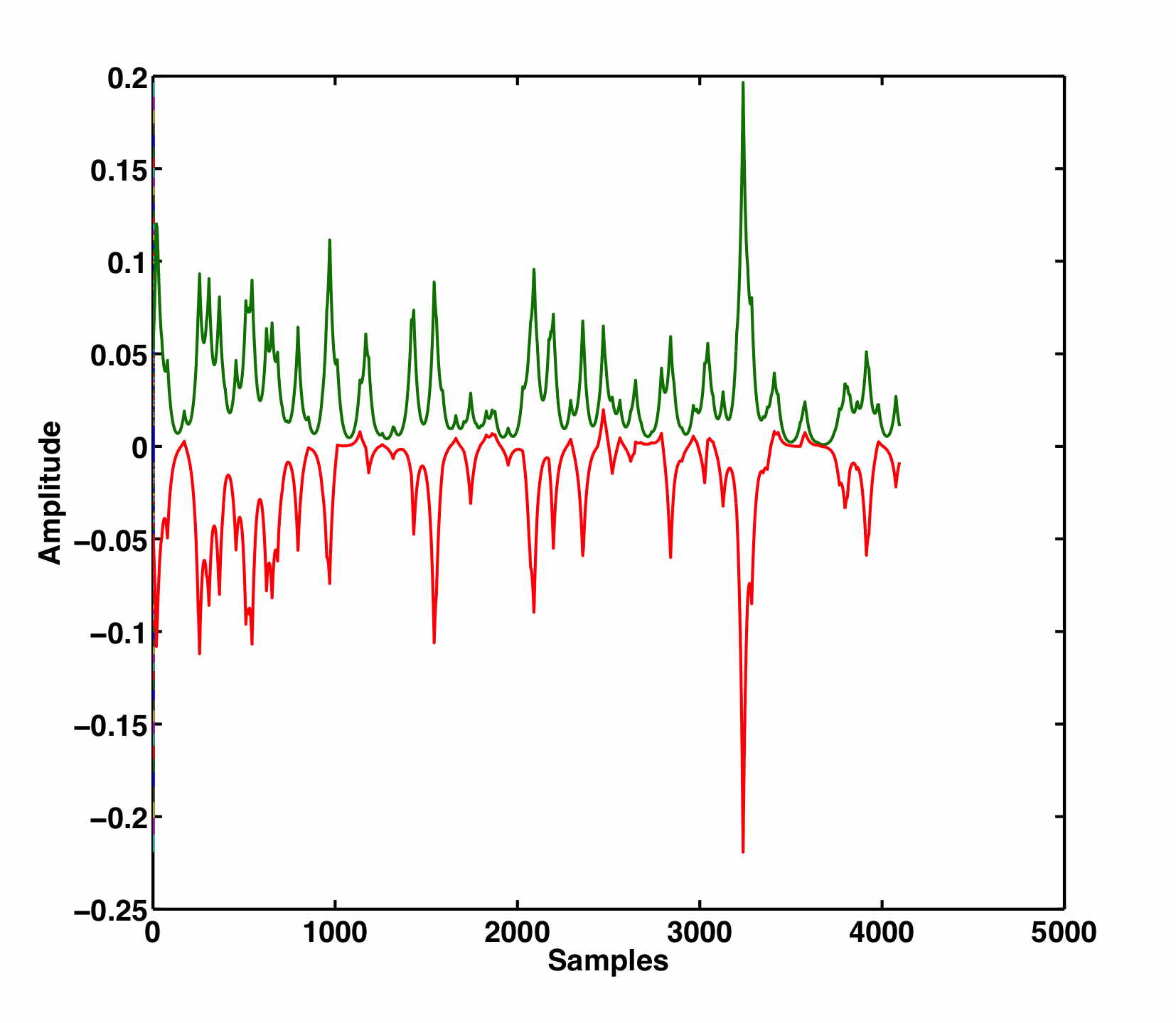}}
\caption{Left: $2$ sources generated from the s.p.c. sources model. Right: $2$ mixtures of s.p.c. sources.}
\label{fig:sources_mix}
\end{figure}

 In this first study, our prime intention is to study the performances of standard sparse BSS methods and the AMCA algorithm to estimate s.p.c. sources. For that purpose, we will first make use of synthetic but rather complex data; this makes Monte-Carlo simulations possible in various settings. To mimic spectroscopy-like data, the (s.p.c.) sources are generated according to the so-called (SPC) model introduced below~:

\begin{itemize}
\item The sources $\{s_j\}_{j=1,\cdots,n}$ are $K$-sparse signals. This means that each source $s_j$ has only $K$ non-zero entries out of $T$ entries. The activation of the entries of the sources will be distributed according to a Bernoulli process $\pi$ with parameter $0 \leq \rho \leq 1$ so that :
\begin{equation}
\pi = \left \{
\begin{array}{cc}
1 & \mbox{ with probability } \rho, \\
0 & \mbox{ with probability } 1 - \rho.
\end{array}
\right.
\end{equation}
\item For each source, $L \leq K$ active entries are shared by all the sources and the remaining $K-L$ are independently drawn. Let us recall that $\Omega_c$ is the set of the $L$ entries which are common to all the sources; the distribution of these entries is Gaussian with mean $0$ and variance $\tau^2$. In this model, common active entries do not have correlated amplitudes.\\
\item The $K-L$ nonzero entries which belong to $\Omega_j \backslash \Omega^c$ are are independently distributed according to a Gaussian law with mean $0$ and variance $1$. The parameter $\tau$ defines some dynamic range between the amplitude of common and independent samples of the sources.\\
\item the sources are convolved with a Laplacian kernel of full width at half height (FWHM) equal to $15$. 
\end{itemize}
An example of $2$ simulated sources ({\it resp.} $2$ mixtures) is shown in the left ({\it resp.} right) panel of Figure~\ref{fig:sources_mix}. The resulting sources can be sparsely represented in a frame of redundant (translation invariant) 1D wavelets\footnote{For that purpose we used the RICE wavelet toolbox : {\it http://dsp.rice.edu/software/rice-wavelet-toolbox}}. Each source is made of $T = 4096$ samples. The mixing matrix will be picked at random from a Gaussian distribution. In the following experiments, the sparsity level $K/T$ is fixed to $0.02$. This means that for $T = 4096$ --- which will be the actual value in the next --- about $81$ spikes will be active in each source.

\paragraph{Evaluation criteria}
Following \cite{Vincent_06_Performancemeasurementin}, the estimated sources can be decomposed into the combination of contributions of different origins:
\begin{equation*}
s^\text{est}=s_\text{target}+s_\text{interf}+s_\text{noise}+s_\text{artefacts},%\label{eq:decomposition}
\end{equation*}
with the following interpretations for the terms: 
\begin{itemize}
\item $s_\text{target}$ is the projection of $s^\text{est}$ on the target (ground-truth) source. In other words, it is the one part of this decomposition which corresponds to what needs to be recovered. The other ones are residues.
\item $s_\text{interf}$ accounts for interferences due to other sources.
\item $s_\text{noise}$ is the part of the reconstruction which is due to noise.
\item $s_\text{artefacts}$ stands for the remaining artefacts which are neither due to interferences nor noise.
\end{itemize}
\medskip
From this decomposition, one can derive a global criterion named Source to Distortion Ratio (SDR) which contains information about each of the different terms of the decomposition:
\begin{equation}
\text{SDR}(s^\text{est})=10~\text{log}_{10}\left(\frac{\|s_\text{target}\|_2^2}{\|s_\text{interf}+s_\text{noise}+s_\text{artefacts}\|_2^2}\right).\label{eq:SDR}
\end{equation}
In \cite{starck:bobin07}, we advocated the use of a global performance criterion that only depends on the mixing matrix. Such a criterion is particularly interesting in the noisy setting as it is not dependent on the denoising effect of the methods to be compared: methods like GMCA or AMCA have the ability to provide denoised estimates of the sources while others are not. For that purpose, the following mixing matrix criterion $C_A$ is introduced:
\begin{equation}
C_A = \|{{\bf P A}^+} {\bf A^\star} - {\bf I}\|_{\ell_1}.
\end{equation}
where ${\bf P A}^+$ is the pseudo-inverse of the estimated mixing matrix corrected -- through $\bf P$ -- for the scale/permutation indeterminacies; ${\bf A^\star}$ is the true mixing matrix. Low values of $C_A$ then indicate better separation performances.\\
In the rest of the paper, all criteria will be computed as the median value of independent Monte-Carlo simulations. Unless specified, the number of simulations for each of the figures was set to $100$.

\subsection{Comparisons with standard sparse BSS methods}\label{sec:numerics:comparisons}

\subsubsection{An example of recovered sources}
\begin{figure}[tb]
\centerline{\includegraphics [scale=0.19]{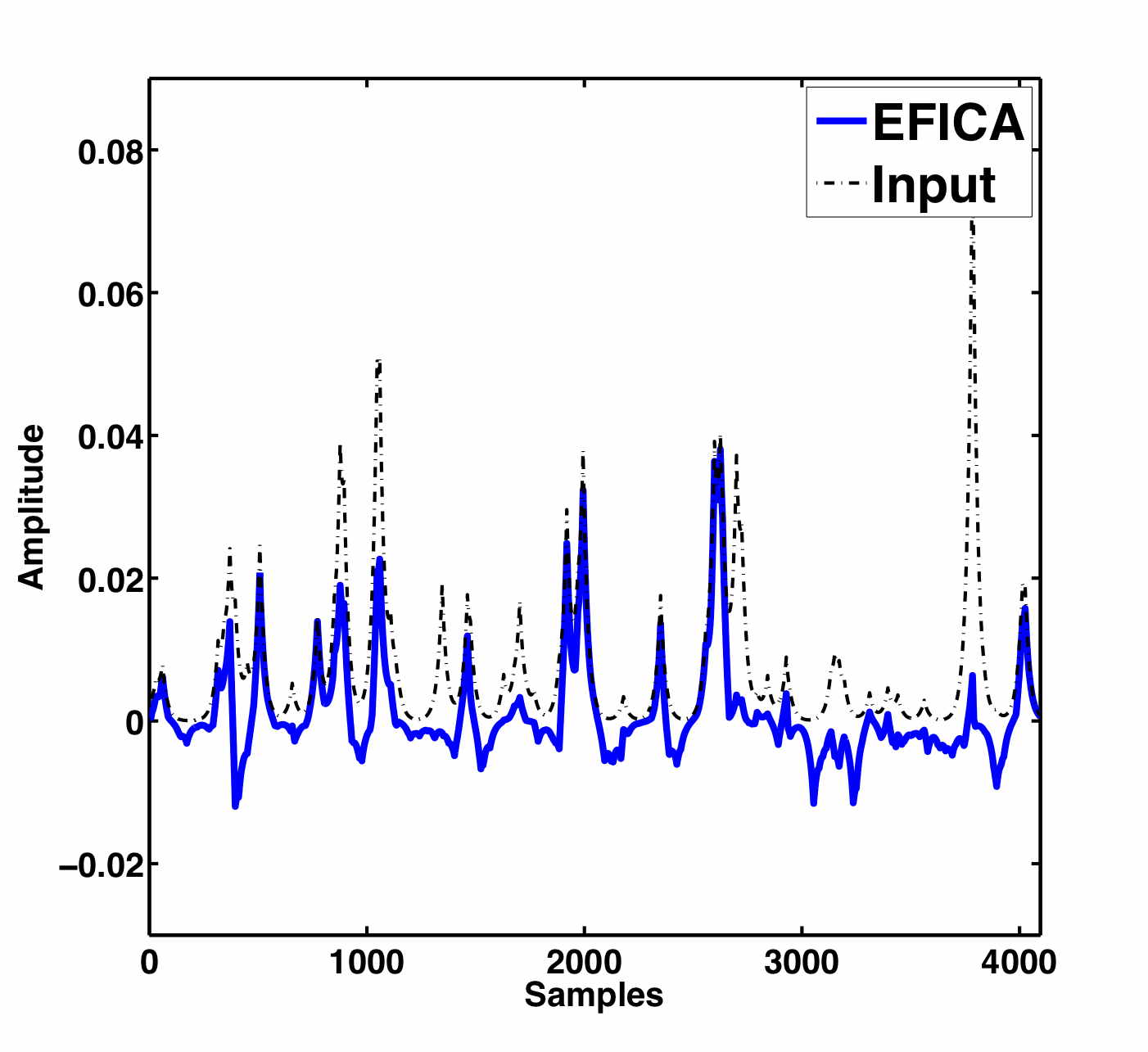}  \includegraphics [scale=0.17]{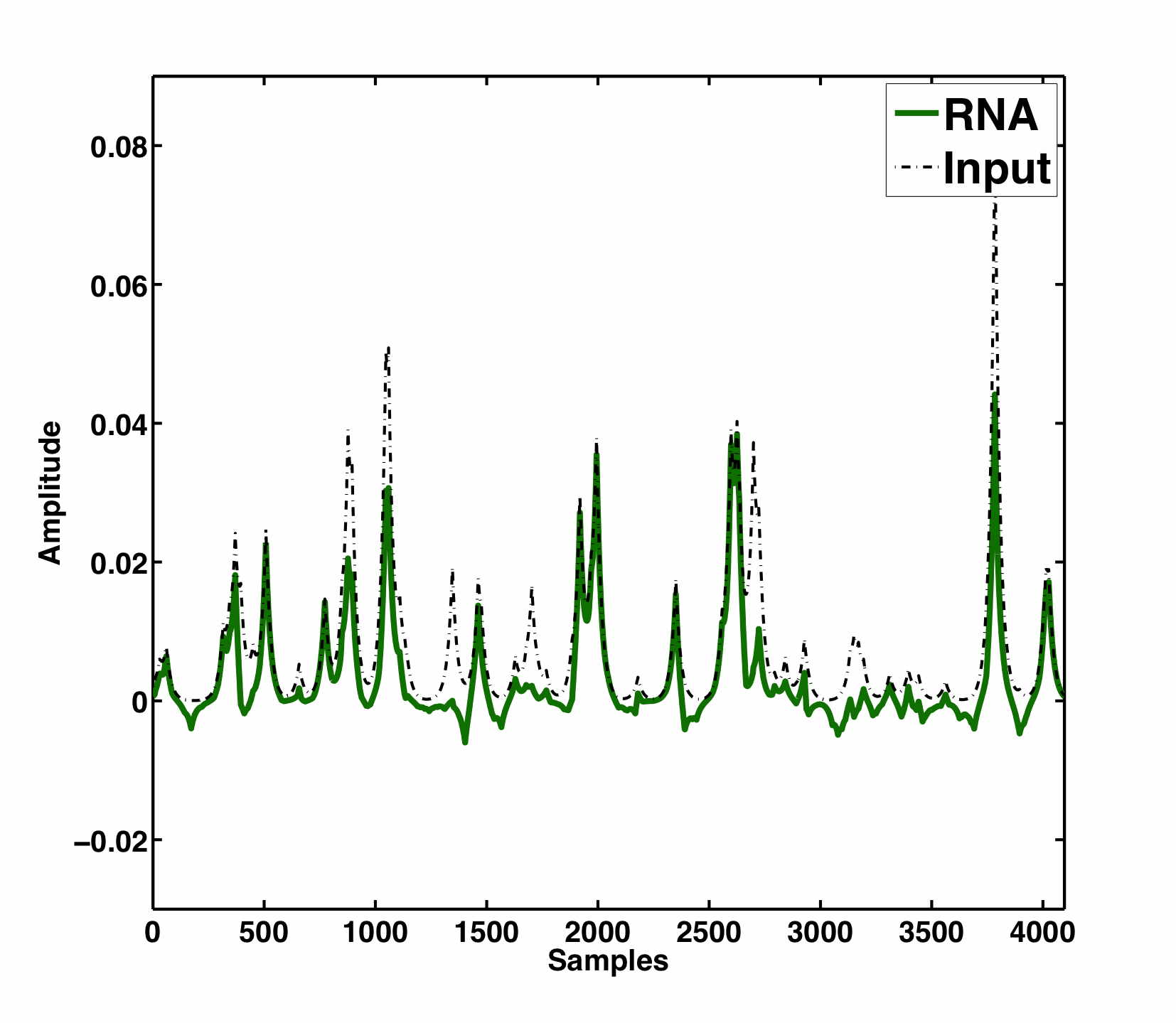}}
\centerline{\includegraphics [scale=0.19]{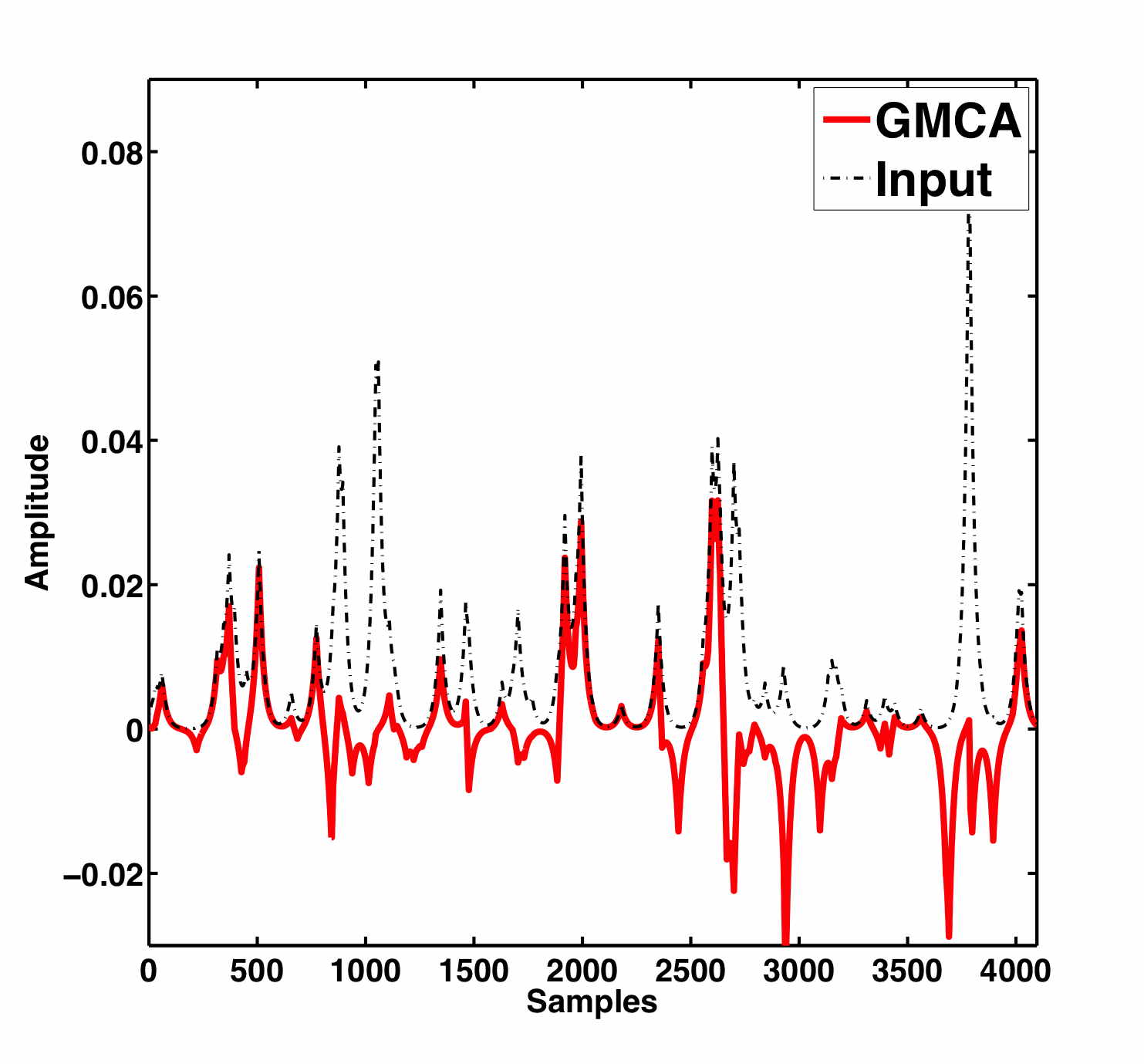} 
\includegraphics [scale=0.18]{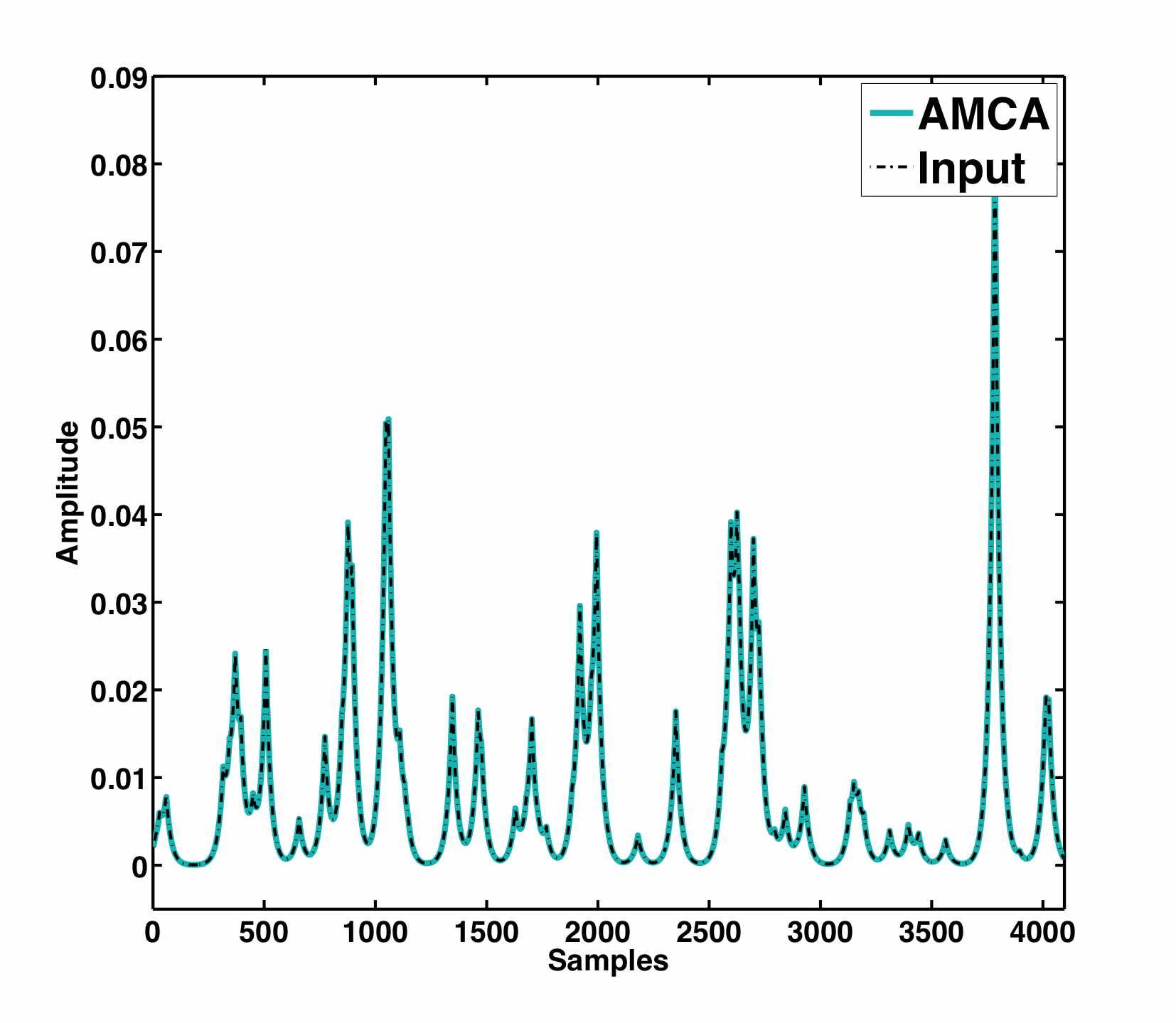}}
\caption{Estimated s.p.c. source. Top left : EFICA, right: RNA. Bottom left: GMCA, right: AMCA.}
\label{fig:sources_estimated}
\end{figure}

To better visualize how sparse BSS algorithms perform on s.p.c. sources, Figure~\ref{fig:sources_estimated} displays $2$ sources out of $10$ estimated from $10$ mixtures with $c = 0.6$ and $\tau = 4$. The number of observations $m$ as well as the number of sources are set to $10$. The number of samples per source is fixed to $4096$. Since $c=0.6$, this experiment turns out to be an arduous problem as $60\%$ of the active spikes are shared by all the sources. Figure~\ref{fig:sources_estimated} shows that standard BSS methods all seem to recover some of the most prominent spikes of the sought after source. However they all exhibit a certain level of interferences which indicates poor separation quality. The source estimated by AMCA does not visually exhibit such interferences and perfectly matches the source to be retrieved.

\subsubsection{Coherence level}
\begin{figure}[tb]
\centerline{\includegraphics [scale=0.25]{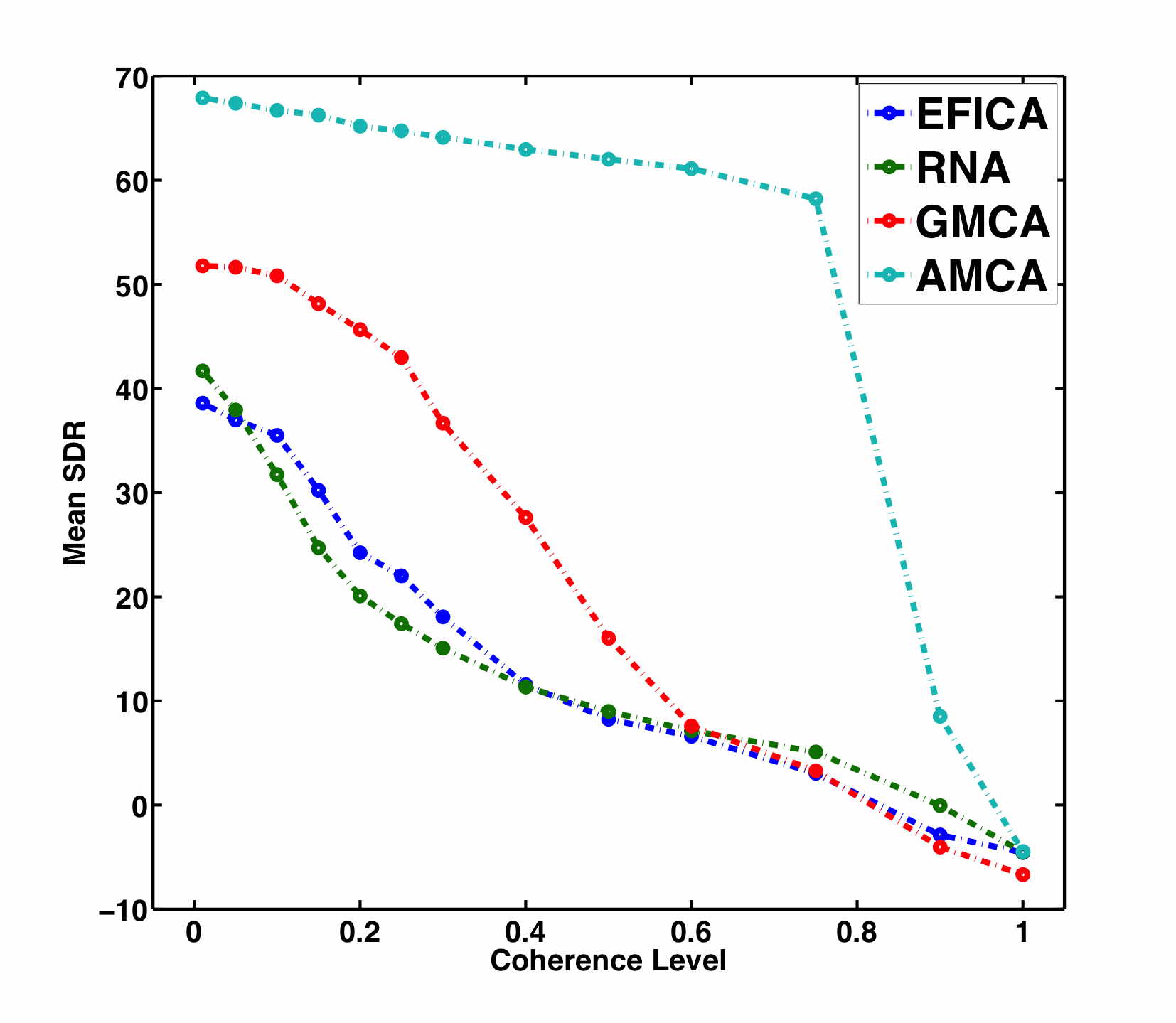}
\hfill
\includegraphics [scale=0.25]{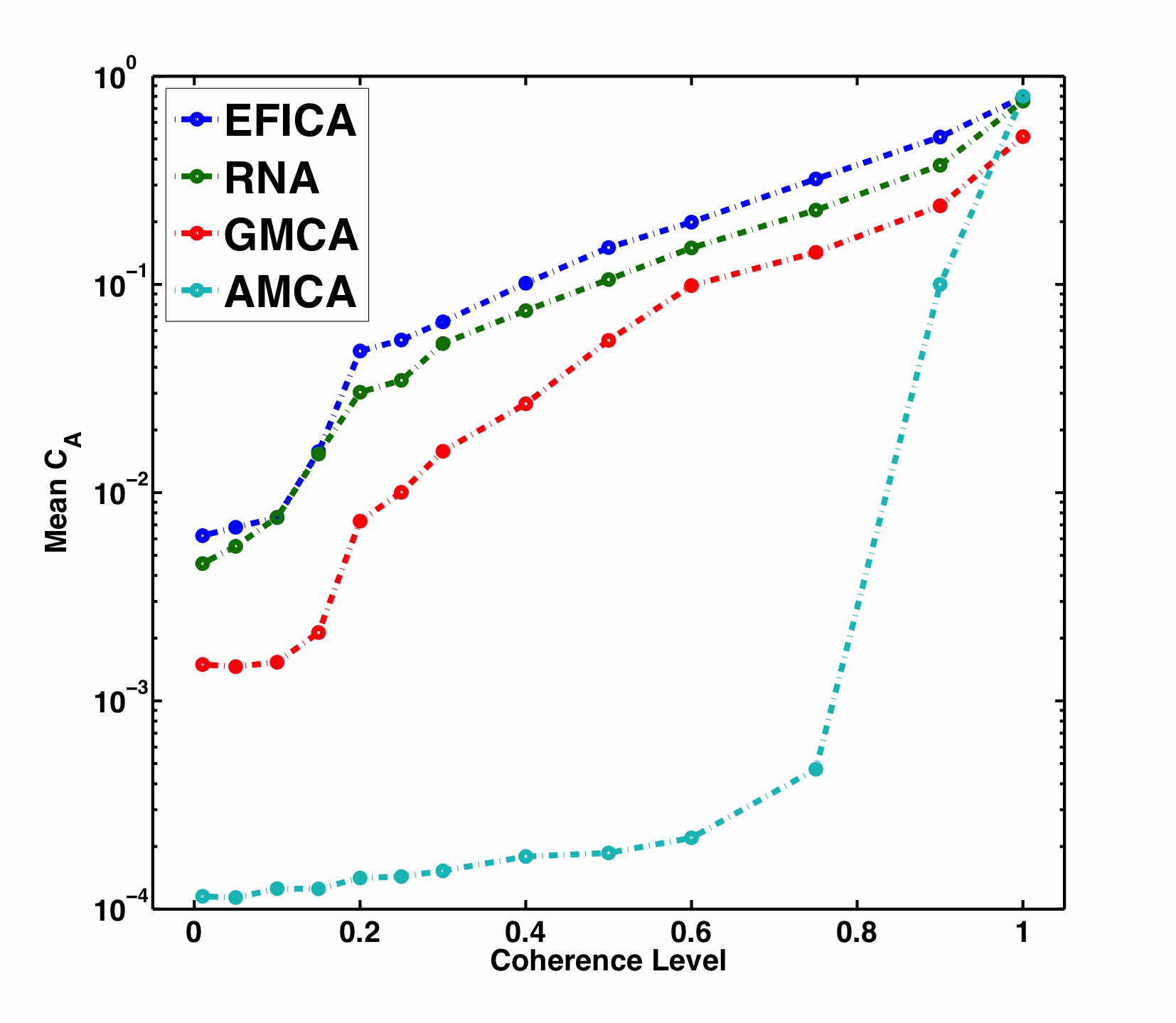}}
\caption{Evolution of the mean SDR (top) and the mixing matrix criterion (bottom) as a function of the coherence level. Each sample is the average over $100$ Monte-Carlo simulations.}
\label{fig:coherence_MeanSDR}
\end{figure}

\begin{figure}[tb]
\centering \includegraphics [scale=0.25]{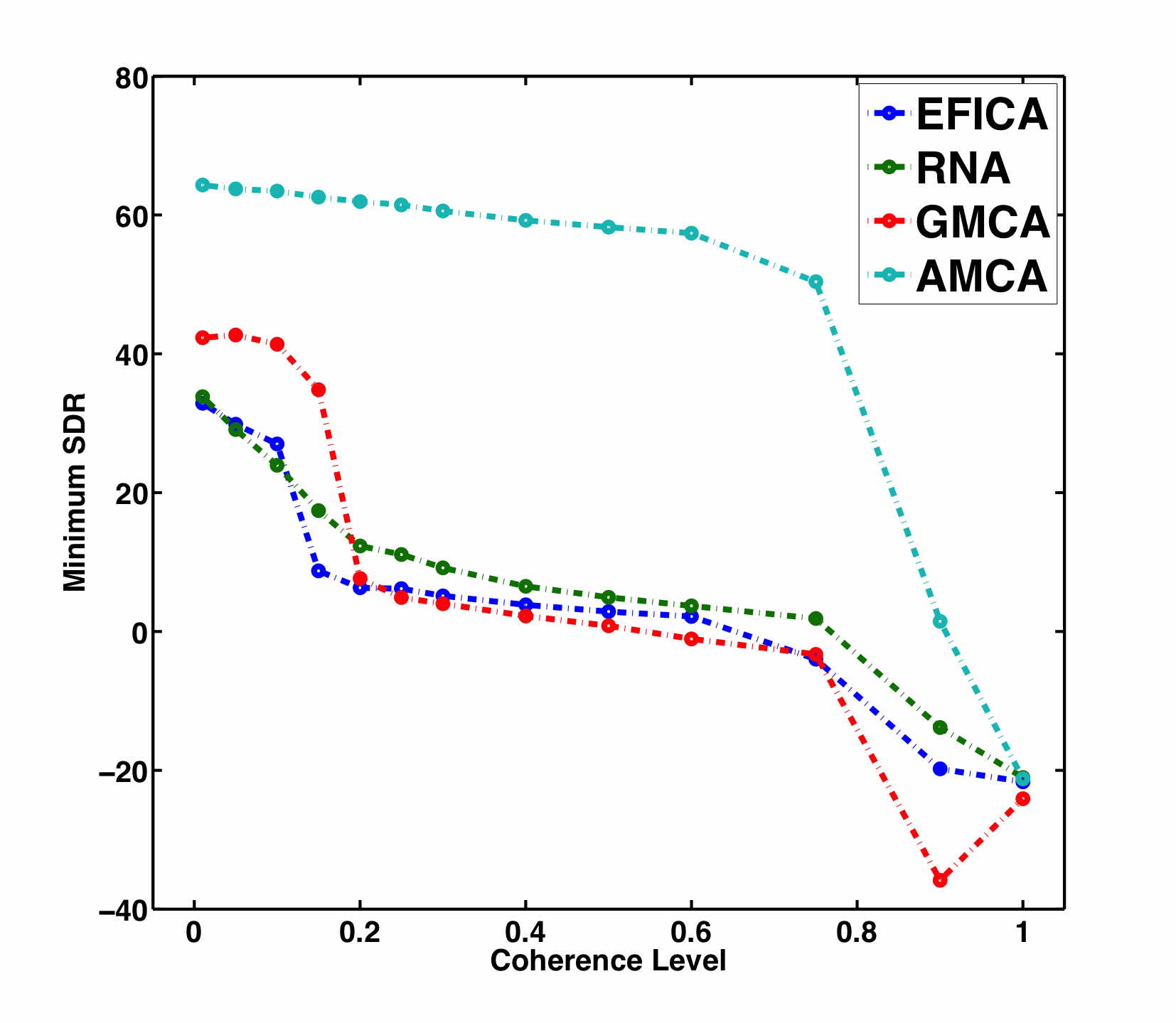}
\caption{Evolution of the minimal value of the SDR across the sources as a function of the coherence level. Each sample is the average over $100$ Monte-Carlo simulations.}
\label{fig:coherence_MinSDR}
\end{figure}

In this paragraph, the performances of the AMCA algorithm are evaluated when the level of correlation or more generally coherence, defined by the ratio $c = L/K$, varies between $0$ (the no-correlation case) and $1$ (the sources all share the same entries). Its values vary in the range $[0,1]$; when $c=0$, all the sources are independently distributed. The opposite extreme case $c=1$ corresponds to sources with exactly the same common active entries which means, according to our model, that the sources are fully correlated sources. The performances of standard sparse BSS methods should decrease for high valued of $c$. In this experiment, the number of sources and the number of channels are both fixed to $10$. The standard deviation of the entries in $\Omega_c$ is fixed to $\tau = 4$. The noise level is fixed to $120$dB. The evolution of the SDR as a function of the coherence level $c$ is displayed in the left panel of Figure~\ref{fig:coherence_MeanSDR}. First, it is interesting to notice that the performances of the standard BSS algorithms (GMCA, EFICA and RNA) are quickly hampered by the partial correlation between the sources. The GMCA algorithm seems to be less sensitive to the correlation of the sources for $c < 0.2$. When more than $20 \%$ of the entries are shared by all the sources, these algorithms behave equally badly. The AMCA algorithm already shows a much higher SDR even when only a few entries are correlated. When less than $80\%$ of the entries are correlated, the SDR of the AMCA algorithm is higher than $60$ dB. Its performances start degrading for $c > 0.8$ to eventually behave similarly to the standard BSS algorithm when $c$ tends towards $1$. The evolution of the mixing matrix criterion is shown in the right panel of Figure~\ref{fig:coherence_MeanSDR}. One can notice that the mixing matrix criterion with the AMCA algorithm is consistently more than $1$ order of magnitudes lower than for the standard BSS methods. Figure~\ref{fig:coherence_MinSDR} displays the average over $100$ random simulations of the minimum SDR over the estimated sources; in other words this quantity measures the SDR the most badly estimated source for each simulation. This figure shows that the GMCA algorithm is very good at recovering some of the sources but not all with the same quality. From that respect, the standard BSS methods behave quite similarly~: for $c > 0.2$ they are not able to identify all the sources (the minimum SDR is close to $0$ dB). The minimum SDR is $50$ dB higher for $c < 0.8$ with the AMCA algorithm. Consequently, the AMCA algorithm is good at recovering all the sources even when the number of common spikes in the sources is equal to $80 \%$.

\subsubsection{Dynamic range}
\begin{figure}[tb]
\centerline{\includegraphics [scale=0.25]{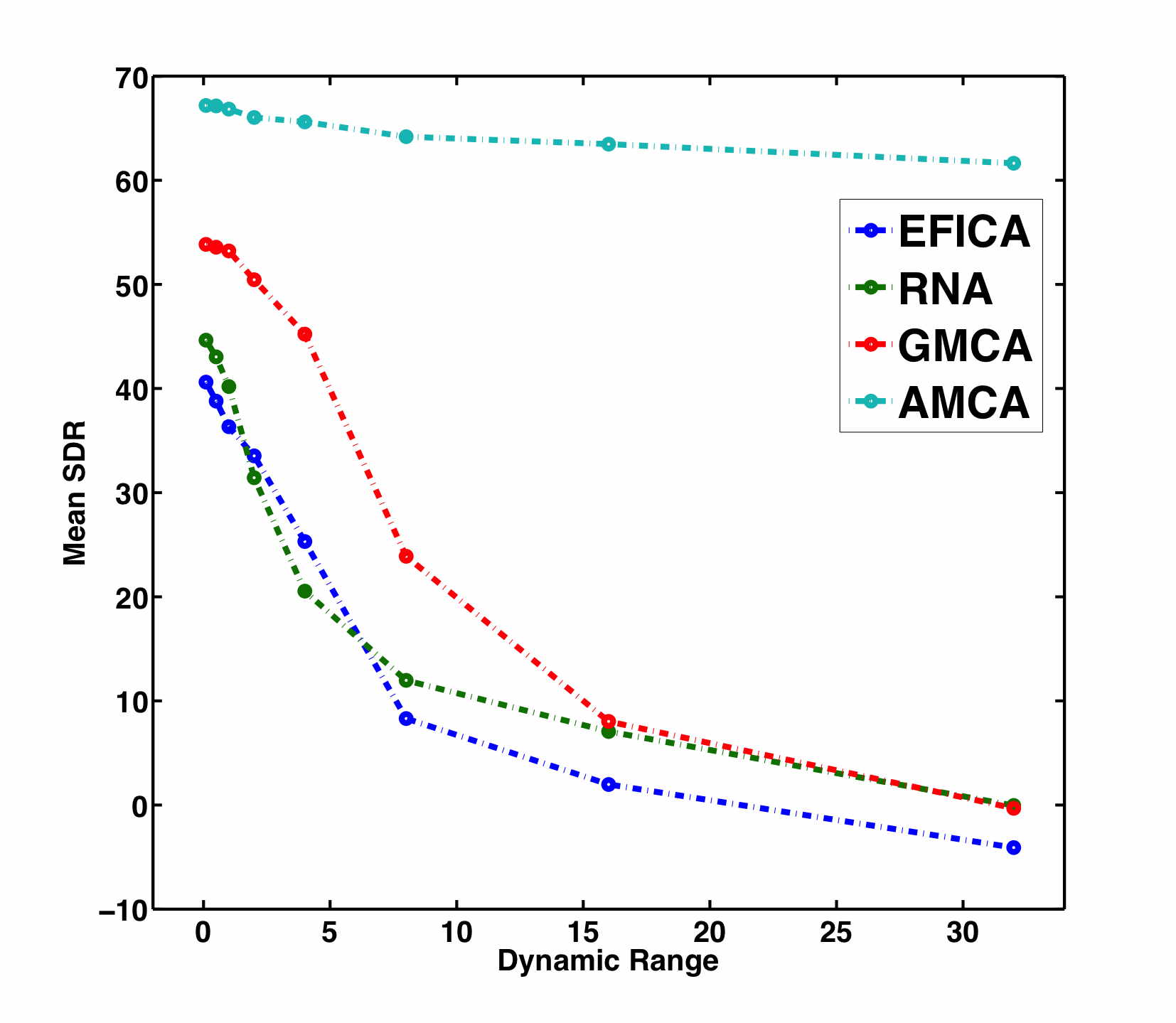}
\hfill
\includegraphics [scale=0.25]{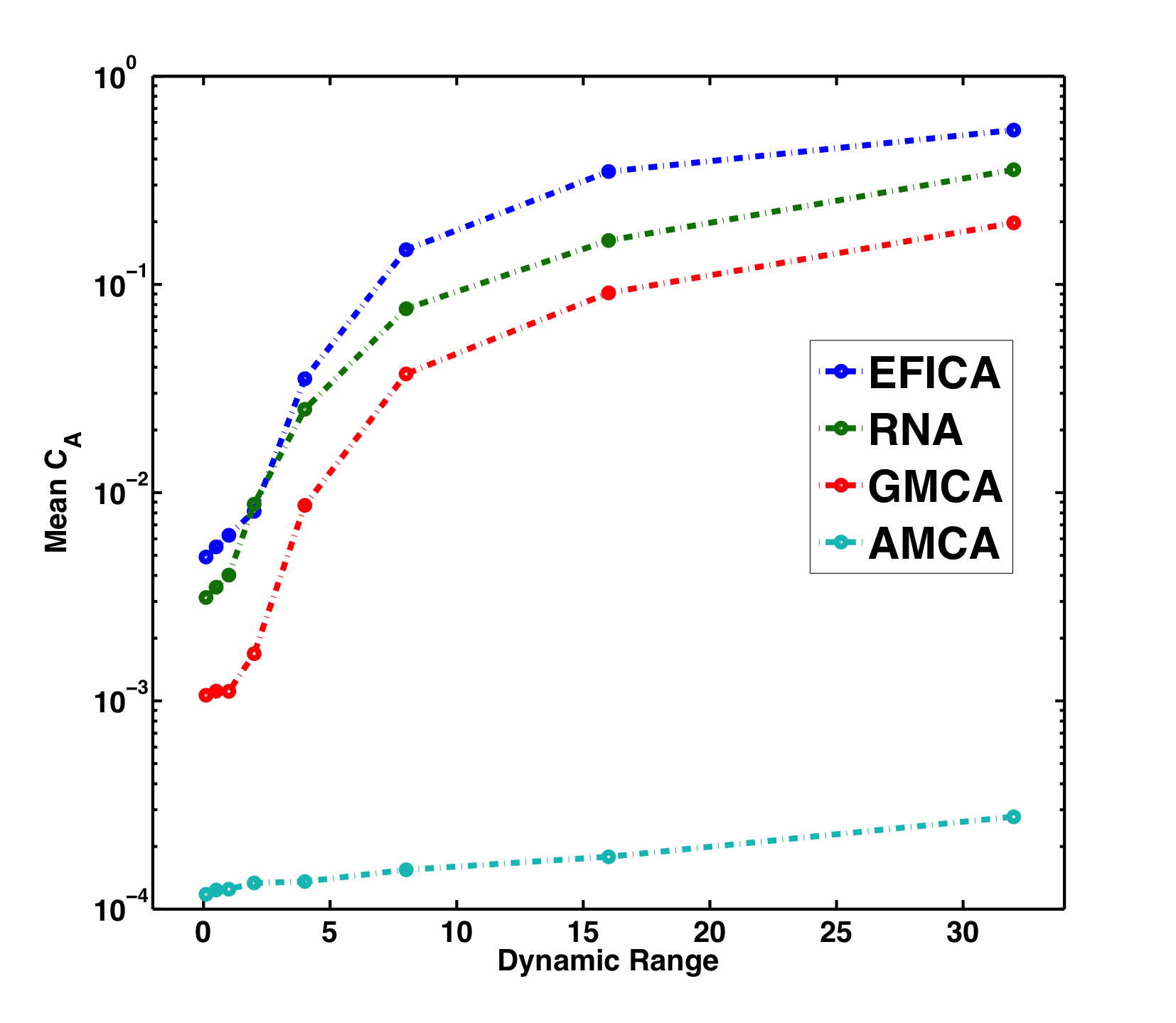}}
\caption{Evolution of the mean SDR (top) and the mixing matrix criterion (bottom) as a function of the dynamic range. Each sample is the average over $100$ Monte-Carlo simulations.}
\label{fig:DR_MeanSDR}
\end{figure}

A second parameter which is very likely to hamper the performances of standard BSS methods is the dynamic range between between the amplitudes of correlated and non-correlated entries. More precisely, standard sparse BSS algorithms are highly sensitive to the coefficients of the sources which have the highest amplitudes. This makes perfect sense when morphological diversity holds true: the most significant entries of the sources are most likely the most discriminant. Furthermore, we claimed earlier in this paper that the weighting strategy of the AMCA algorithm penalizes more high amplitude coefficients which are shared by several sources. The AMCA algorithm should therefore be less sensitive to the dynamic range between correlated and non-correlated samples of the sources than the standard BSS methods. In this experiment, the number of sources and the number of channels are both fixed to $10$. The proportion of correlated entries between the sources is fixed to $c = L/K = 0.2$. The noise level is fixed to $120$dB. Figure~\ref{fig:DR_MeanSDR} features the evolution of the SDR ({\it resp.} mixing matrix criterion $C_A$) in the left ({\it resp.} right) panel as a function of the standard deviation of the correlated entries of the sources $\tau$ in the interval $[0.1,32]$. For $\tau < 2$ ({\it i.e.} on average, correlated entries have twice the amplitude of the non-correlated entries), standard BSS methods all behave quite well with SDR values higher than $40$ dB. For higher values of $\tau$ their performances dramatically decrease towards low SDR values. As expected, the AMCA algorithm is only slightly impacted by the variation of the dynamic range. Even for $\tau = 32$, the SDR of the sources with the AMCA algorithm is higher than $60$dB and the corresponding mixing matrix criterion is about $3$ orders of magnitude lower than with the standard BSS algorithms.\\

\subsubsection{Number of sources}

\begin{figure}[tb]
\centerline{\includegraphics [scale=0.25]{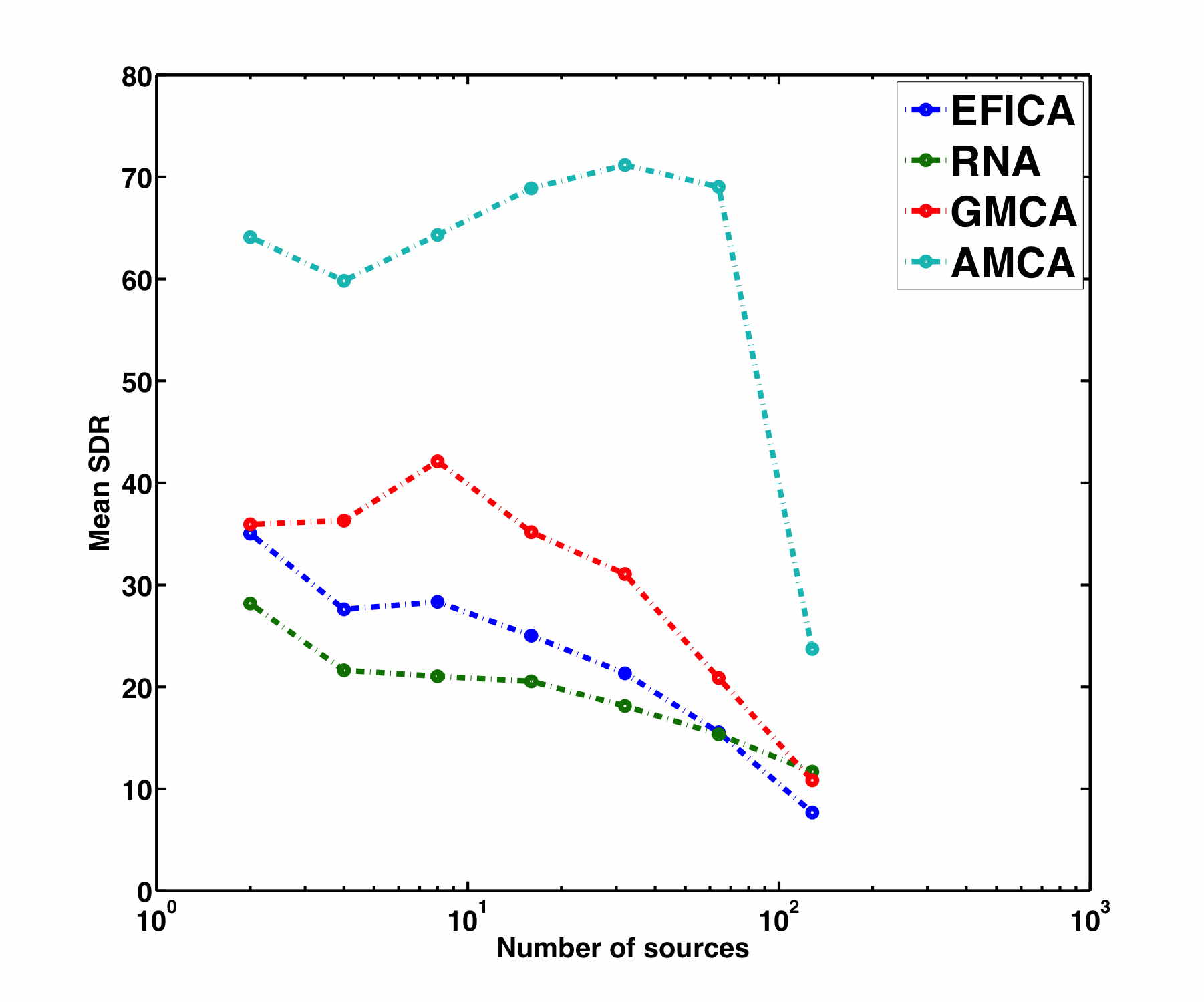}
\hfill
\includegraphics [scale=0.25]{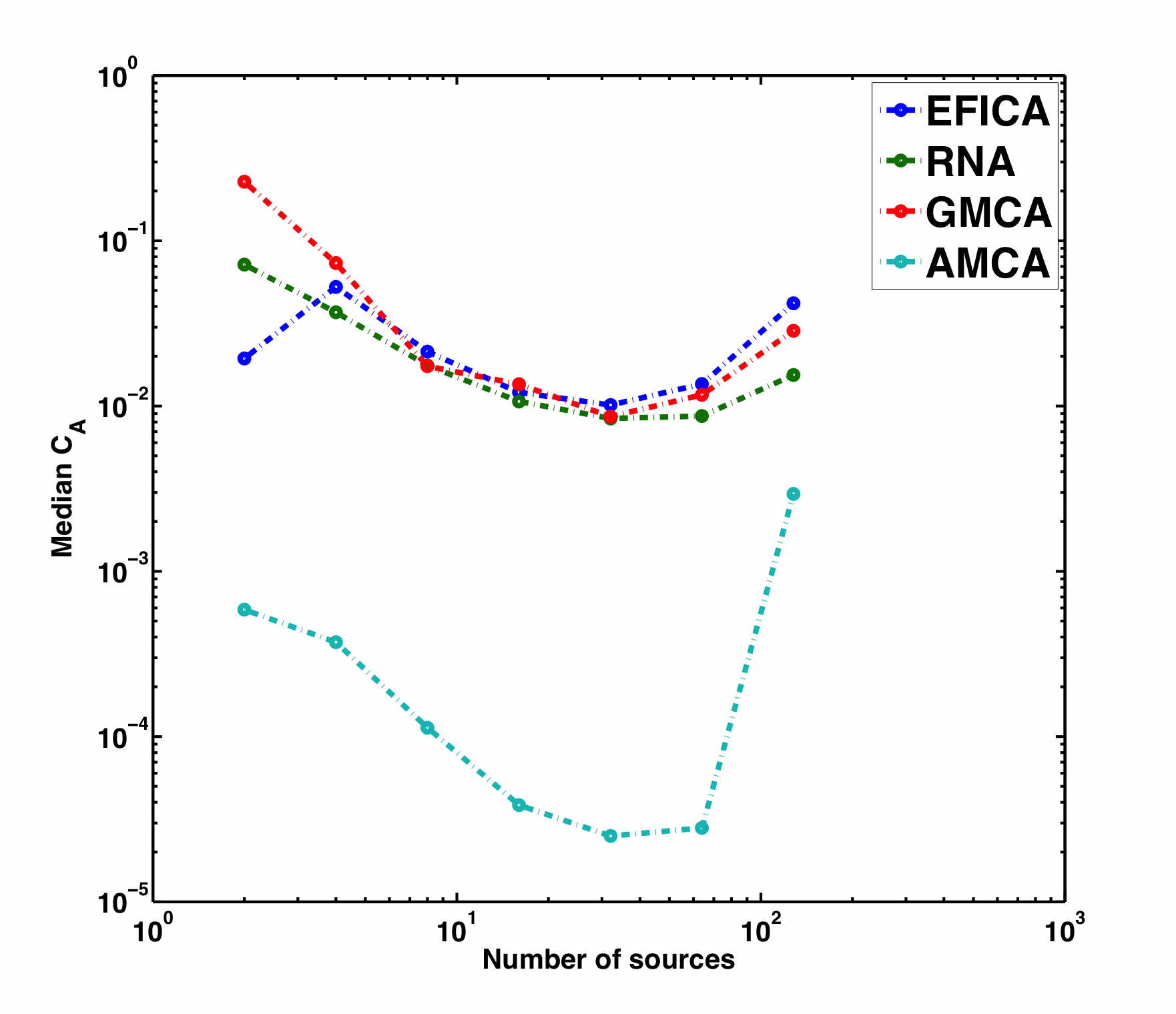}}
\caption{Evolution of the mean SDR (top) and the mixing matrix criterion (bottom) as a function of the number of sources. Each sample is the average over $100$ Monte-Carlo simulations.}
\label{fig:NS_MeanSDR}
\end{figure}

Recovering a large number of sources is generally a complicated task. Indeed, for a fixed number of samples $T$, the ability of source separation techniques to disentangle a growing number of sources is rapidly limited. This is true in the case of s.p.c. sources as the number of entries of the sources which are most relevant for the separation are also limited. In the next experiment, the number of sources and the number of channels are both fixed to the same value. The proportion of correlated entries is fixed to $c = L/K = 0.2$. The noise level is fixed to $120$dB and $\tau = 4$. Figure~\ref{fig:NS_MeanSDR} displays the evolution of the SDR on the left panel and $C_A$ on the right panel as a function of the number sources in a range $2 \leq n_s \leq 128$. As expected the performances of the standard sparse BSS algorithms, as measured by the SDR, decay as the number of sources to be recovered grows. The AMCA algorithm provides good results (the SDR is higher than $60$dB) for less than roughly $64$ sources. For a higher number of sources, the performances of the AMCA decrease quite rapidly. However, one has to keep in mind that the number of samples per source is kept fixed to $T=4096$ from which $2\%$ are active, that is to say on average $~81$. Amongst these active samples, only $80\%$ have been drawn independently. This means that each source has about $65$ likely discriminant samples out of $4069$. When the number of sources increases, the probability for two independently drawn samples to take simultaneously significant amplitudes grows. This might become likely when the number of sources is large which explains why the performances of AMCA decay for $n > 64$. The evolution of the mixing matrix criterion has to be analyzed with care as the performances of all the methods seem to improve as the number of sources increases while the evolution of the SDR shows a quite different behavior. It has to be noticed that the mixing matrix criterion averages the scalar product of the rows of the inverse of the estimated mixing matrix and the true mixing matrix. In other words, it looks at a quantity that only involves interferences between two sources. On the other hand, the SDR is a global measure of distortion that impacts each of the sources. Precisely, for a fixed value of the mixing matrix criterion, the distortion will increase with the number of sources which explains why the actual SDR of the standard sparse BSS methods decrease with $n$. 

\subsection{The AMCA algorithm and noise}\label{sec:numerics:noise}

\paragraph*{Discussion about the impact of noise}
In this section, we discuss the impact of the re-weighting scheme on the performances of the AMCA in the noisy setting. First, in the AMCA algorithm, the weights $w_q$ are estimated from the estimated sources. These sources are obtained via Step 1 of the AMCA algorithm.\\
In the low noise limit, one interesting feature of the proposed re-weighting scheme is that it is inversely proportional to the amplitude of the columns of $\bf S$. More precisely, this entails that large entries of $\bf S$ which are shared by several sources are more penalized than small entries with the same relative distribution across the sources. Strongly penalizing large and correlated entries is desirable since they are detrimental to the estimation of the mixing matrix and the sources.\\
In the noisy setting, the situation turns out to be rather different since small-amplitude samples are more likely perturbed by noise than large amplitude entries. On the one hand, the proposed re-weighting procedure might be disastrous for the separation of the sources whether they are partially correlated or not. Indeed, since the weights $w_q$ are inversely proportional to the amplitude of the columns of the sources, the proposed procedure will tend to favor small entries which are more affected by the presence of noise. On the other hand, Step 1 of the AMCA algorithm rejects entries with amplitudes smaller than some prescribed noise-dependent level. This should lower the impact of noise on the performances of the AMCA algorithm.

\paragraph*{Comparisons}

\begin{figure}[tb]
\centerline{\includegraphics [scale=0.25]{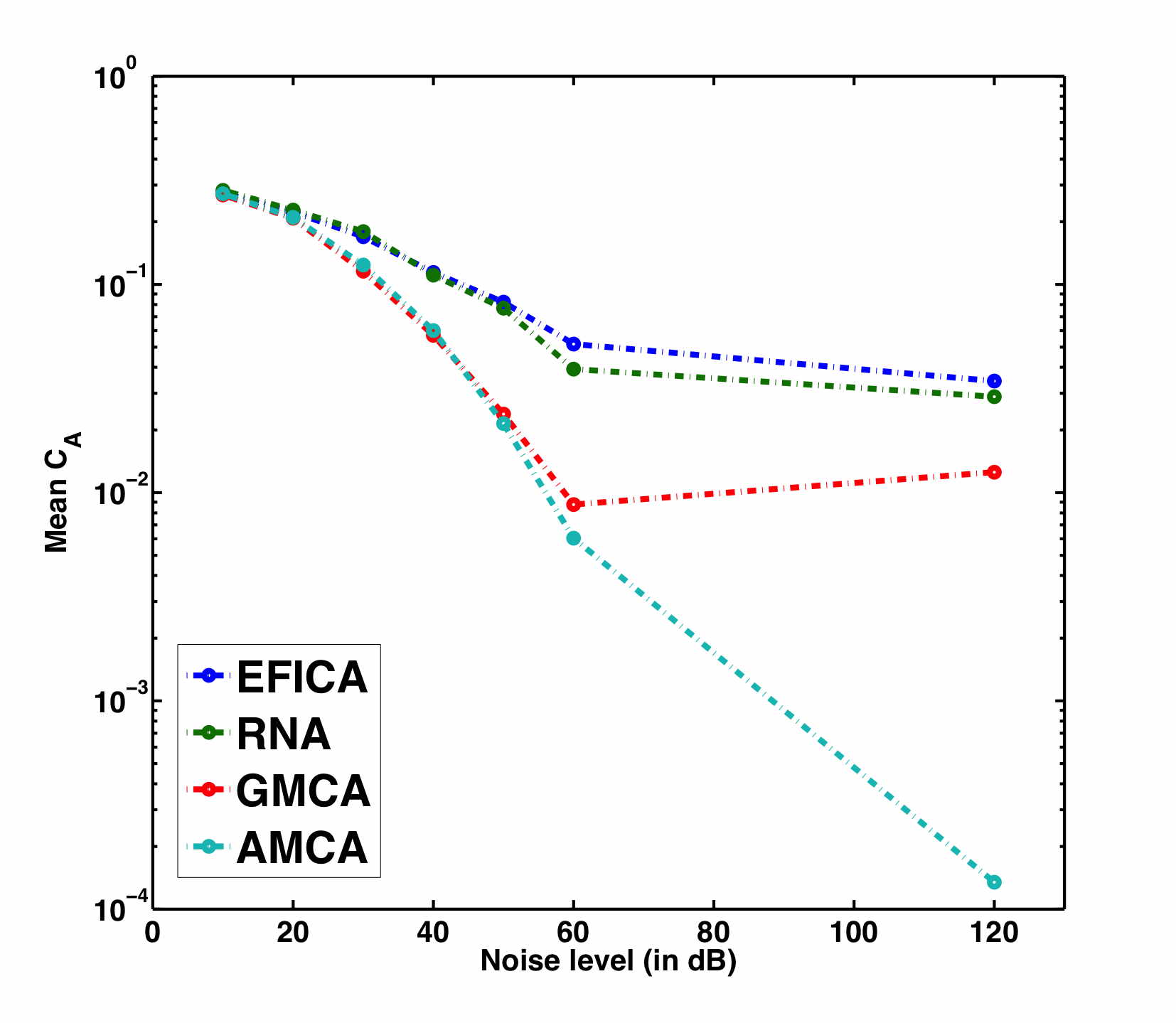}}
\caption{Evolution of $C_A$ as a function of the noise level (in dB). Each sample is the average over $100$ Monte-Carlo simulations.}
\label{fig:NoiseLevel_CA}
\end{figure}

In \cite{starck:bobin07}, the authors demonstrated that the GMCA algorithm is robust to additive noise contamination. This is especially true whenever morphological diversity holds; in that case the most discriminant sources are the entries of the sources with the most significant amplitudes. It turns out these entries are also the least contaminated by additive noise. In the case of s.p.c. sources, the most discriminant sources are not necessarily the large-amplitude samples. A first consequence is that noise will very likely have a strong impact on the quality of the separation.\\
 In this experiment, the number of sources and the number of channels are both fixed to $10$. The proportion of correlated entries of the sources is fixed to $c = L/K = 0.2$, $\tau$ is set to $4$. The noise level varies in the range $[20,120]$dB. The evolution of the mixing matrix criterion is featured in Figure~\ref{fig:NoiseLevel_CA}. It first reveals that the GMCA-based methods have slightly better performances when the noise level increases. When the noise level tends towards $20$dB all these methods tend to all perform badly. More surprisingly, the performances of the AMCA algorithm do not seem to be more impacted by the presence of noise than standard BSS methods. To better understand this behavior, let us recall that, in the AMCA algorithm, the mixing matrix is updated from the current estimate of the sources which is naturally denoised in Step 1 of the algorithm. This means that coefficients of the sources which are of the order of or below the noise level are naturally rejected in Step 1; the estimation of the mixing matrix won't therefore account for these entries. In other words the re-weighting procedure won't ``amplify" the impact of the noise. Eventually, for high noise levels, only high amplitude coefficients will be detected. In this regime, the GMCA and AMCA algorithms will be similarly handicapped by correlated large-amplitude coefficients.

%\begin{figure}[tb]
%\centerline{\includegraphics [scale=0.3]{Figures/vsGMCA/NaturalImages_CA.jpg}}
%\caption{Evolution of $C_A$ as a function of the noise level (in dB). Each sample is the average over $100$ Monte-Carlo simulations.}
%\label{fig:NoiseLevel_CA_NI}
%\end{figure}

%
%
%  I AM HERE
%
%

\subsection{Application in Astrophysics}\label{sec:appli_astro}
\begin{figure}[tb]
\centerline{\includegraphics [scale=0.23]{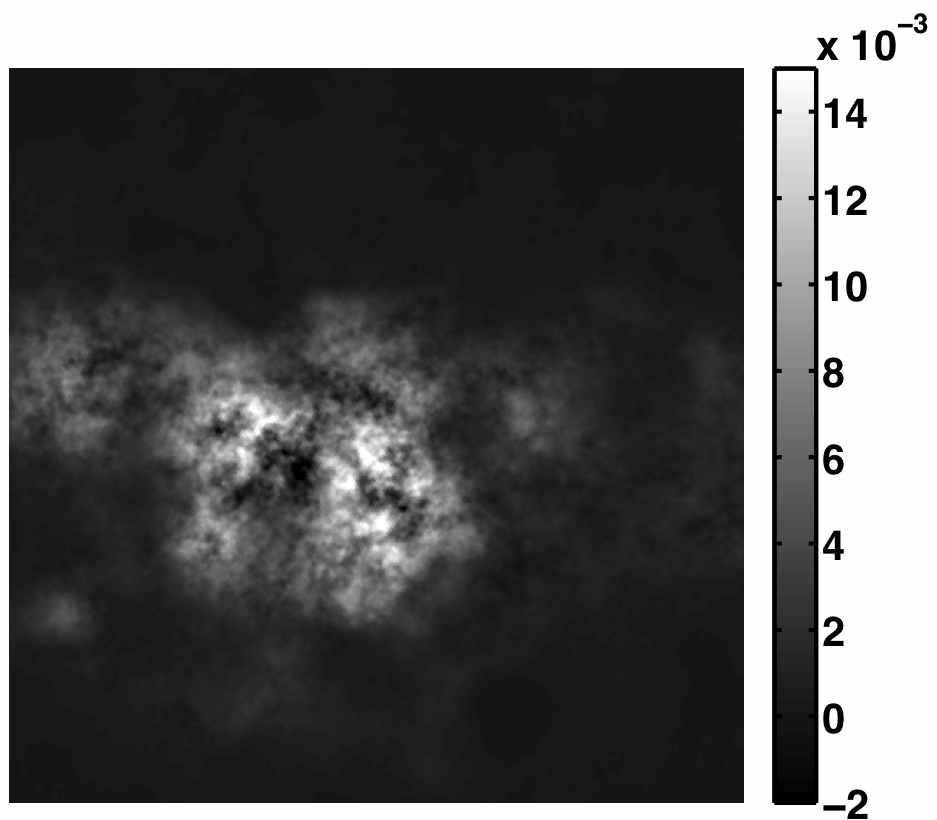} \includegraphics [scale=0.23]{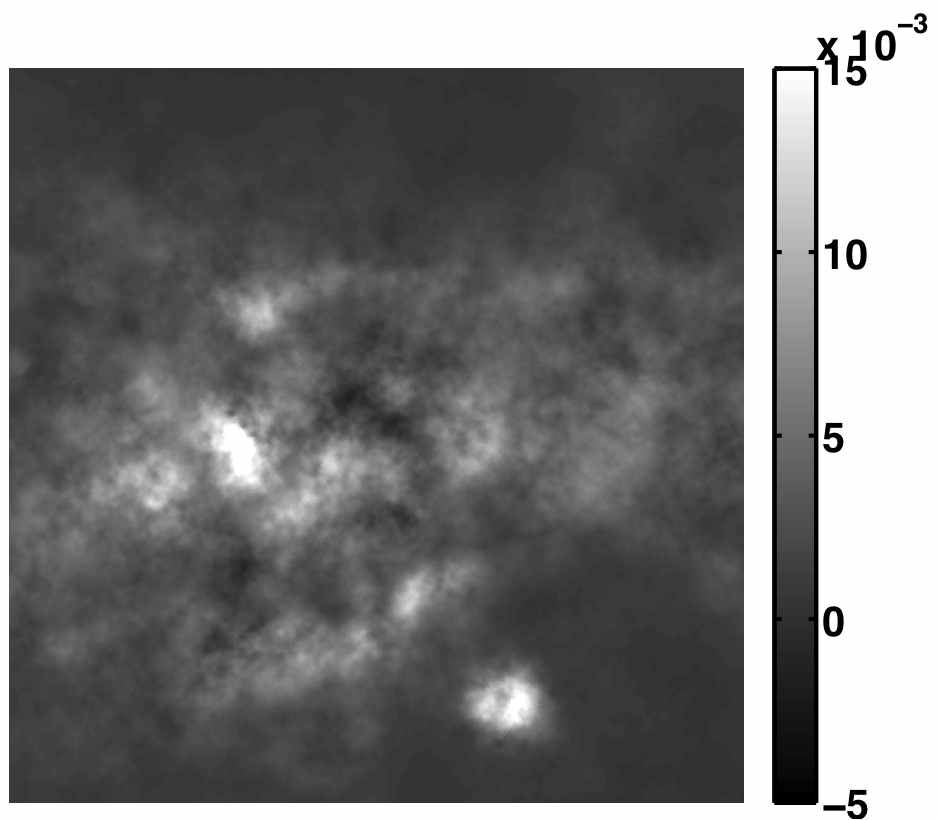}}
\centerline{\includegraphics [scale=0.23]{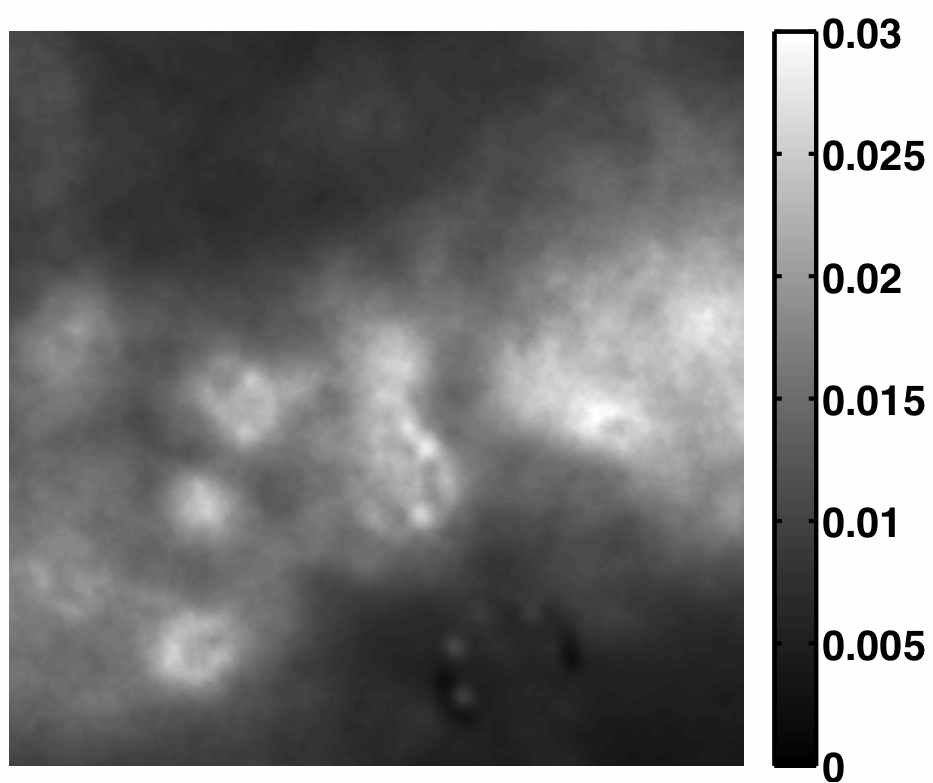} \includegraphics [scale=0.23]{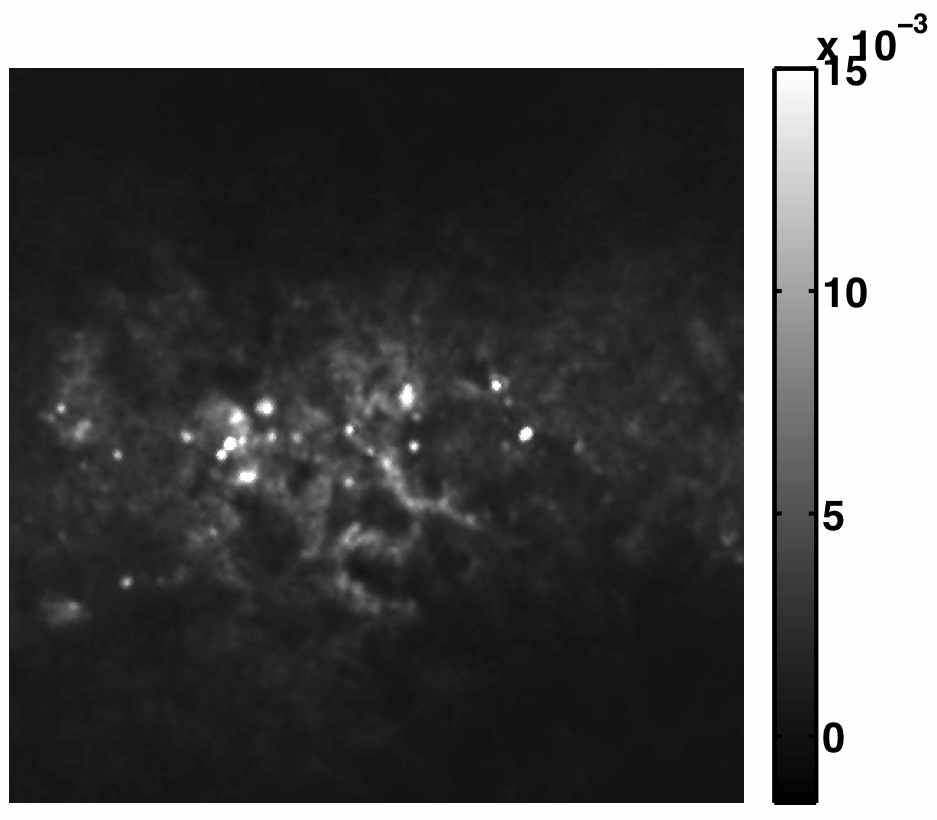}}
\caption{WMAP data : the $4$ input sources.}
\label{fig:wmap_inputsources}
\end{figure}

\begin{figure}[tb]
\centerline{\includegraphics [scale=0.23]{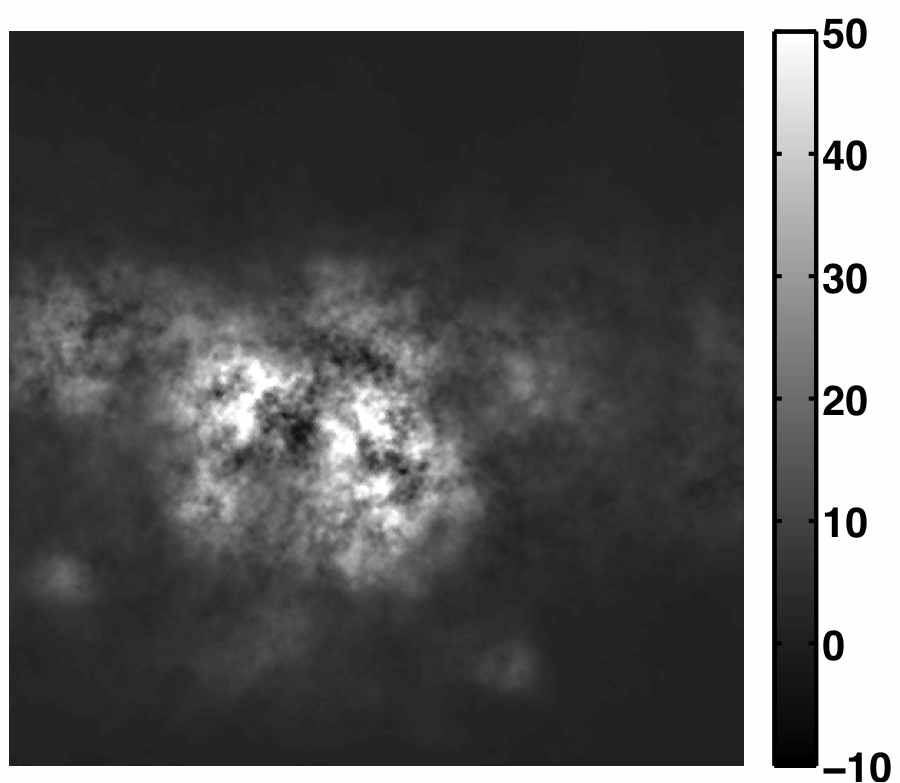} \includegraphics [scale=0.23]{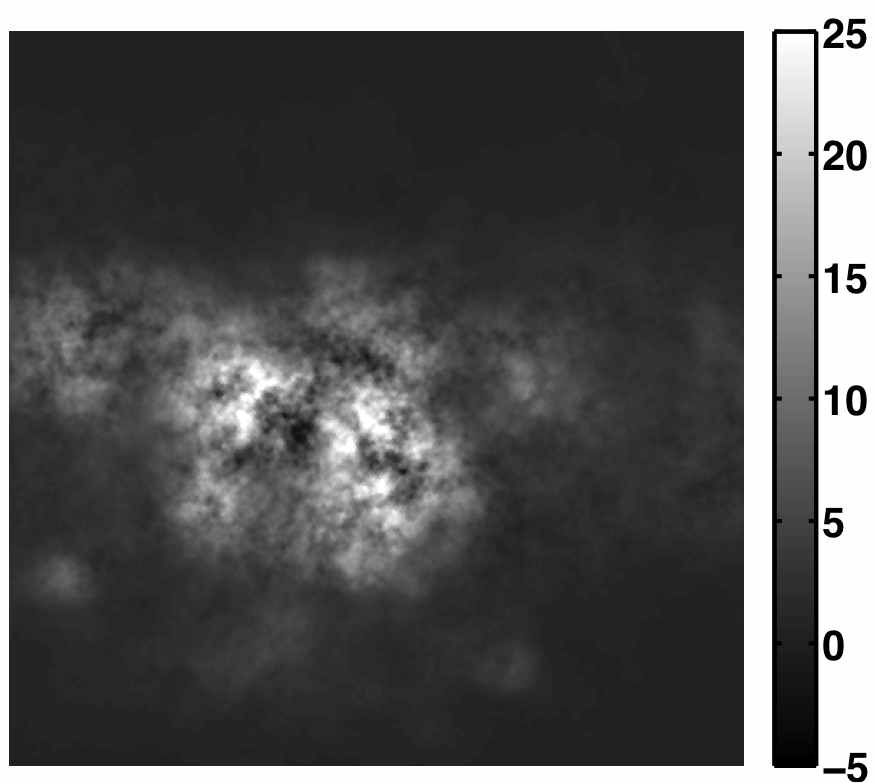}}
\centerline{\includegraphics [scale=0.23]{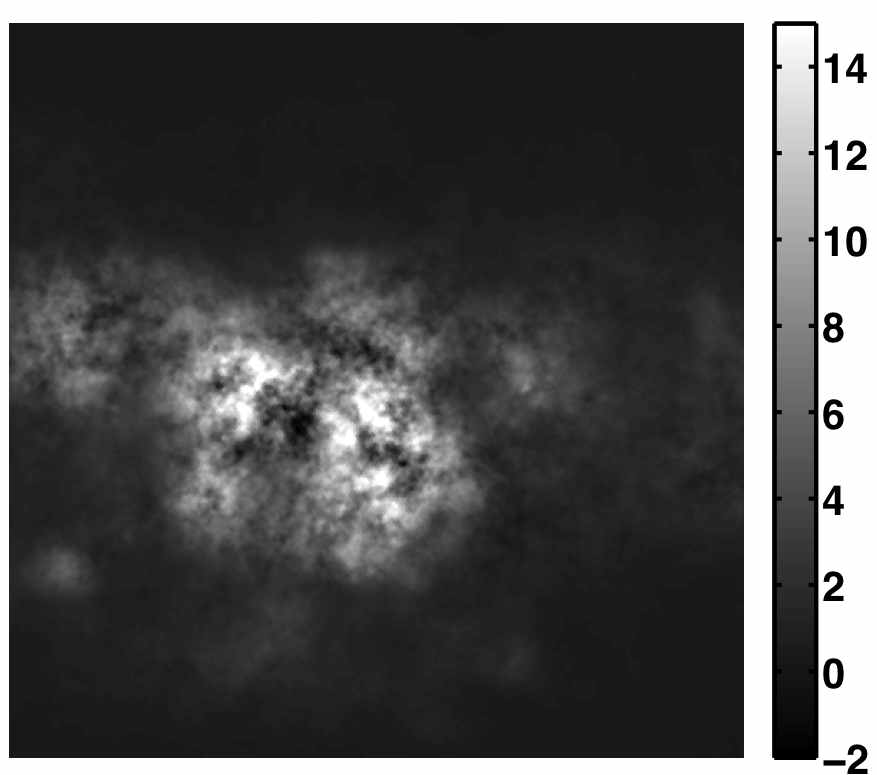} \includegraphics [scale=0.23]{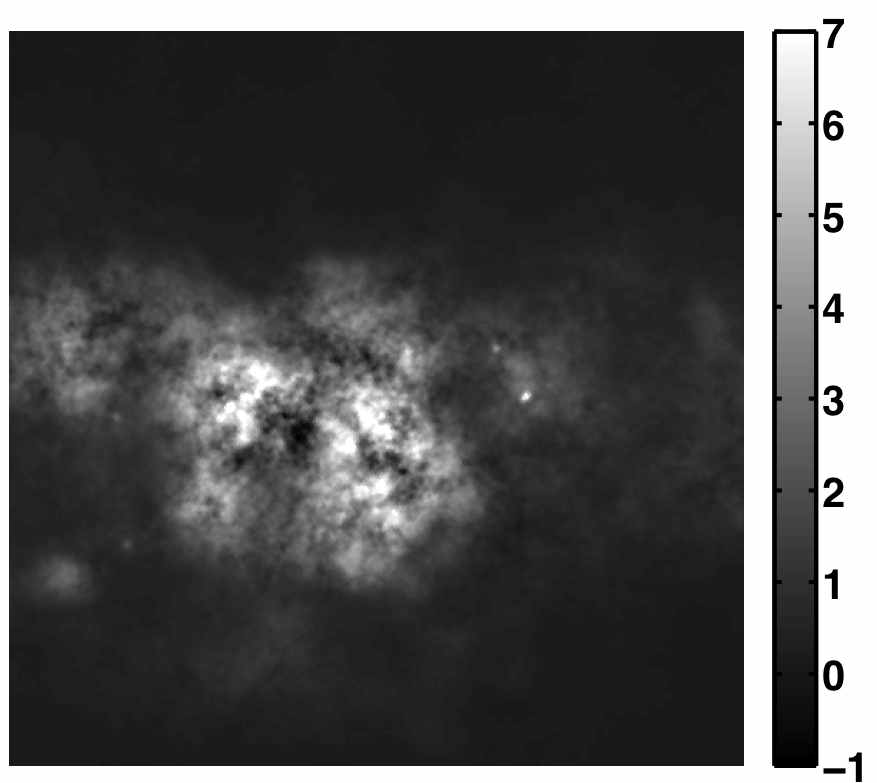}}
\centerline{\includegraphics [scale=0.23]{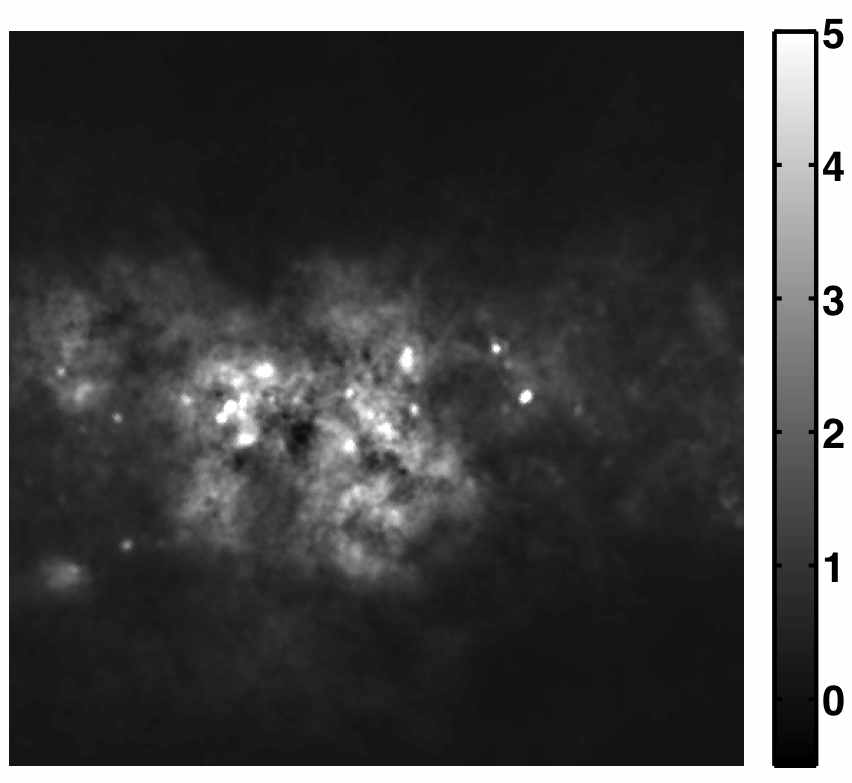} }
\caption{WMAP data : the $5$ observations.}
\label{fig:wmap_data}
\end{figure}

\begin{figure}[tb]
\centerline{\includegraphics [scale=0.23]{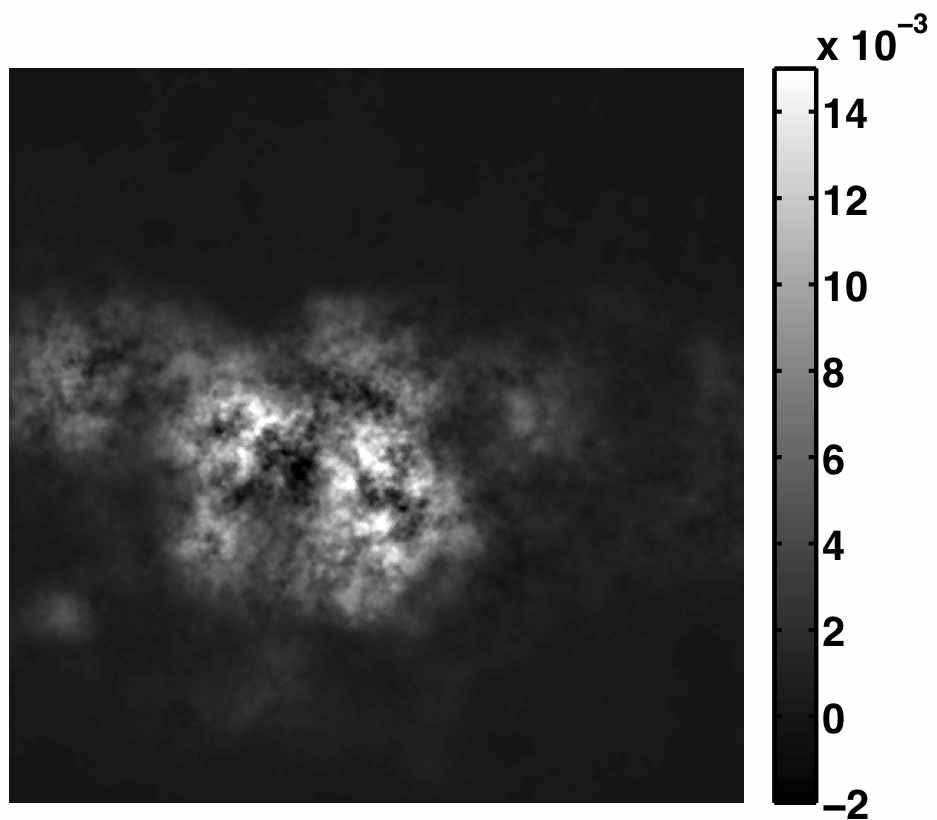} \includegraphics [scale=0.23]{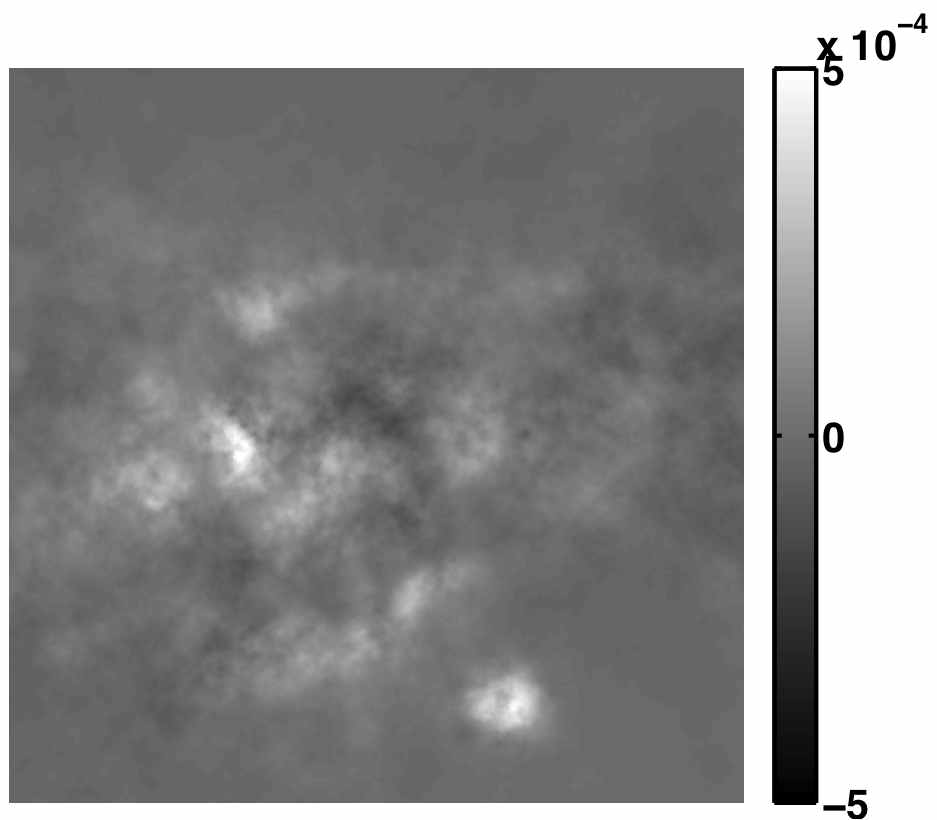}}
\centerline{\includegraphics [scale=0.23]{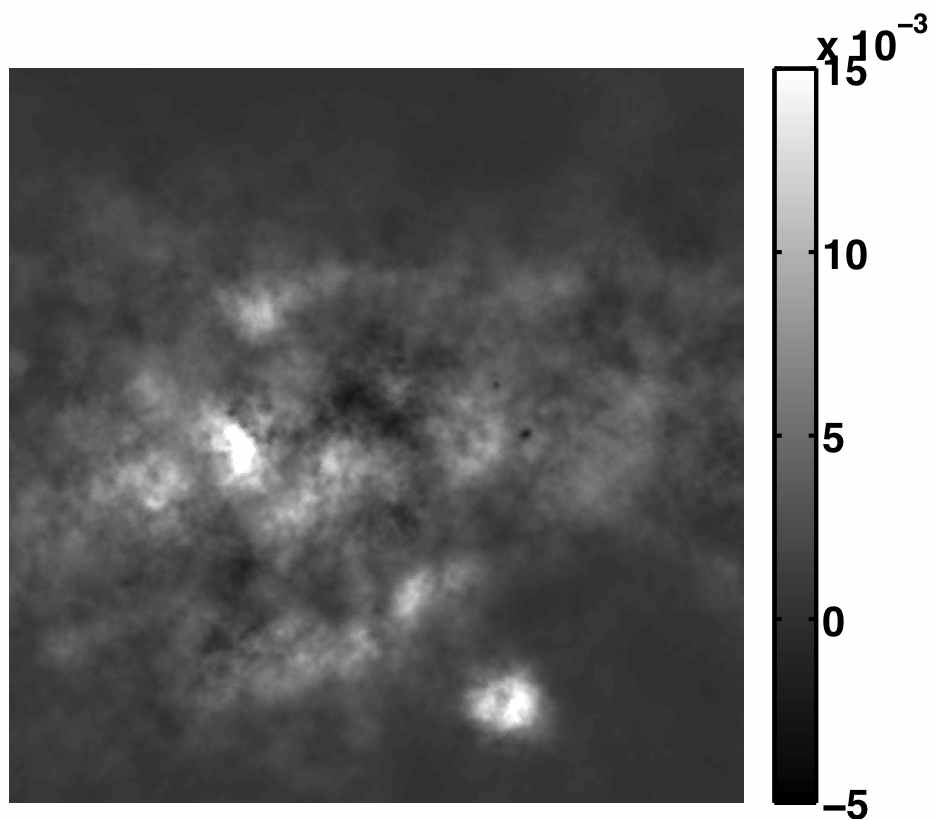} \includegraphics [scale=0.23]{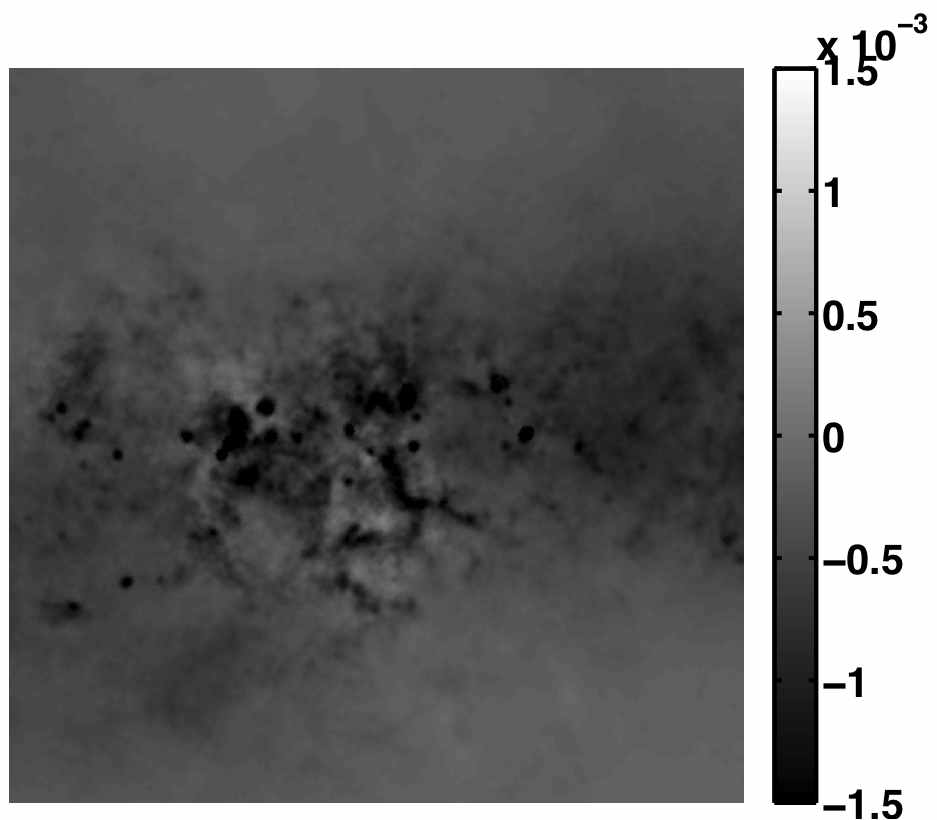}}
\centerline{\includegraphics [scale=0.23]{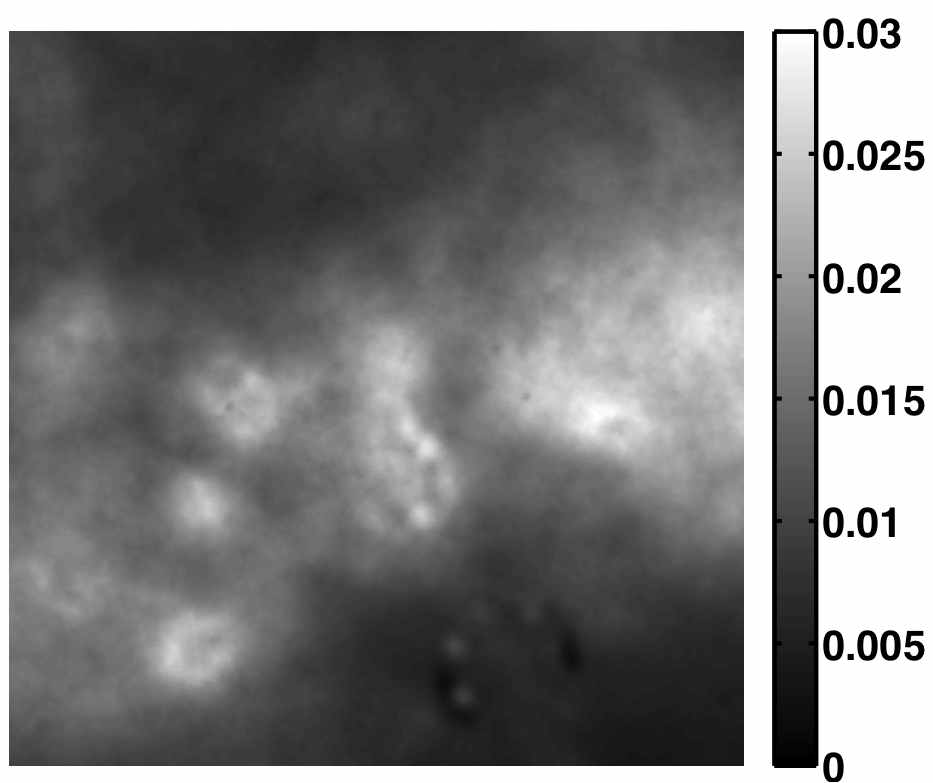} \includegraphics [scale=0.23]{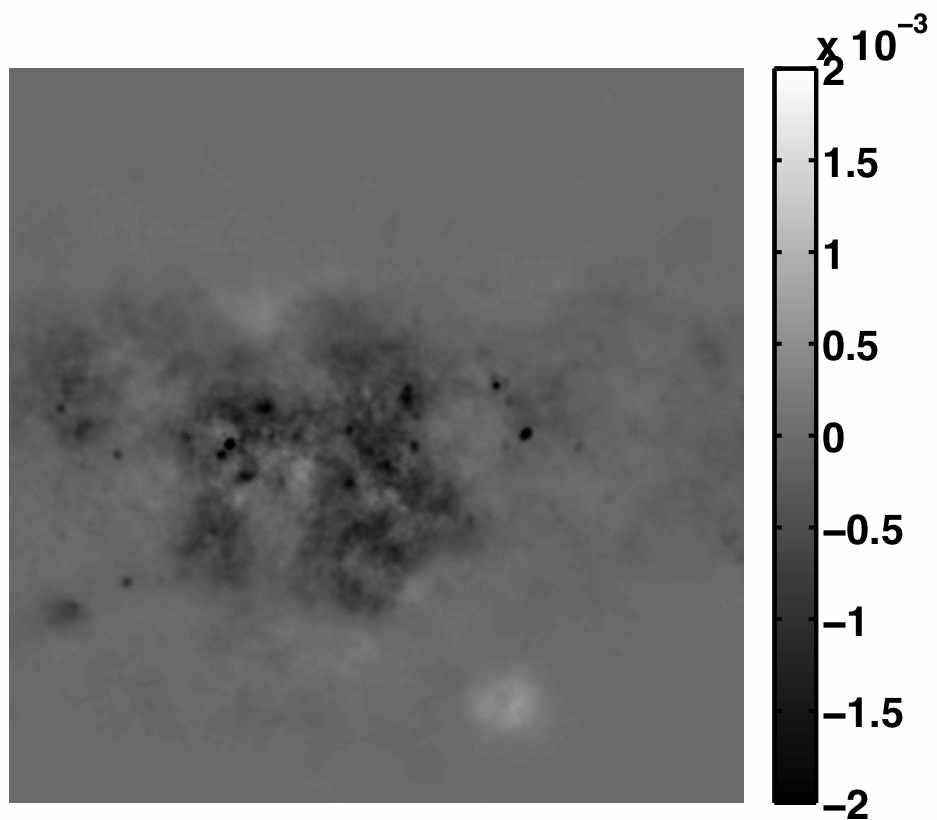}}
\centerline{\includegraphics [scale=0.23]{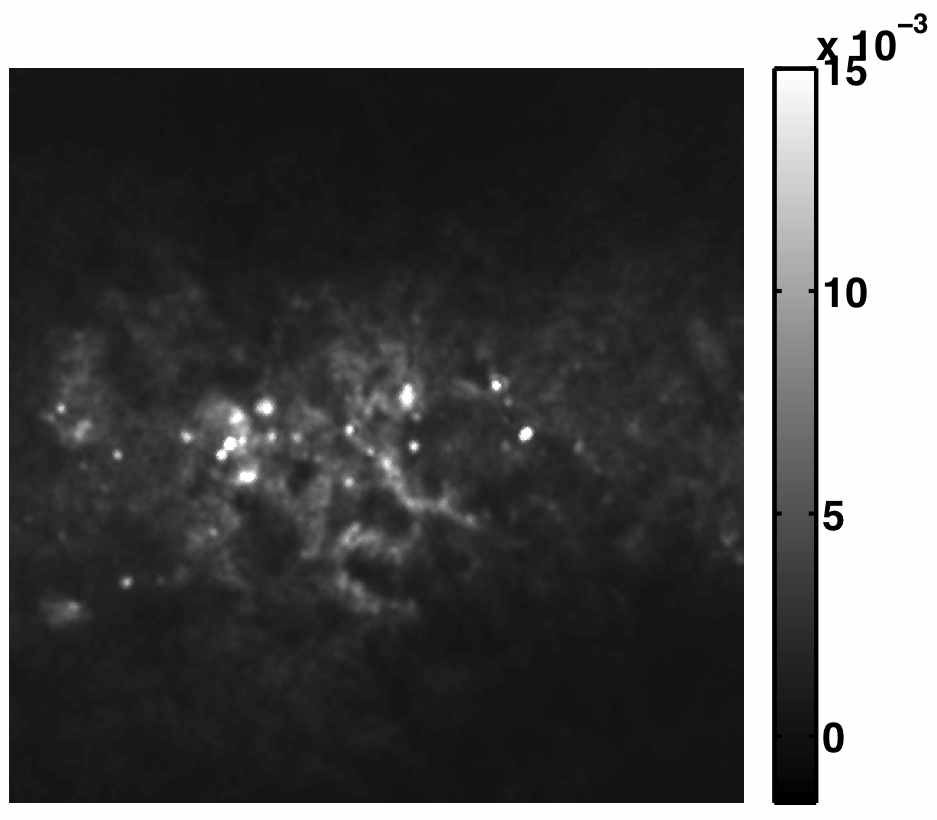} \includegraphics [scale=0.23]{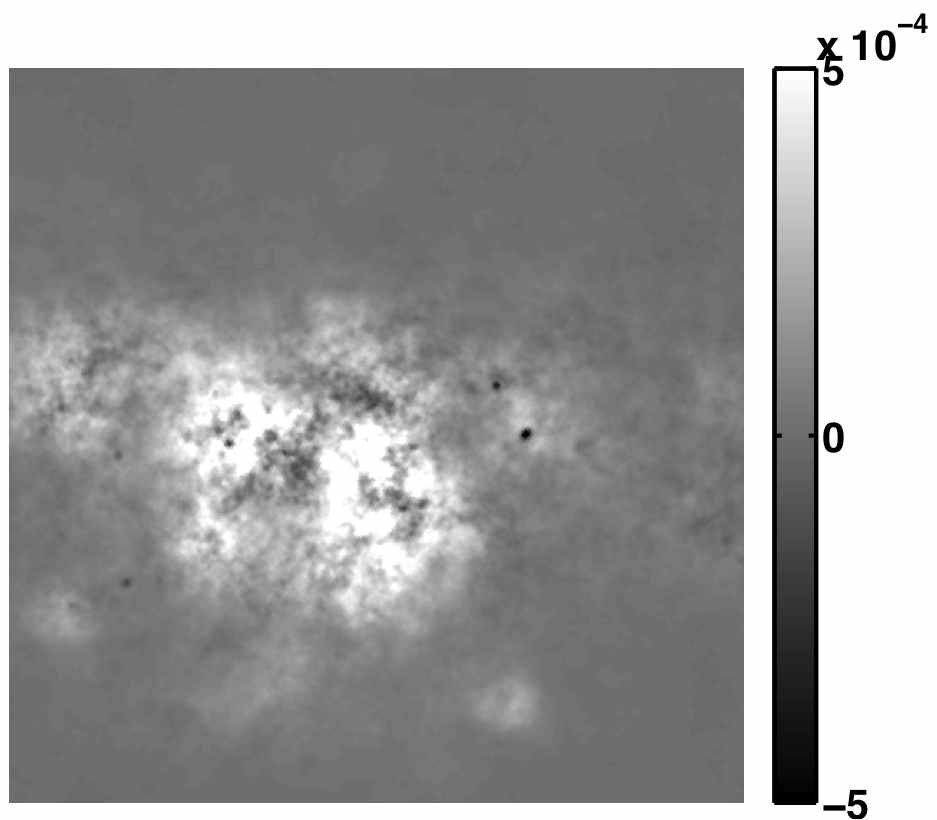}}
\caption{Left : estimated EFICA sources. Right: estimation error defined by the difference between the estimated and original sources. From top to bottom: free-free, spinning dust, synchrotron and thermal dust.}
\label{fig:wmap_estimated_efica}
\end{figure}

\begin{figure}[tb]
\centerline{\includegraphics [scale=0.23]{Efica_FF.jpg} \includegraphics [scale=0.23]{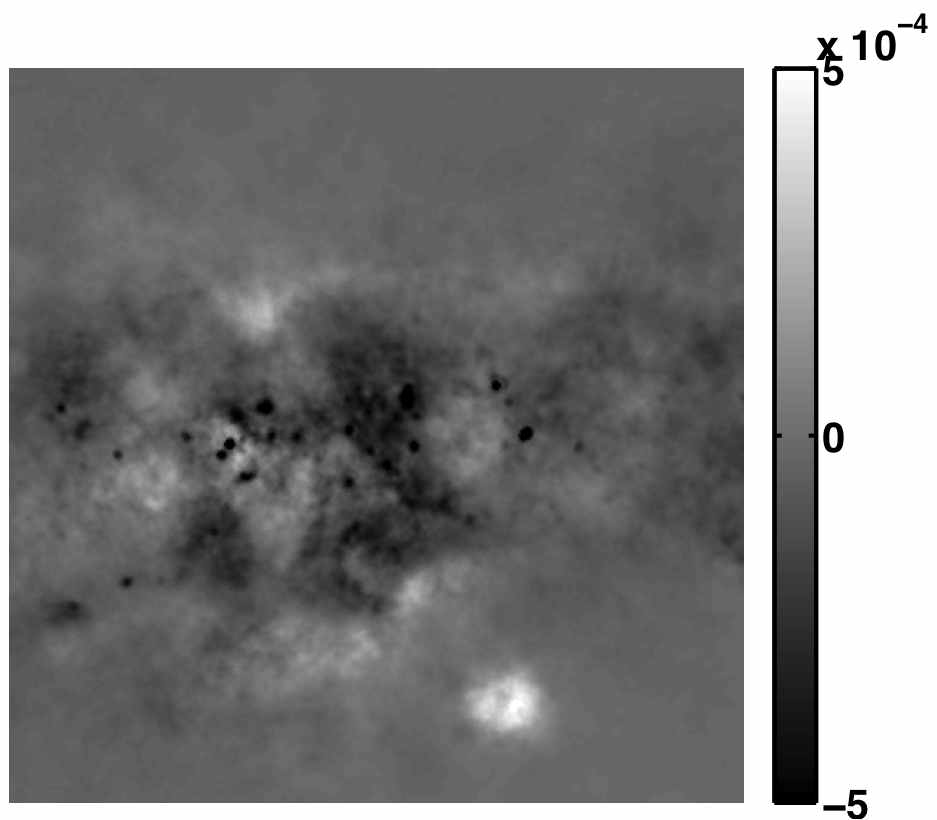}}
\centerline{\includegraphics [scale=0.23]{Efica_SpinningDust.jpg} \includegraphics [scale=0.23]{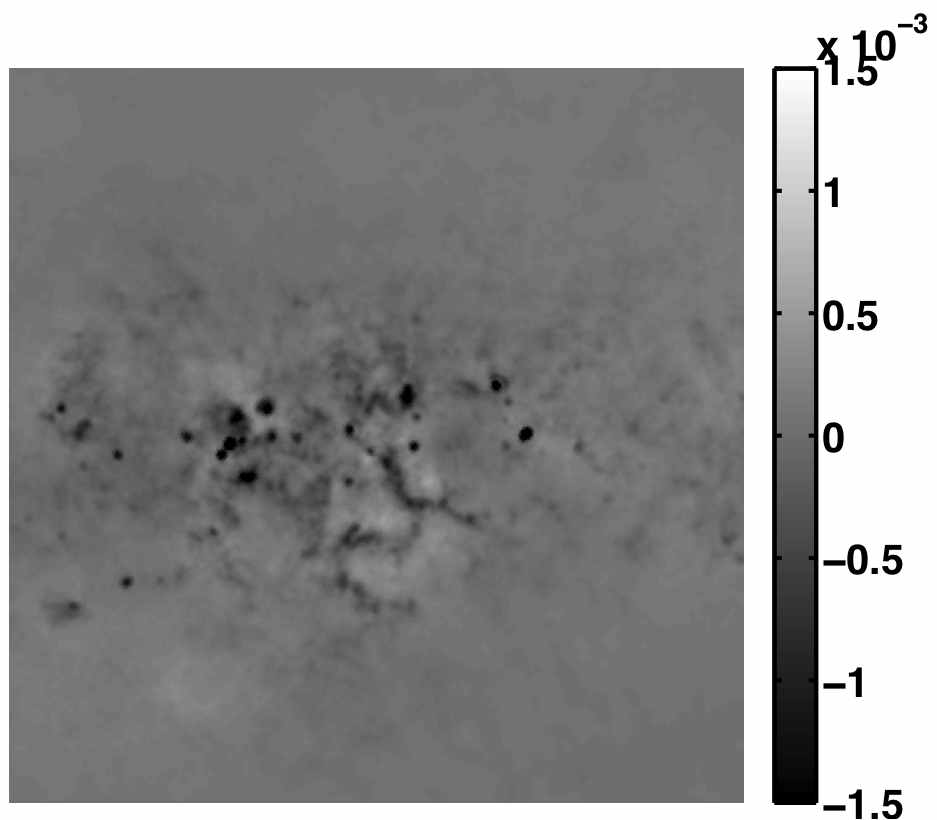}}
\centerline{\includegraphics [scale=0.23]{Efica_Synchrotron.jpg} \includegraphics [scale=0.23]{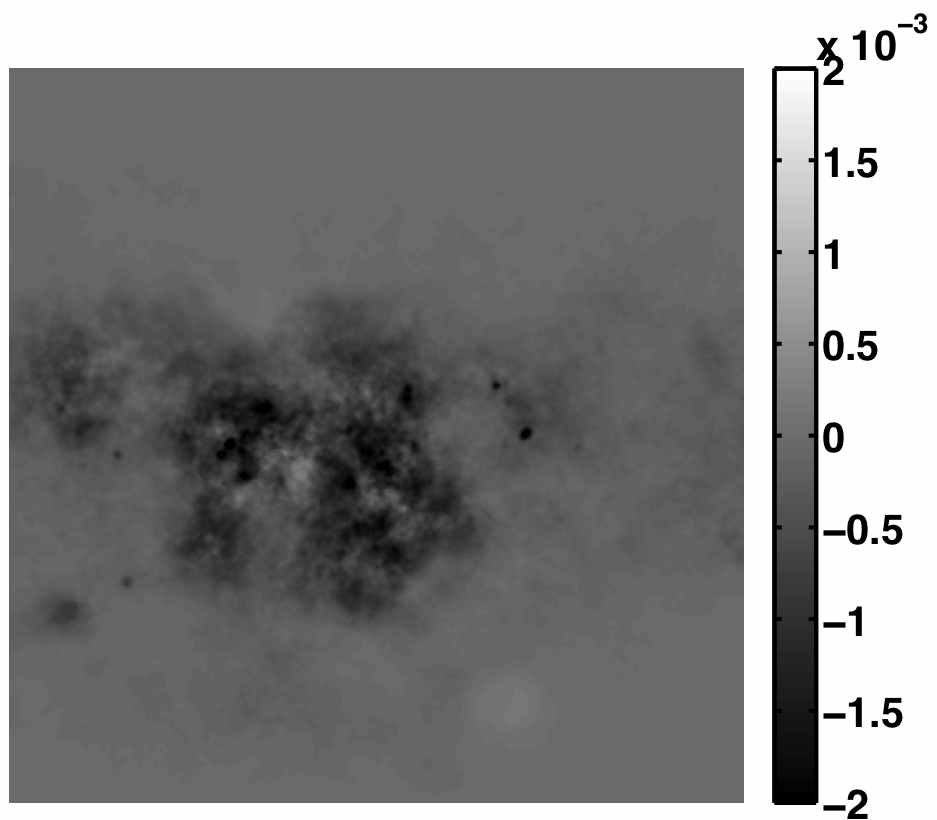}}
\centerline{\includegraphics [scale=0.23]{Efica_ThermalDust.jpg} \includegraphics [scale=0.23]{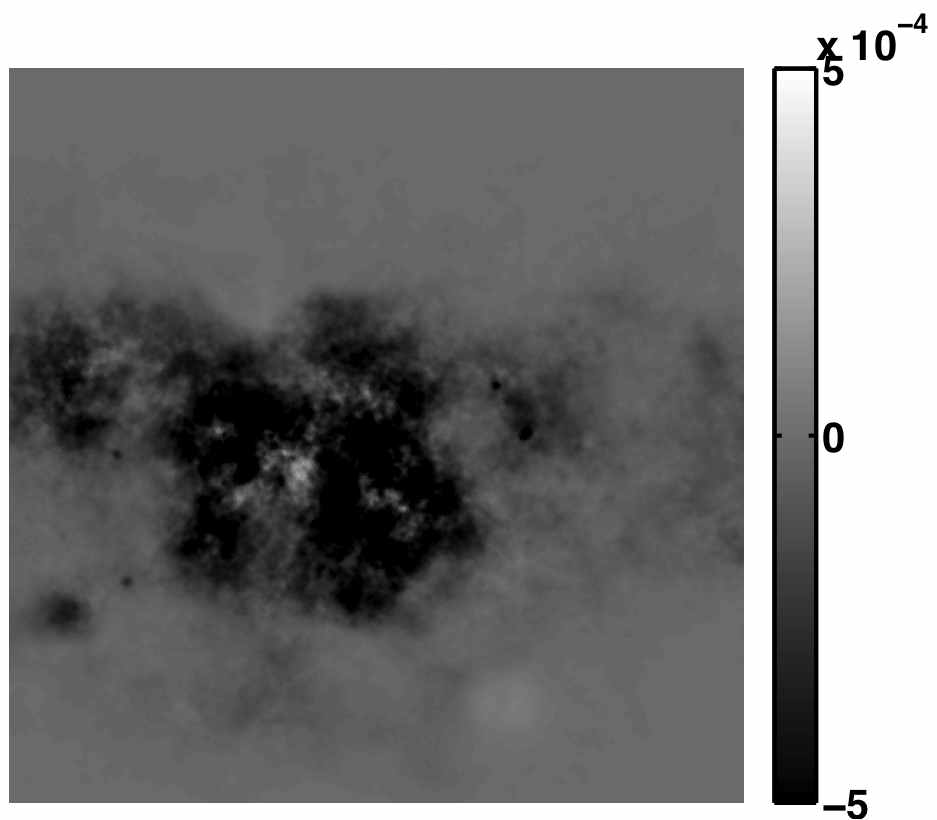}}
\caption{Left : estimated RNA sources. Right: estimation error defined by the difference between the estimated and original sources. From top to bottom: free-free, spinning dust, synchrotron and thermal dust.}
\label{fig:wmap_estimated_RNA}
\end{figure}

\begin{figure}[tb]
\centerline{\includegraphics [scale=0.23]{Efica_FF.jpg} \includegraphics [scale=0.23]{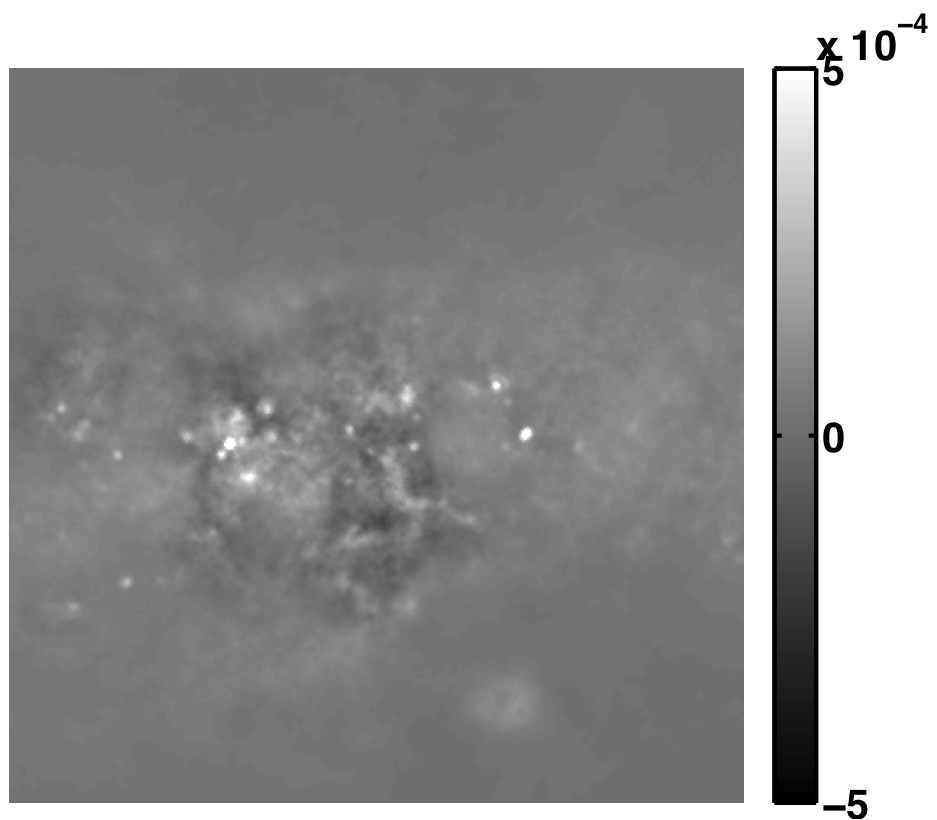}}
\centerline{\includegraphics [scale=0.23]{Efica_SpinningDust.jpg} \includegraphics [scale=0.23]{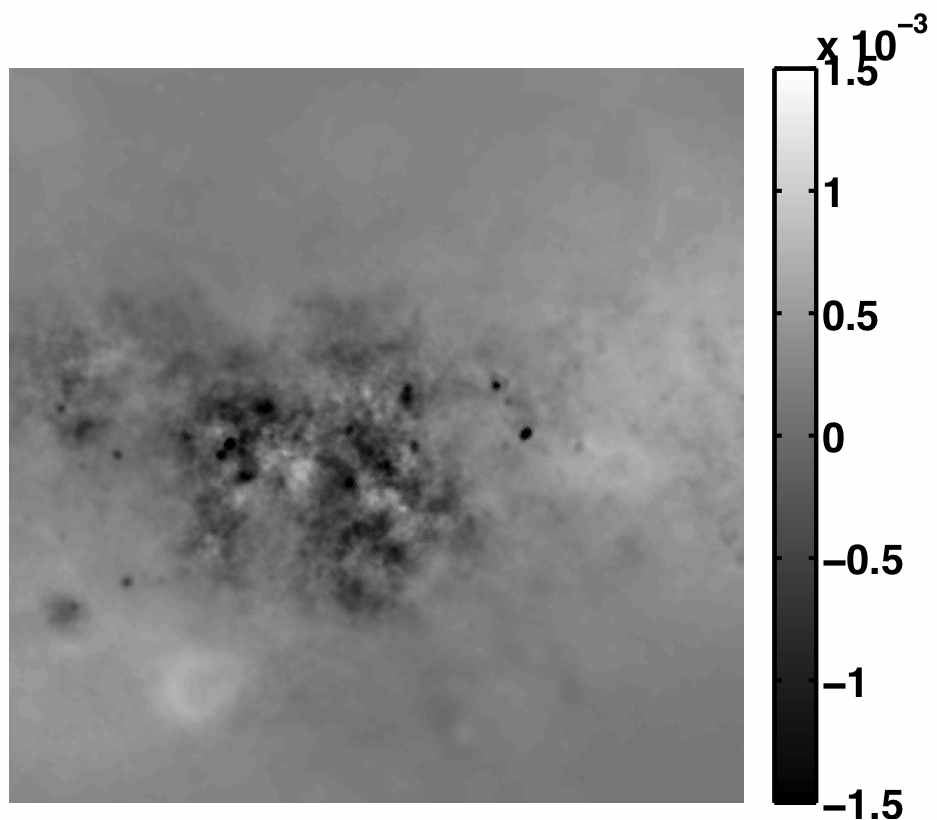}}
\centerline{\includegraphics [scale=0.23]{Efica_Synchrotron.jpg} \includegraphics [scale=0.23]{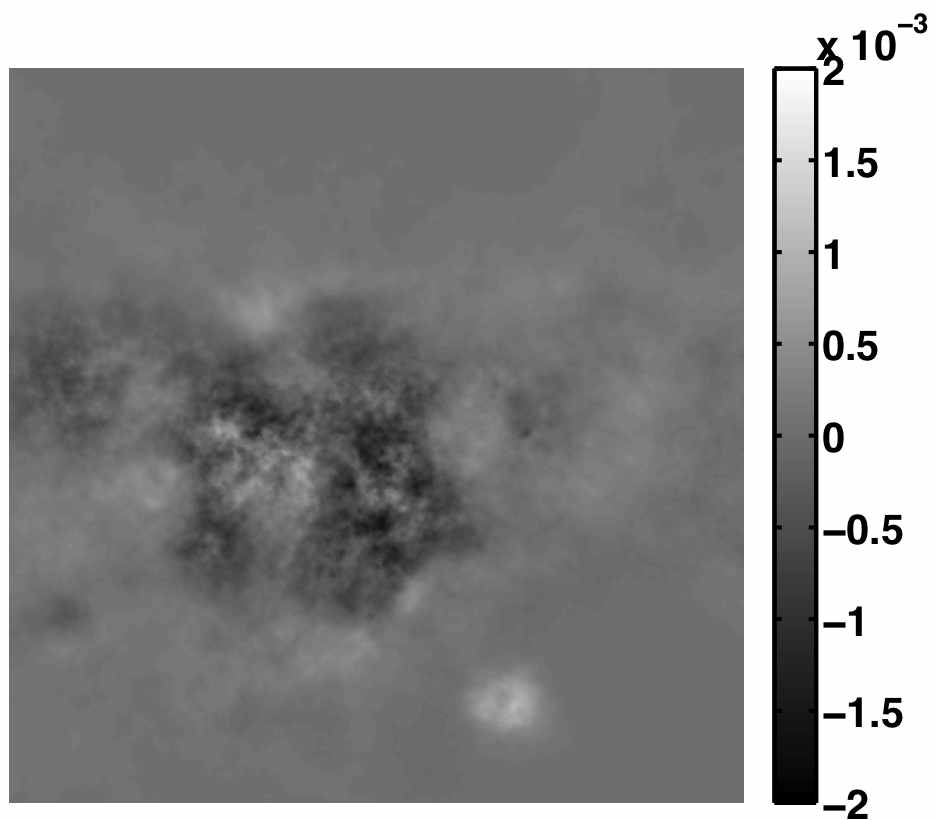}}
\centerline{\includegraphics [scale=0.23]{Efica_ThermalDust.jpg} \includegraphics [scale=0.23]{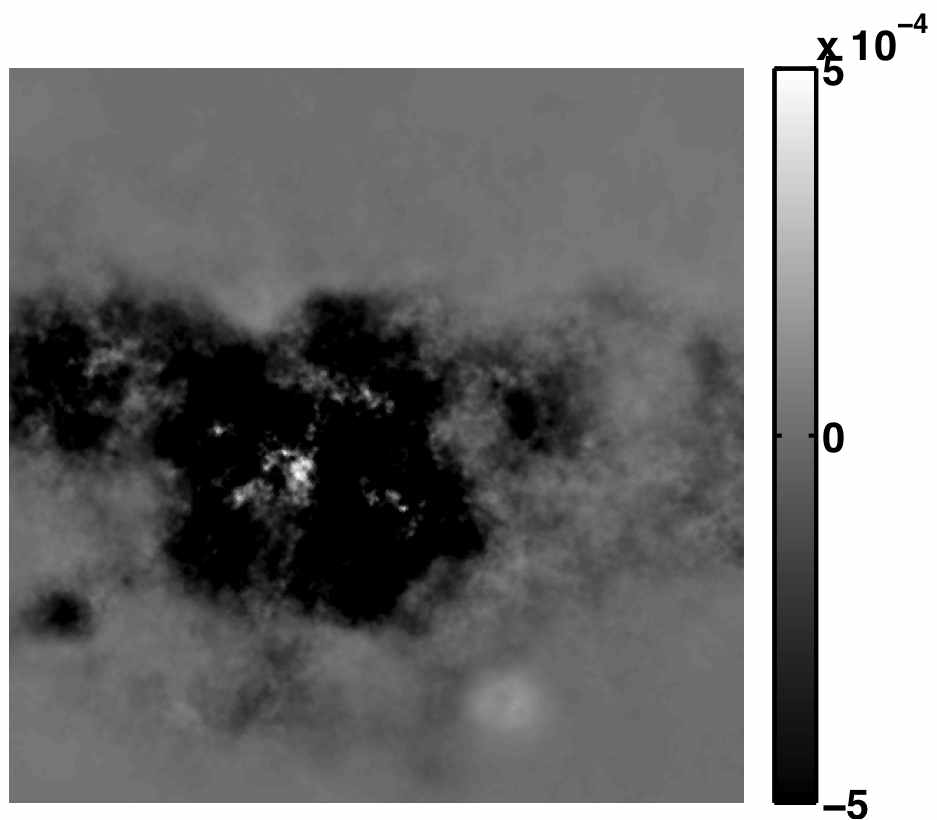}}
\caption{Left : estimated GMCA sources. Right: estimation error defined by the difference between the estimated and original sources. From top to bottom: free-free, spinning dust, synchrotron and thermal dust.}
\label{fig:wmap_estimated_GMCA}
\end{figure}

\begin{figure}[tb]
\centerline{\includegraphics [scale=0.23]{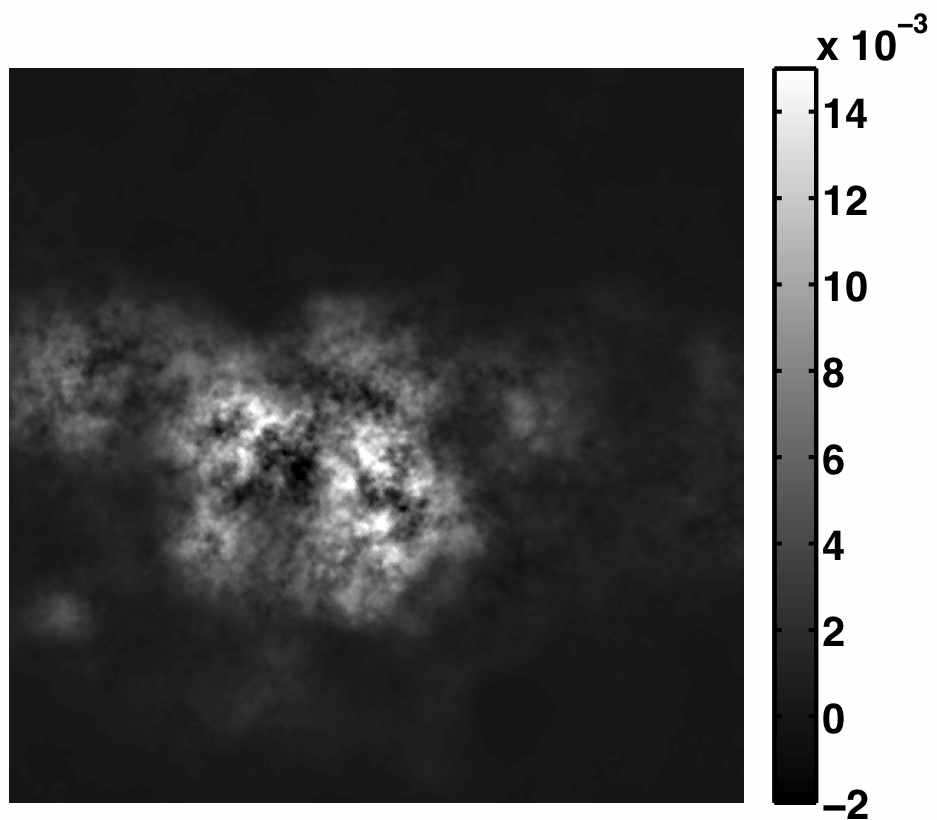} \includegraphics [scale=0.23]{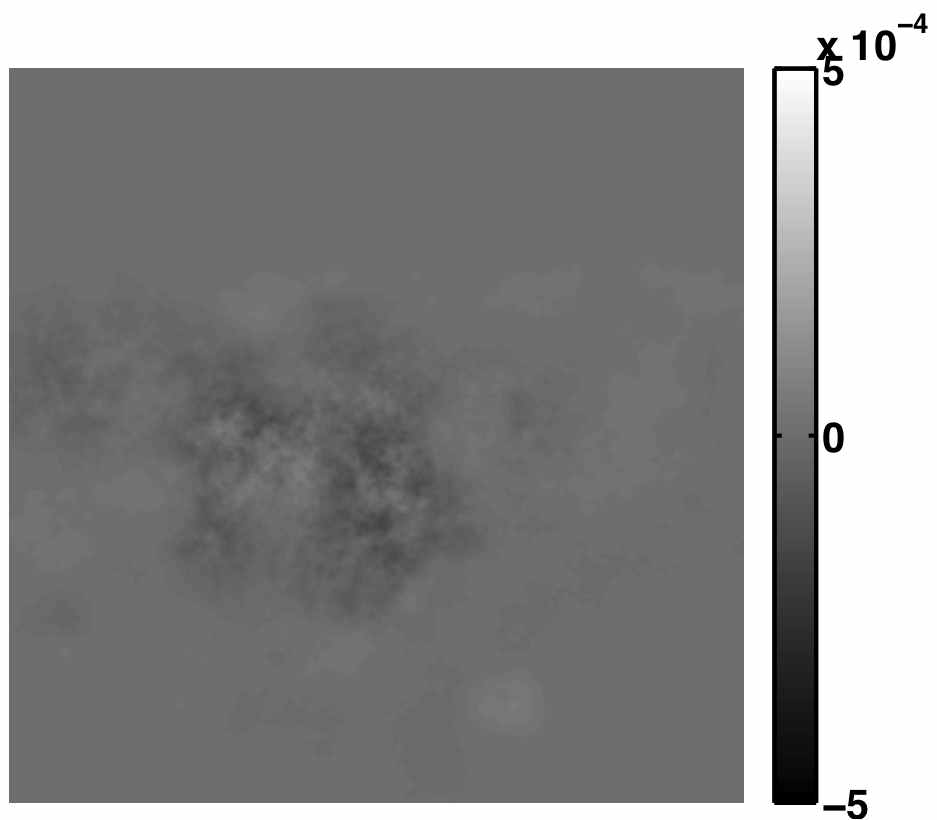}}
\centerline{\includegraphics [scale=0.23]{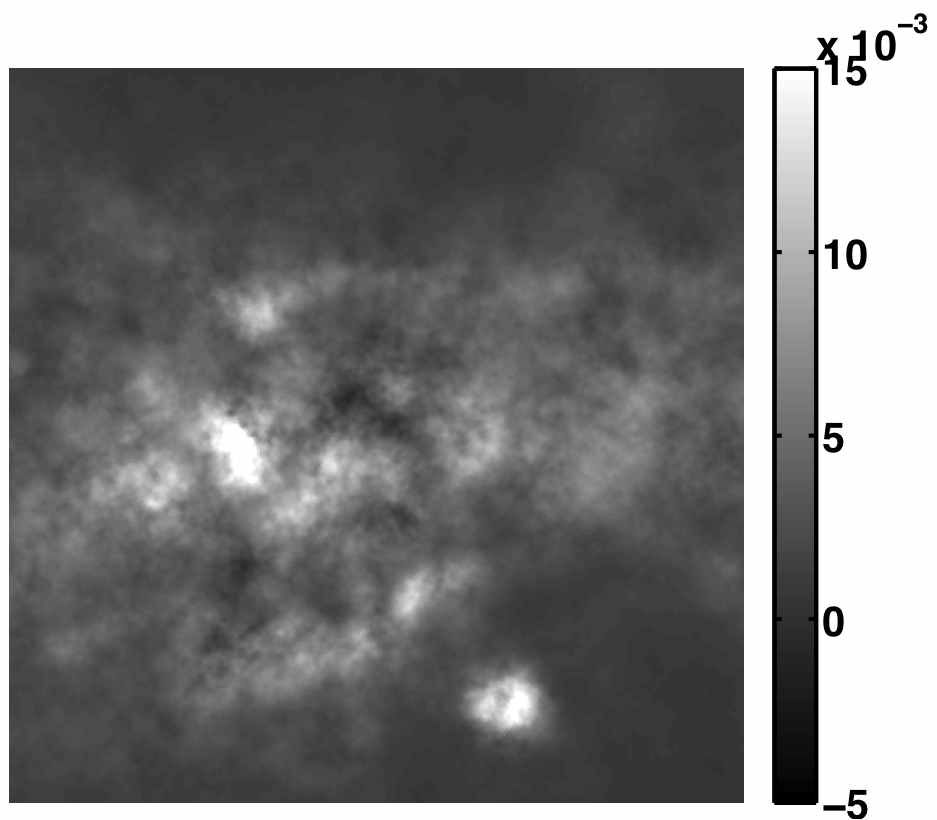} \includegraphics [scale=0.23]{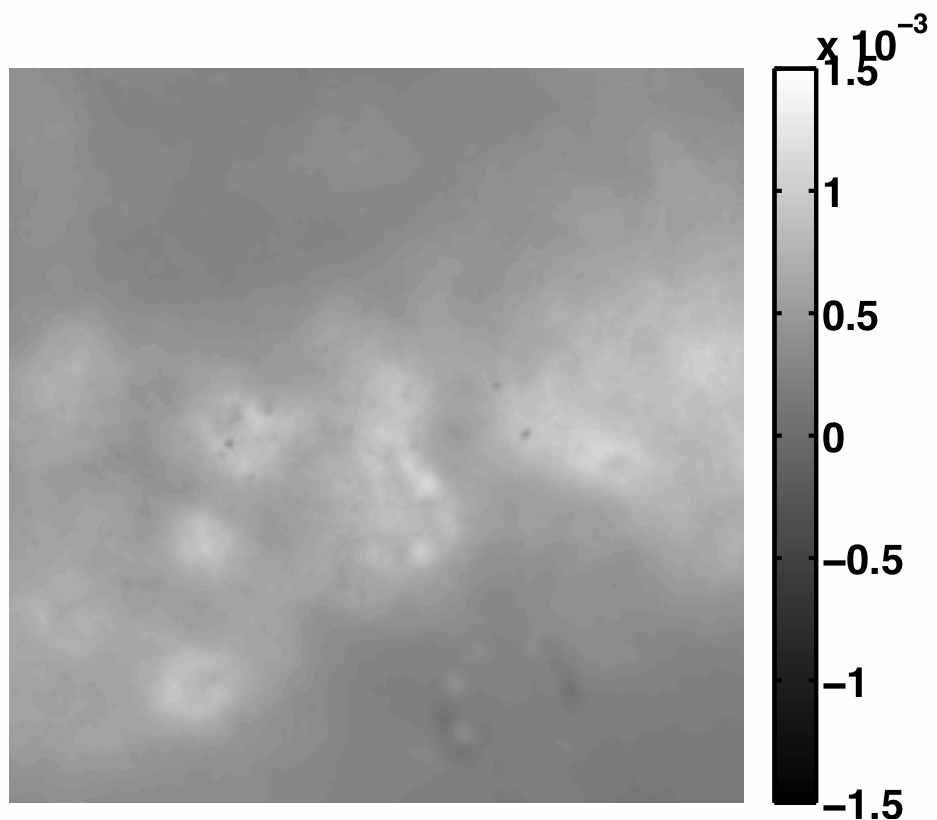}}
\centerline{\includegraphics [scale=0.23]{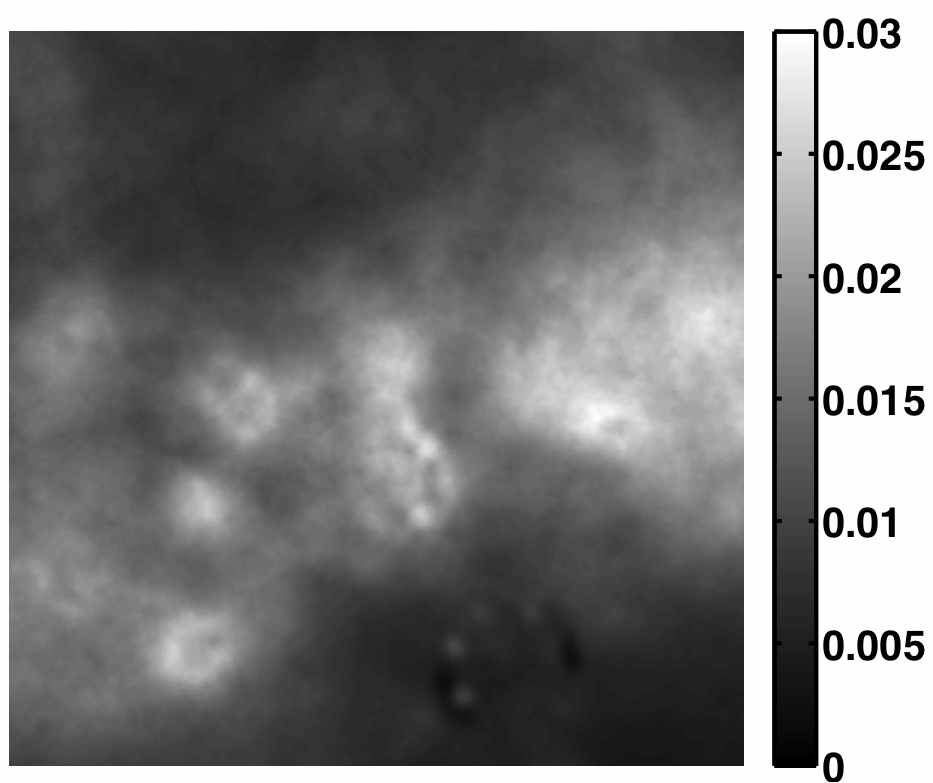} \includegraphics [scale=0.23]{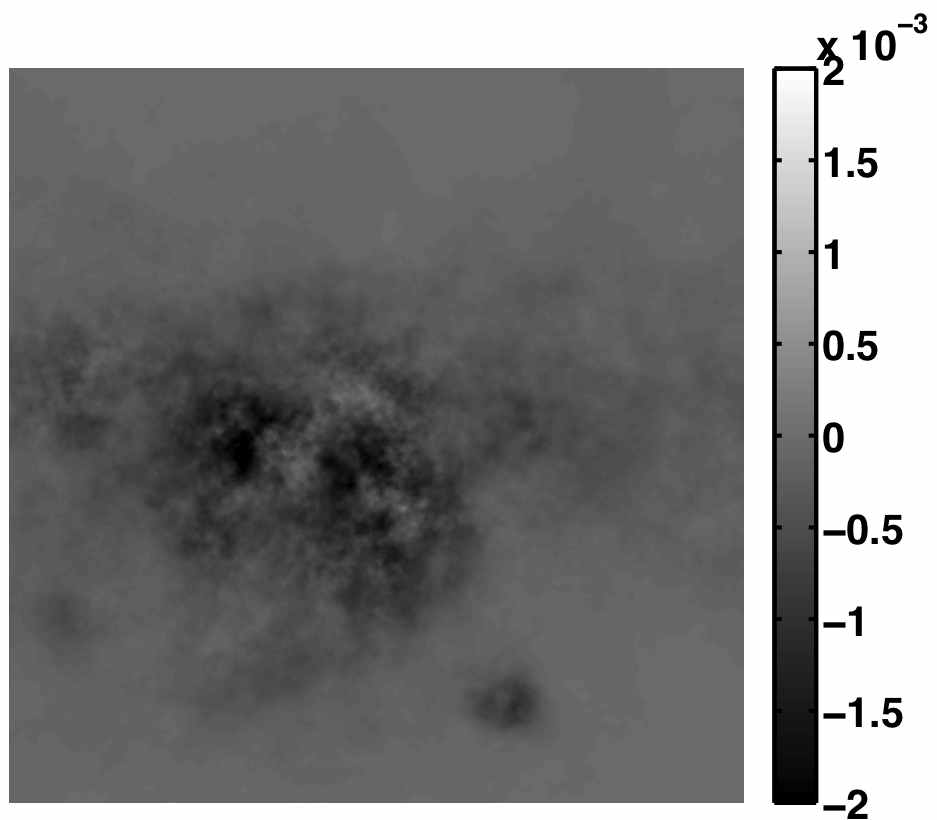}}
\centerline{\includegraphics [scale=0.23]{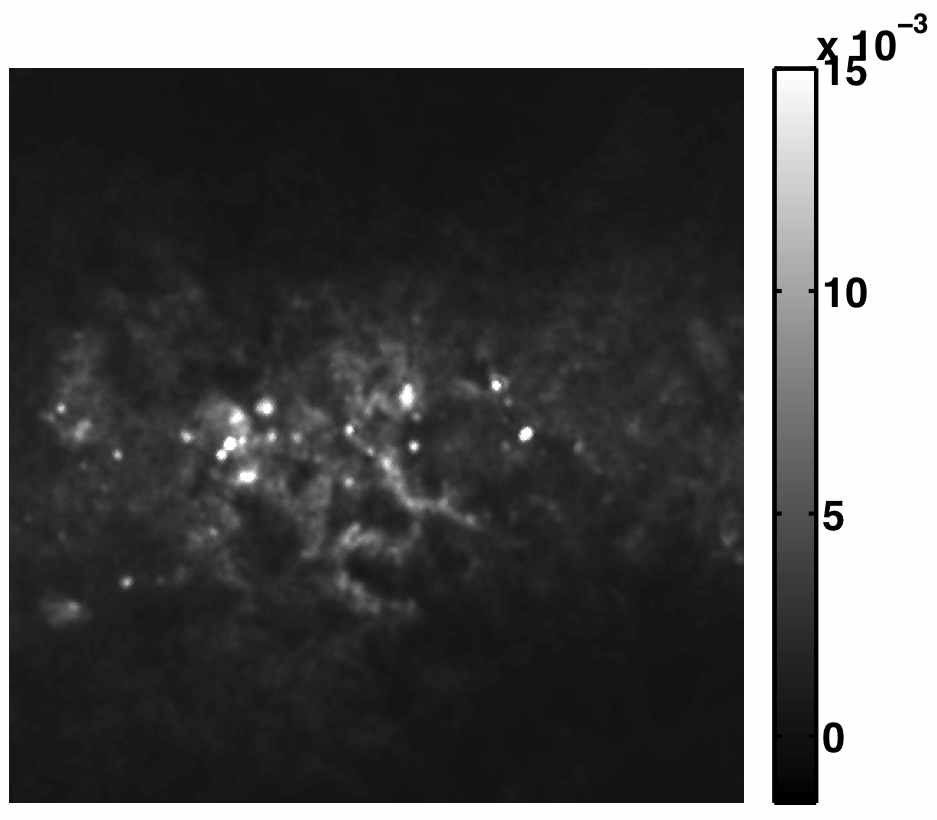} \includegraphics [scale=0.23]{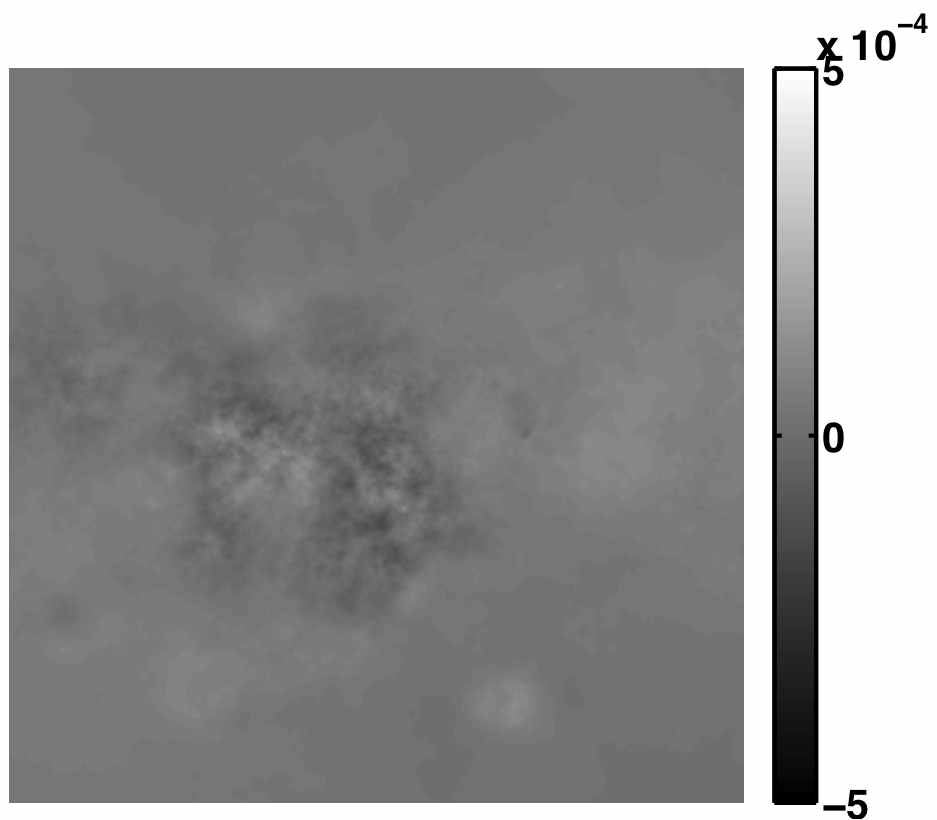}}
\caption{Left : estimated AMCA sources. Right: estimation error defined by the difference between the estimated and original sources. From top to bottom: free-free, spinning dust, synchrotron and thermal dust.}
\label{fig:wmap_estimated_amca}
\end{figure}

\begin{figure}[tb]
\centerline{\includegraphics [scale=0.27]{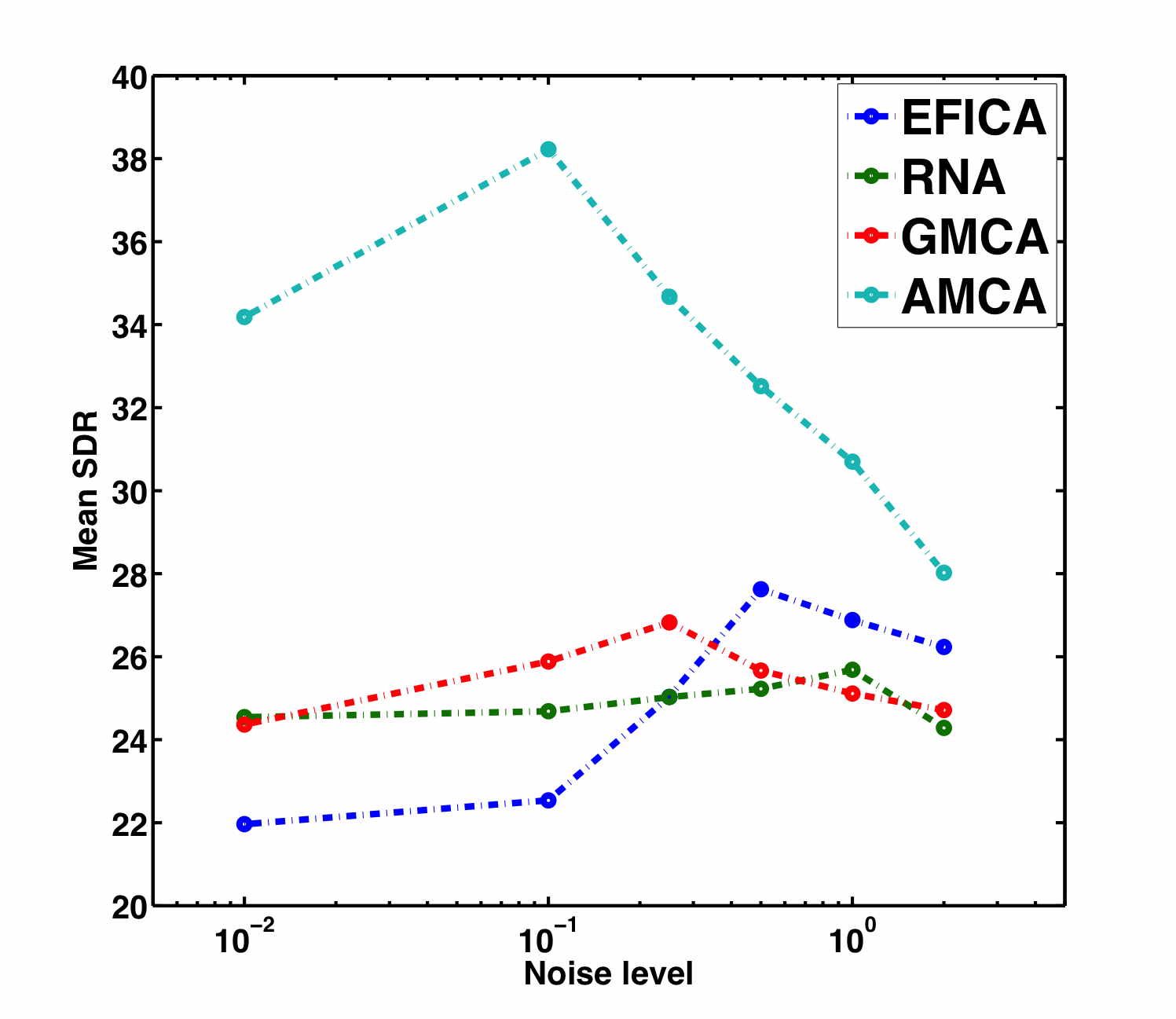}}
\hfill
\centerline{\includegraphics [scale=0.27]{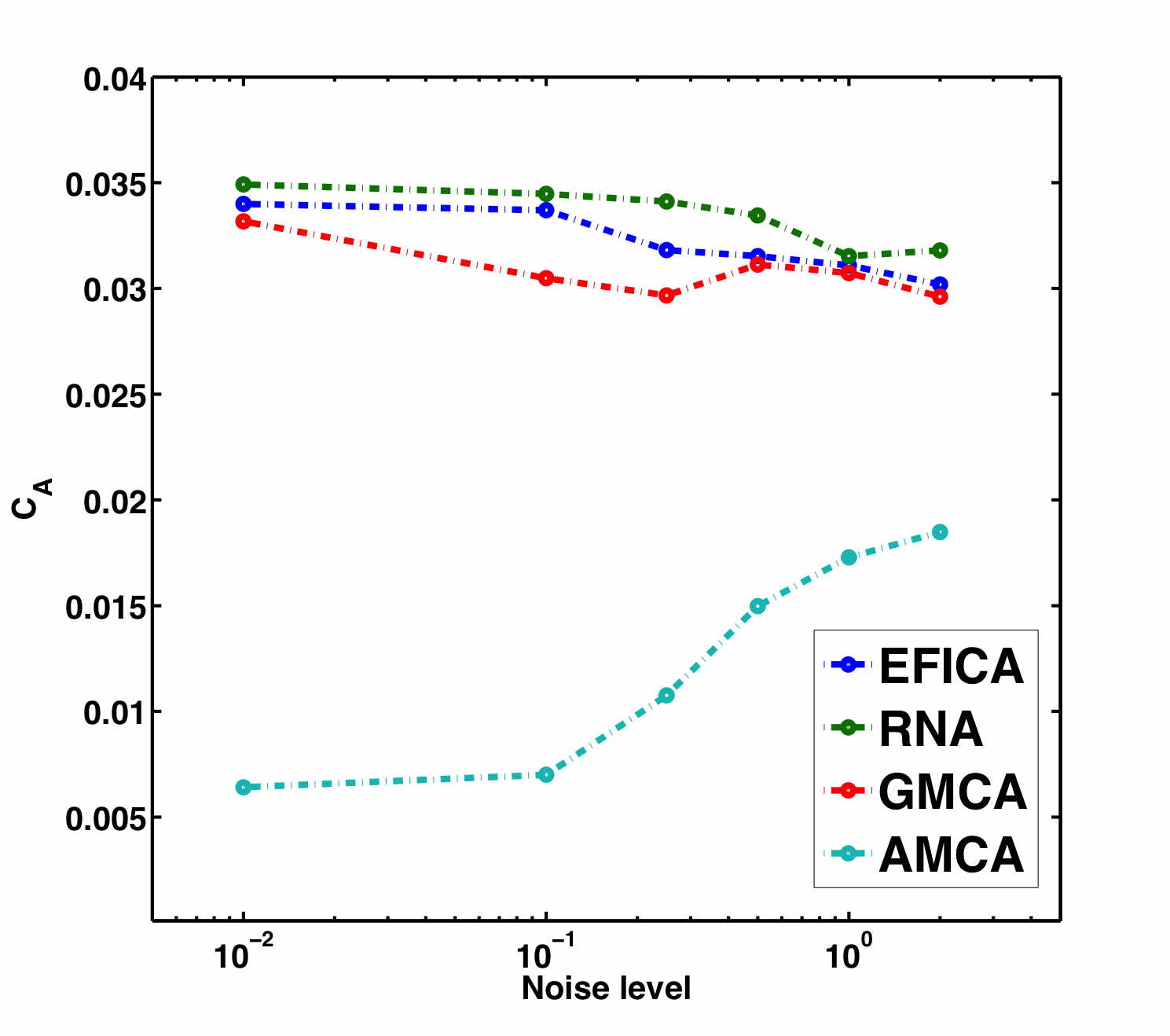}}
\caption{Evolution of the mean SDR (top) and the mixing matrix criterion (bottom) as a function of the noise level. Each sample is the average over $100$ Monte-Carlo simulations.}
\label{fig:wmap_CA}
\label{fig:wmap_SDR}
\end{figure}

Multi-wavelength instruments and  sensors have long been used in Astrophysics. However, it is only recently that the design of modern and sophisticated data analysis methods, such as blind source separation, play a significant role in this field. The recent results of the ESA/Planck space mission in 2013 has shed light on the crucial role that blind source separation has played to provide a very accurate map of the Cosmological Microwave Background (CMB) from multi-wavelength microwave data \cite{PR1_compsep,LGMCA_PR1}. In this context, each observation measures the mixture of several astrophysical components including the CMB and galactic foreground emissions to only name two. Galactic emissions can further be decomposed into several physically relevant sources with different spectral contributions. As emphasized in the introduction, these galactic sources are emitted from similar regions of the sky which entails that these components are, by nature, partially correlated. Disentangling between the different galactic sources is a challenging task which has been rarely tackled using blind source separation so far. In the rest of this section, we focus on the estimation of the various galactic sources from CMB-free data -- the CMB is customarily removed from the data before studying galactic components. In the range of the microwave wavelengths observed by instruments such as WMAP \cite{WMAP9_1,WMAP9_LGMCA} and Planck, the most prominent galactic emissions are : the free-free emission, galactic synchrotron emission, spinning dust and thermal dust emissions -- see \cite{PSM12} for more details about astrophysical emissions in the range of microwave wavelengths and their simulations.\\
In this section, our goal is to study the impact of the partial correlation of astrophysical components on the performances of blind source separation methods. This study will therefore be carried out on simulations of the $5$ WMAP\cite{WMAP9_1} observations in the range $23 - 94$ Ghz generated by the PSM (Planck sky model\footnote{http://www.apc.univ-paris7.fr/~delabrou/PSM/psm.html} -- see \cite{PSM12}). Simulations of these four major galactic emissions at the vicinity of the galactic center are displayed in Figure~\ref{fig:wmap_inputsources}. In these simulations, for the sake of simplicity, the linear mixture model \eqref{eq:mmodel} holds true. As displayed in Figure~\ref{fig:wmap_inputsources}, the most prominent emissive areas of all four sources are located in similar regions of the sky. This makes them partially correlated, especially at large scale or low frequencies. The $5$ simulated WMAP observations are featured in Figure~\ref{fig:wmap_data}. The sources displayed in Figure \ref{fig:wmap_inputsources} have an approximately sparse distribution in a bidimensional translation-invariant wavelet transform \cite{SFM:unde}; this will be our choice for the signal representation $\bf \Phi$ in this section. All the sparse BSS algorithms used in the forthcoming experiments have been performed in the wavelet domain using default parameters.\\
In the context of astrophysics, the impact of noise is usually deemed an important contribution that can hamper the quality of separation. The performances of sparse BSS are therefore evaluated for different levels of noise using the SDR and the mixing matrix criterion $C_A$ defined above. The left column of Figure~\ref{fig:wmap_estimated_efica} ({\it resp.} \ref{fig:wmap_estimated_RNA} and \ref{fig:wmap_estimated_GMCA}) features the sources estimated by the EFICA algorithm ({\it resp.} RNA and GMCA) for a normalized noise level of $1$. The normalized noise level corresponds to the nominal level of the WMAP data. It corresponds to a signal-to-noise ratio of $62$dB. This value might seem very large but only large-scale features are emerging above the noise level for mild values of the SNR.\\
With the exception of the estimated free-free emission which shows some spurious point sources residuals that originally belong to the thermal dust emission, all the four sources seem to be correctly estimated. The column on the left of Figure ~\ref{fig:wmap_estimated_efica} ({\it resp.} \ref{fig:wmap_estimated_RNA} and \ref{fig:wmap_estimated_GMCA}) features the residual error defined by the difference between the estimated and original sources. Figure ~\ref{fig:wmap_estimated_amca} displays exactly the same pictures for the AMCA algorithm. The display levels of the residual errors are kept the same to better reveal the differences between the results from all tested algorithms. The level of estimation error provided by the AMCA algorithm seem to be significantly lower for the free-free, spinning dust and thermal dust emissions while tending to be slighter better for the synchrotron emission.\\
Figure~\ref{fig:wmap_SDR} shows the evolution of the SDR for the four tested sparse BSS methods when the normalized noise level varies from $0.01$ ({\it i.e.} the noise level is actually $100$ lower than for that real WMAP data) to $2$.  Each entry of this figure has been obtained as the median value of $100$ independent Monte-Carlo simulations. This figure shows that the performances of the standard sparse BSS methods seem to be quite similar at all levels of noise with values of SDR in the range $22-28$ dB. The AMCA algorithm provides estimates with a level of SDR that is almost $10$ dB higher in the low noise regime. We already emphasized that the evolution of the SDR yields a partial view of the performances of these algorithms when the noise level increases. In this regime, the mixing matrix criterion $C_A$ is better suited to evaluate the quality of separation independently of the level of noise that contaminates the estimates of the sources. The evolution of $C_A$ is provided in Figure~\ref{fig:wmap_CA}. This figure confirms that the standard sparse BSS algorithms perform in a rather similar manner with values of the mixing matrix criterion of about $0.03$. In the meantime, the AMCA algorithm performs slightly better, especially in the low noise regime, with a value of $C_A$ that is about $7$ times lower in this case. In this high noise level regime, only the large-scale features of the sources are detected. Since these features are likely to be at the origin of the partial correlation of the sources, all algorithms, including AMCA, are handicapped in this setting.\\

%------------------------------------------------------------------------------------
\section*{Software}\label{sec:software}
%------------------------------------------------------------------------------------
Following the philosophy of reproducible research \cite{Buckheit_95_WaveLabandReproducible}, the codes and algorithms introduced in this article will be available at \textit{http://www.cosmostat.org/GMCALab}.

%------------------------------------------------------------------------------------
\section{Conclusion}\label{sec:conclusion}
%------------------------------------------------------------------------------------
The blind separation of sparse and partially correlated (s.p.c.) sources is a challenging task which standard sparse BSS methods generally hardly tackle with success. In this article, we introduce a novel sparse blind source separation coined Adaptive Morphological Component Analysis which is designed to retrieve sparse and partially correlated sources. We emphasize that recovering s.p.c. sources can be approached by identifying and privileging the most discriminant entries of the sources. We then propose learning adaptively these entries via a re-weighting scheme jointly with the estimation of the sources and the mixing matrix yielding the proposed AMCA (Adaptive Morphological Component Analysis) algorithm. We further evaluate the performances of the AMCA algorithm with respect to standard sparse BSS methods in various experimental scenarios using Monte-Carlo simulations of s.p.c. sources. The AMCA algorithm is shown to be much less sensitive to partial correlations of the sources even when the sources are highly correlated ({\it i.e.} when about $80\%$ of the active samples are common to all the sources). The proposed algorithm is slightly impacted by the dynamic of the amplitudes of correlated and independent entries. It is competitive when the number of sources to be retrieved is large. Finally, we applied the AMCA algorithm to the separation of astrophysical components from simulated microwave data. This application illustrates that the AMCA algorithm is well suited to estimate physical components which are, by nature, partially correlated. 

%------------------------------------------------------------------------------------
\section*{Acknowledgment}
%------------------------------------------------------------------------------------
This work was supported by the French National Agency for Research (ANR) 11-ASTR-034-02-MultID.

% -------------------------------------------------------------------------
\bibliographystyle{IEEEtran}
\bibliography{ASD}
%------------------------------------------------------------------------------------

\end{document}